\documentclass[11pt]{article}
\usepackage{epsfig}
\usepackage{axodraw}
\usepackage{epsfig}
\usepackage{graphicx}
\usepackage{rotate}
\usepackage{latexsym}
\usepackage{amssymb}
\newcommand\pubnumber{}
\newcommand\pubdate{September 19, 2002}
\newcommand\hepnumber{hep-ph/0209219}

\def\csuma{Department of Physics, New York University, U.S.A.}
\def\csumb{Dipartimento di Fisica G.~Galilei, Universit\`a di Padova, Italy\\
INFN, Sezione di Padova, Italy}
\def\csumc{Dipartimento di Fisica Teorica, Universit\`a di Torino, Italy\\
INFN, Sezione di Torino, Italy}
\def\support{\footnote{Work supported by the
European Union under contract HPRN-CT-2000-00149 and by MIUR under contract
2001023713${}_{}$006.}}
\def\change{\footnote{From October 1 2002: Fakult\"at f\"ur Physik,
Albert-Ludwigs Universit\"at, Freiburg, Germany.}}
\def\Title#1{\begin{center} {\LARGE\bf #1 } \end{center}}
\def\Author#1{\begin{center}{ \sc #1} \end{center}}
\newcommand{\Authors}[2]{\begin{center}{ \sc #1 \hspace{0.1cm} {\rm and}
\hspace{0.1cm} #2} \end{center}}
\def\Address#1{\begin{center}{ \it #1} \end{center}}

\newcommand\pubblock{\rightline{\begin{tabular}{l} \pubnumber\\
         \pubdate\\ \hepnumber \end{tabular}}}
\newenvironment{Abstract}{\begin{quotation}  }{\end{quotation}}

\def\Acknowledgments{\bigskip  \bigskip \begin{center}
          \large\bf Acknowledgments\end{center}}

\makeatletter
\def\section{\@startsection{section}{0}{\z@}{5.5ex plus .5ex minus
 1.5ex}{2.3ex plus .2ex}{\large\bf}}
\def\subsection{\@startsection{subsection}{1}{\z@}{3.5ex plus .5ex minus
 1.5ex}{1.3ex plus .2ex}{\normalsize\bf}}
\def\subsubsection{\@startsection{subsubsection}{2}{\z@}{-3.5ex plus
-1ex minus  -.2ex}{2.3ex plus .2ex}{\normalsize\sl}}

\renewcommand{\@makecaption}[2]{%
   \vskip 10pt
   \setbox\@tempboxa\hbox{\small #1: #2}
   \ifdim \wd\@tempboxa >\hsize     
       \small #1: #2\par          
     \else                        
       \hbox to\hsize{\hfil\box\@tempboxa\hfil}
   \fi}

 \def\citenum#1{{\def\@cite##1##2{##1}\cite{#1}}}
\def\citea#1{\@cite{#1}{}}

\newcount\@tempcntc
\def\@citex[#1]#2{\if@filesw\immediate\write\@auxout{\string\citation{#2}}\fi
  \@tempcnta\z@\@tempcntb\m@ne\def\@citea{}\@cite{\@for\@citeb:=#2\do
    {\@ifundefined
       {b@\@citeb}{\@citeo\@tempcntb\m@ne\@citea\def\@citea{,}{\bf }\@warning
       {Citation `\@citeb' on page \thepage \space undefined}}%
    {\setbox\z@\hbox{\global\@tempcntc0\csname b@\@citeb\endcsname\relax}%
     \ifnum\@tempcntc=\z@ \@citeo\@tempcntb\m@ne
       \@citea\def\@citea{,}\hbox{\csname b@\@citeb\endcsname}%
     \else
      \advance\@tempcntb\@ne
      \ifnum\@tempcntb=\@tempcntc
      \else\advance\@tempcntb\m@ne\@citeo
      \@tempcnta\@tempcntc\@tempcntb\@tempcntc\fi\fi}}\@citeo}{#1}}
\def\@citeo{\ifnum\@tempcnta>\@tempcntb\else\@citea\def\@citea{,}%
  \ifnum\@tempcnta=\@tempcntb\the\@tempcnta\else
  {\advance\@tempcnta\@ne\ifnum\@tempcnta=\@tempcntb \else\def\@citea{--}\fi
    \advance\@tempcnta\m@ne\the\@tempcnta\@citea\the\@tempcntb}\fi\fi}
\makeatother

\newcommand{\nl}{\nonumber\\}

\newcommand{\ds}{\displaystyle}

\newcommand{\lpar}{\left(}                            
\newcommand{\rpar}{\right)}

\newcommand{\bq}{\begin{equation}}                    
\newcommand{\eq}{\end{equation}}
\newcommand{\bqa}{\arraycolsep 0.14em\begin{eqnarray}}
\newcommand{\eqa}{\end{eqnarray}}
\newcommand{\ba}[1]{\begin{array}{#1}}
\newcommand{\ea}{\end{array}}
\newcommand{\ben}{\begin{enumerate}}
\newcommand{\een}{\end{enumerate}}
\newcommand{\bei}{\begin{itemize}}
\newcommand{\eei}{\end{itemize}}
\newcommand{\eqn}[1]{Eq.(\ref{#1})}
\newcommand{\eqns}[2]{Eqs.(\ref{#1})--(\ref{#2})}

\newcommand{\fig}[1]{Fig.~\ref{#1}}

\newcommand{\sect}[1]{Section~\ref{#1}}
\newcommand{\sects}[2]{Section~\ref{#1} and \ref{#2}}

\def\Re{\mathop{\operator@font Re}\nolimits}
\def\Im{\mathop{\operator@font Im}\nolimits}
\newcommand{\ord}[1]{{\cal O}\lpar#1\rpar}

\newcommand{\asums}[1]{\sum_{#1}}

\newcommand{\mone}{m_1}
\newcommand{\mtwo}{m_2}
\newcommand{\mtre}{m_3}
\newcommand{\mfor}{m_4}
\newcommand{\mfiv}{m_5}

\newcommand{\spro}[2]{{#1}\cdot{#2}}
\newcommand{\li}[2]{\mathrm{Li}_{#1}\lpar\displaystyle{#2}\rpar} 

\newcommand{\egam}[1]{\Gamma\lpar#1\rpar}               

\newcommand{\intfx}[1]{\int_{\scriptstyle 0}^{\scriptstyle 1}\,d#1}
\newcommand{\intfxy}[2]{\int_{\scriptstyle 0}^{\scriptstyle 1}\,d#1\,
                        \int_{\scriptstyle 0}^{\scriptstyle #1}\,d#2}

\newcommand{\sign}[1]{{\rm{sign}}\lpar{#1}\rpar}
\newcommand{\lpi}{\ln\pi}

\newcommand{\Ddr}{{\ds\frac{1}{{\bar{\varepsilon}}}}}

\newcommand{\dre}{\varepsilon}

\newcommand{\ep}{\epsilon}

\newcommand{\Reb}{{\rm{Re}}}

\newcommand{\pmom}{p}

\newcommand{\pone}{p_1}
\newcommand{\ptwo}{p_2}
\newcommand{\ptre}{p_3}
\newcommand{\pfor}{p_4}

\newcommand{\upar}[1]{u}

\newcommand{\ssA}{{\scriptscriptstyle{A}}}
\newcommand{\ssB}{{\scriptscriptstyle{B}}}

\newcommand{\ssG}{{\scriptscriptstyle{G}}}

\newcommand{\ssI}{{\scriptscriptstyle{I}}}

\newcommand{\ssL}{{\scriptscriptstyle{L}}}

\newcommand{\ssN}{{\scriptscriptstyle{N}}}

\newcommand{\ssR}{{\scriptscriptstyle{R}}}
\newcommand{\ssS}{{\scriptscriptstyle{S}}}
\newcommand{\ssT}{{\scriptscriptstyle{T}}}
\newcommand{\ssU}{{\scriptscriptstyle{U}}}

\newcommand{\bqas}{\begin{eqnarray*}}
\newcommand{\eqas}{\end{eqnarray*}}

\newcommand{\eilc}{\gamma}

\def\app#1#2 {{\it Acta. Phys. Pol.} {\bf#1},#2}
\def\cpc#1#2 {{\it Computer Phys. Comm.} {\bf#1},#2}
\def\np#1#2 {{\it Nucl. Phys.} {\bf#1},#2}
\def\pl#1#2 {{\it Phys. Lett.} {\bf#1},#2}
\def\prep#1#2 {{\it Phys. Rep.} {\bf#1},#2}
\def\prev#1#2 {{\it Phys. Rev.} {\bf#1},#2}
\def\prl#1#2 {{\it Phys. Rev. Lett.} {\bf#1},#2}
\def\zp#1#2 {{\it Zeit. Phys.} {\bf#1},#2}
\def\sptp#1#2 {{\it Suppl. Prog. Theor. Phys.} {\bf#1},#2}
\def\mpl#1#2 {{\it Modern Phys. Lett.} {\bf#1},#2}
\def\jetp#1#2 {{\it Sov. Phys. JETP} {\bf#1},#2}
\def\fpj#1#2 {{\it Fortschr. Phys.} {\bf#1},#2}
\def\afp#1#2 {{\it Acta.Phys. Polon.} {\bf#1},#2}
\def\err#1#2 {{\it Erratum} {\bf#1},#2}
\def\ijmp#1#2 {{\it Int. J. Mod. Phys} {\bf#1},#2}
\def\nc#1#2 {{\it Nuovo Cimento} {\bf#1},#2}
\def\ap#1#2 {{\it Ann. Phys.} {\bf#1},#2}
\def\cmp#1#2 {{\it Comm. Math. Phys.} {\bf#1},#2}
\def\el#1#2 {{\it Europhys. Lett.} {\bf#1},#2}
\def\hpa#1#2 {{\it Helv. Phys. Acta} {\bf#1},#2}
\def\yf#1#2 {{\it Yad. Fiz.} {\bf#1},#2}
\def\nim#1#2 {{\it Nucl. Instrum. Meth.} {\bf#1},#2}
\def\spz#1#2 {{\it Sov. Pisma Zhetf} {\bf#1},#2}
\def\jetpl#1#2 {{\it JETP Lett.} {\bf#1},#2}
\def\sjnp#1#2 {{\it Sov. J. Nucl. Phys.} {\bf#1},#2}
\def\ptp#1#2 {{\it Progr. Theor. Phys. (Kyoto)} {\bf#1},#2}
\def\rmp#1#2  {{\it Rev. Mod. Phys.} {\bf#1},#2}
\def\zhetf#1#2 {{\it ZhETF} {\bf#1},#2}
\def\prs#1#2 {{\it Proc. Roy. Soc.} {\bf#1},#2}
\def\phys#1#2 {{\it Physica} {\bf#1},#2}

\newcommand{\egams}[1]{\Gamma^2\lpar#1\rpar}               

\newcommand{\intfxx}[2]{\int_{\scriptstyle 0}^{\scriptstyle 1}\,d#1\,
                        \int_{\scriptstyle 0}^{\scriptstyle 1}\,d#2}
\def\bfi{\begin{figure}}
\def\efi{\end{figure}}

\newcommand{\hyper}[4]{{}_2F_1(#1\,,\,#2\,;\,#3\,;\,#4)}

\newcommand{\dsimp}[1]{\int\,dS_{#1}}
\newcommand{\dcub}[1]{\int\,dC_{#1}}

\begin{document}
\begin{titlepage}
\pubblock
\vfill
\def\thefootnote{\fnsymbol{footnote}}
\Title{All-Purpose Numerical Evaluation of \\[3mm]
One-Loop Multi-Leg Feynman Diagrams\support}
\vfill
\Author{Andrea Ferroglia\change}
\Address{\csuma}
\Author{Massimo Passera}
\Address{\csumb}
\Authors{Giampiero Passarino}{Sandro Uccirati}
\Address{\csumc}
\vfill
\begin{Abstract}
A detailed investigation is presented of a set of algorithms which form the
basis for a fast and reliable numerical integration of one-loop multi-leg
(up to six) Feynman diagrams, with special attention to the behavior around
(possibly) singular points in phase space. No particular restriction is 
imposed on kinematics, and complex masses (poles) are allowed.
\end{Abstract}
\vfill
\vfill
\begin{center}
\end{center}
\end{titlepage}
\def\thefootnote{\arabic{footnote}}
\setcounter{footnote}{0}
\small
\thispagestyle{empty}
\tableofcontents
\setcounter{page}{1}
\normalsize
\clearpage
\section{Introduction}
In this work we present a systematic approach to the numerical evaluation of
one-loop multi-leg Feynman diagrams. One may wonder why or what is new and
in what respect there is something to be gained after the seminal work of 
't Hooft and Veltman~\cite{'tHooft:1979xw}, where the analytic answer to the
same problem was given.

There are several reasons for attempting a systematization of the
problem.\footnote{Alternative approaches, based on differential equation
methods, have been originated in~\cite{Kotikov:1990kg}.}  First of all we
want a fast and reliable way of computing one-loop multi-leg Feynman
diagrams for all kinematical configurations, in particular around the
(possibly) singular points of the diagrams which are determined by the
solution of the corresponding Landau equations~\cite{Landau:1959fi}. For
each graph we have a leading Landau singularity, usually called anomalous
threshold for the triangle and higher functions, which for almost all 
practical situations lies outside the physical region. However, there are 
important examples where the external legs of the one-loop diagrams are not 
representing the in/out states of the process under examination and, for a 
general treatment, we must go beyond the class of $1 \to 2(3)$ or $2 \to 2$ 
processes, to a situation where classifying a priori the physical nature of 
the anomalous threshold is a tremendous task.

Furthermore we have sub-leading singularities which are the leading ones for
the reduced diagrams originating from the primary one~\cite{elop}.  In
principle, all of them are fully accounted for by the analytic answer but,
in practice, the corresponding numerical evaluation of the special functions
that form the answer can be rather unstable.  For complicated processes,
like full one-loop corrections to $2 \to 4(6)$ scatterings, we have a rich
phase space at our disposal and any automatized calculation must be designed
in such a way to foresee all possible sources of numerical problems without
having to stop the full evaluation process.

A particularly severe example is represented by the decomposition of
pentagons into a sum of boxes~\cite{vanNeerven:1983vr}. For a real
multi-scale problem these boxes are to be evaluated in regions where the
analytic result requires cumbersome analytic continuations~\cite{cac}. 
Our goal is to derive a numerical answer without having to worry about 
kinematics. Even more important, our approach does not suffer from the 
introduction of complex masses which, instead, is a burdensome aspect in the 
analytic way. At most some care is needed when we take into account that 
complex masses are to be understood as complex poles lying on the second 
Riemann sheet~\cite{Beenakker:1996kn}.

Another well known drawback of the standard analytic
approach~\cite{Passarino:1979jh} is the appearance of negative powers of
Gram's determinants in the reduction of tensor integrals, which represents 
a particular annoyance since their zeros do not correspond, under normal 
circumstances, to true singularities of Feynman diagrams. Clearly, in any
approach and after a certain number of symbolic manipulations, certain 
denominators will show up in the answer. It seems therefore reasonable to 
design an approach where the zeros of these denominators correspond to true 
singularities of the diagrams. In any case, a new strategy for the so called 
problem of reduction of tensor integrals is highly desirable.

Furthermore, in any realistic calculation infrared divergent (IR) virtual
corrections will emerge, and we have found it very attractive to use the
same tools that are needed to solve the problem of numerical evaluation of
multi-scale diagrams for classifying and computing residues of infrared
poles and IR finite parts of IR divergent diagrams. In this respect, the
interplay between the special class of IR singularities and the general one
of Landau singularities is essential.

In this work we make use of a recent proposal for numerical evaluation of
multi-loop integrals~\cite{Passarino:2001wv} and extend the work
of~\cite{Bardin:2000cf} to all one-loop diagrams (up to six external legs).
The global strategy is based on the use of the Bernstein-Tkachov
theorem~\cite{Tkachov:1997wh} which, away from the leading and sub-leading
Landau singularities, allows us to cast any one-loop integrand in a form
well suited for numerical treatment. As it will be shown in \sect{btt}, a
unique factor $B_{\ssG}$ is associated with each one-loop diagram
$G$.\footnote{This factor is proportional to the determinant of the
coefficients in the system of the corresponding Landau equations;
S.~Uccirati, PhD dissertation (2002).}  For each family of one-loop diagrams
we then consider three different cases: $B_{\ssG} \ne 0$, $B_{\ssG} = 0$ but
$G$ regular, and $B_{\ssG} = 0$ with $G$ singular. For the last case, the
method is based on a particular combination of Mellin-Barnes~\cite{ellip}
and sector decomposition~\cite{Binoth:2000ps} techniques, as a result of
which we are able to write a Laurent expansion for $G$ around the
singularity.  Alternatively, we have been able to derive new integral
representations that are particularly suited for numerical treatment around
the singularities.

To a large extent, the use of complex masses (poles) represents a
significant simplification of the whole approach. For physical (real) values
of the Mandelstam invariants describing the process, none of the $B_{\ssG}$
factors can be zero, and a straightforward iteration of the BT procedure is
all what we need.

To return to our original question on why devoting such an effort to the
evaluation of one-loop diagrams, we may add that the results of our work
must be seen as one of the many building blocks towards the computation of
physical observables at the two-loop level. Here our strategy has been
already outlined in~\cite{Passarino:2001wv}: full numerical analysis of the
two-loop content. Clearly we will have to include the one-loop part, and it
is rather obvious that the two pieces should be treated on equal
footing. Furthermore, our main emphasis has been on a general treatment,
therefore extending previous works on the same subject~\cite{Fujimoto:1991bm}.

We do not present numbers in this paper for a very simple reason: in all
cases where the analytic answer is known, we have found perfect agreement
with the one-loop library of TOPAZ0~\cite{Montagna:1993ai}.
The resulting comparison, basically a long list of identical numbers, is not
very illuminating.

The outline of the paper will be as follows: in \sect{btt} we introduce the
definitions and discuss the Bernstein-Tkachov theorem, the Mellin-Barnes
techniques, and provide a short discussion about the physical/unphysical
role of the Landau singularities.  In \sect{gtpf} the generalized two-point
function is introduced, while in \sect{C0} we start the presentation of the
one-loop, three-point, $C$-family,  and in \sect{IRC0} we discuss
its infrared divergent configurations.  The four-point, $D$-family, is
discussed in \sect{D0} and the corresponding infrared configurations are
presented in \sect{IRD0}.  The five- and six-point functions are analyzed in
Sections~\ref{E0}, \ref{IRE0} and \ref{F0}.
\section{Prolegomena\label{btt}}
\begin{itemize}
\item[--] The Bernstein-Tkachov theorem
\end{itemize}
The Bernstein-Tkachov theorem~\cite{Tkachov:1997wh} (hereafter BT) tells us
that for any finite set of polynomials $V_i(x)$, where $x = \,\lpar
x_1,\dots, x_k\rpar$ is a vector of Feynman parameters, there exists an 
identity of the following form:
\bq
{\cal P}\,\lpar x,\partial\rpar \prod_i\,V_i^{\mu_i+1}(x) = B\,
\prod_i\,V_i^{\mu_i}(x).
\label{functr}
\eq
where ${\cal P}$ is a polynomial of $x$ and $\partial_{j} =
\partial/\partial x_j$; $B$ and all coefficients of ${\cal P}$ are
polynomials of $\mu_i$ and of the coefficients of $V_i(x)$.  Any multi-loop
Feynman diagram can be cast into the form of \eqn{functr}.  Iterative
applications of the BT functional relations, followed by integration by
parts, allows us to express the integrand in parametric space as a
combination of (polynomials)${}^\omega\,\times\,$ logarithms of the same
polynomials, with any given integer $\omega\ge0$, therefore achieving a
result that is well suited for numerical integration. The $B$ coefficients
of \eqn{functr} are connected to the leading Landau singularity of the
corresponding diagram while repeated applications of \eqn{functr}, after
integration by parts, will bring the sub-leading ones into the final answer.

For general one-loop diagrams we have an explicit solution for the
polynomial ${\cal P}$ which is due to F.~V.~Tkachov~\cite{Tkachov:1997wh}.
Any one-loop Feynman diagram $G$, irrespective of the number $N$ of
vertices, can be expressed as
\bq
G = \int_{\ssS}\,dx\,V^{-\mu}(x),
\eq
where the integration region is $x_j \ge 0, \,\asums{j}\,x_j \le 1$, with
$j=1,\ldots,N-1$, and $V(x)$ is a quadratic polynomial in $x$,
\bq
V(x) = x^t\,H\,x + 2\,K^t\,x + L.
\label{defHKL}
\eq
The solution to the problem of determining the polynomial ${\cal P}$ is as
follows:
\bq
{\cal P} = 1 - {{\,\lpar x - X\rpar^t\,\partial_x} \over {2\,\lpar\mu+1\rpar}},
\qquad
X^t = -\,K^t\,H^{-1}, \qquad B = L - K^t\,H^{-1}\,K,
\label{rol}
\eq
where the matrix $H$ is symmetric.  
In our conventions a multi-leg function is specified by the number 
of its internal lines, without regard to the actual number of external lines
(see \fig{fig:threepoint} for an illustration).
\begin{itemize}
\item[--] Mellin-Barnes transform
\end{itemize}
Another tool that will be needed in the following is represented by the 
Mellin-Barnes~\cite{ellip} splitting~\cite{Boos:1990rg},
\bq
( Q + \lambda )^{-\alpha} = \frac{1}{2\,\pi\,i}\,
\int_{-i\,\infty}^{+i\,\infty}\,ds\,B(s,\alpha-s)\,\rho^{\alpha-s}\,
Q^{-s}, \qquad \alpha > 0,
\eq
where $\rho = 1/\lambda$, $B$ is the Euler beta-function and
\bq
0 < \Reb\,s < \alpha, \qquad \mid {\rm arg}\,Q - {\rm arg}\,\lambda\mid \,
< \, \pi.
\eq
The $i\,\delta$ prescription is essential in deriving the correct analytic
continuation. In all cases $Q$ will be a quadric in $n$ variables and
$\lambda$ will be a constant, essentially a BT factor. Therefore $Q$ is
always understood as $Q - i\,\delta$ with $\delta \to 0_+$. Since
$\mid {\rm arg}\,Q - {\rm arg}\,\lambda\mid \, < \, \pi$ is required we will
assign a small negative imaginary parts to both $Q$ and $\lambda$.
\begin{itemize}
\item[--] Anomalous threshold
\end{itemize}
For a given one-loop diagram $G$, let us denote with $B_{\ssG}$ the
corresponding BT factor of \eqn{functr} which is a function of the internal
masses and external momenta of $G$. At the leading Landau singularity of
$G$, the so-called anomalous threshold (AT)~\cite{oldies}, we have $B_{\ssG} =
0$. Conversely, $B_{\ssG} = 0$ is the condition to have a proper solution
for the system of Landau equations corresponding to $G$ (see \sect{IRLEQ}
for additional discussions). Note that AT is not directly related through
unitarity to physical processes (cut diagrams).  Most of our work is devoted
to analyze specific algorithms for the numerical evaluation of $G$ around
$B_{\ssG} = 0$ for arbitrary values of the internal masses and external
momenta.  However, for a large class of applications these parameters are
bounded to the physical region ${\cal R}$ and the relevant question is
whether or not AT $\,\in {\cal R}$. A complete discussion of this issue is
technically rather complicated and beyond the scope of our work; here we
illustrate a simple case, namely the one-loop vertex of \fig{fig:threepoint}
with the following configuration, $m_i = m$, $p^2_{2,3} = - M^2 \le 0$
and $p^2_1 = - r$.  For real vectors~\footnote{In our metric spacelike $p$
implies positive $p^2$.  Further $p_4 = i\,p_0$ with $p_0$ real for a
physical four-momentum.}  $p_{2,3}$ it follows that ${\cal R}$ is defined by
either $r \ge 4\,M^2$ ($s$-channel) or $r \le 0$ ($t$-channel). The diagram
has an AT at
\bq
r = r_{\ssA\ssT} = 4\,M^2\,\left( 1 - \frac{M^2}{4\,m^2} \right), \qquad
\mbox{iff} \quad M^2 > 2\,m^2.
\eq
The kind of singularity depends on the value of $M^2$: if $0 < M^2 < 2\,m^2$
there is no AT, if $2\,m^2 < M^2 < 4\,m^2$ there is an unphysical AT ($0 <
r_{\ssA\ssT} < 4\,m^2$) and, finally, if $M^2 > 4\,m^2$ there is a physical
AT at $r_{\ssA\ssT} < 0$.
\begin{itemize}
\item[--] Notations
\end{itemize}
Finally, to keep our results as compact as possible, we introduce the 
following notations:
\bq
\dsimp{n} \equiv \int_0^1\,dx_1\,\int_0^{x_1}\,dx_2\,\cdots\,\int_0^{x_{n-1}}\,
dx_n,
\quad
\dcub{n} \equiv \int_0^1\,\prod_{i=1}^{n}\,dx_i,
\quad
\int\,d\bar{S}_2 \equiv \intfxy{x_2}{x_1}.
\eq
Likewise we also need
\bq
\dsimp{n}(X) \equiv 
\int_{-X_1}^{1-X_1}\,dx_1\,\int_{-X_2}^{x_1+X_1-X_2}\,dx_2\,\,\cdots\,
\int_{-X_n}^{x_{n-1}+X_{n-1}-X_n}\,dx_n,
\label{somedef}
\eq
where $X$ is an $n$-dimensional vector.
\section{Generalized two-point function\label{gtpf}}
One of the building blocks for the evaluation of all one-loop diagrams is
the generalized two-point function, i.e.~a self-energy graph with arbitrary
powers for the two internal propagators.
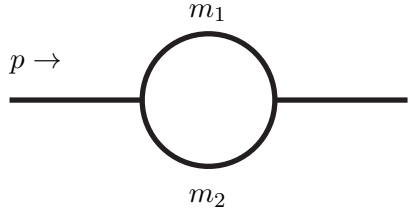
\begin{figure}[h]
\vspace*{-5mm}
\[
  \vcenter{\hbox{
  \begin{picture}(150,90)(0,0)
  \SetWidth{2.}
  \Line(0,50)(50,50)
  \CArc(75,50)(25,-180,180)
  \Line(100,50)(150,50)
  \Text(10,60)[cb]{$\pmom\to$}
  \Text(75,80)[cb]{$\mone$}
  \Text(75,10)[cb]{$\mtwo$}
  \end{picture}}}
\]
\vspace{-0.75cm}
\caption[]{The two-point Green function.}
\label{fig:twopoint}
\end{figure}
The generalized two-point function can be written in terms of the functions
\bq
B_n(\alpha) = \intfx{x}\,x^n\,\Bigl[ \chi_{\ssB}(x) - i\,
\delta\Bigr]^{-\alpha},
\label{iexam}
\eq
where $n$ is a non-negative integer and $\chi_{\ssB}(x)$ is a quadratic
polynomial whose coefficients depend on $p^2$, the external squared
four-momentum, and $m_{1,2}$, the internal masses (see
Fig.~\ref{fig:twopoint}).  Let $x_{\pm}$ be the roots of the equation
$\chi_{\ssB}-i\,\delta = 0$.  If $\lambda(-p^2,m^2_1,m^2_2) = 0$, where
$\lambda(x,y,z)=x^2+y^2+z^2-2(xy+xz+yz)$ is the usual K\"allen function, the
integration contour may be pinched between the two singularities $x_{\pm}$
as $\delta \to 0_+$. If this pinching occurs, the integral is singular
with a branch point of the two-particle cut~\cite{Cutkosky:1960sp}.
Consider 
\bq
\chi_{\ssB}(x) = a\,x^2 + b\,x + c = a\,(x - X)^2 + \lambda ,
\eq
where we have introduced the auxiliary quantities
\bq
X = -\,\frac{b}{2a},
\qquad
\lambda = - \frac{\Delta}{4a},
\qquad
\Delta = b^2 - 4ac = \lambda(-p^2,m^2_1,m^2_2).
\eq
In the following we describe a solution for $B_n(\alpha)$ which allows to
extract its behavior around $\lambda=0$. 

We start our derivation by analyzing $B_0(\alpha)$. First we assume $0 \le X
\le 1$ and obtain
\bq
B_0(\alpha) =
\sum_{l=1}^{2}\,(-1)^l\,X_l\,\intfx{x}\,
(a\,X_l^2\,x^2 + \lambda - i\,\delta)^{-\alpha} =
\sum_{l=1}^{2}\,(-1)^l\,X_l\,J_l(\alpha,\lambda),
\eq
where $X_1 = -X$, $X_2 = 1-X$ and where the $i\,\delta$ prescription is
equivalent to the replacement $a,\lambda \to a,\lambda - i\,\delta$. In the 
following it must be understood that both $a$ and $\lambda$ are given an
infinitesimal imaginary part. Note that with this prescription we have
\bqa
(\frac{\lambda}{a})^{1/2} &\equiv&
\Bigl[ \theta(\frac{\lambda}{a}) - i\,\sign{a}\,\theta(-\frac{\lambda}{a})
\Bigr]\,\mid \frac{\lambda}{a}\mid^{1/2} = \lambda^{1/2}\,a^{-1/2}.
\eqa
Next we split $J_l$ into two pieces 
\bq
J_l(\alpha,\lambda) =
\int_0^\infty\,dx\,(a\,X_l^2\,x^2 + \lambda)^{-\alpha}
- \int_1^\infty\,dx\,(a\,X_l^2\,x^2 + \lambda)^{-\alpha},
\label{twoint}
\eq
where the first one contains the divergent part for $\lambda \to 0$, while
the second is finite.  In the latter we perform the transformation $x \to
x^{-1/2}$ and we obtain:
\bq
J_l(\alpha,\lambda) =
\int_0^\infty\,dx\,(a\,X_l^2\,x^2 + \lambda)^{-\alpha}
- \frac{1}{2}\,\intfx{x}\,x^{\alpha-3/2}\,(a\,X_l^2 + \lambda\,x)^{-\alpha}.
\eq
The first integral in \eqn{twoint} can be cast into the following form,
\bqa
\int_0^\infty\,dx\,(a\,X_l^2\,x^2 + \lambda)^{-\alpha} &=&
\frac{1}{2}\,\frac{\lambda^{1/2-\alpha}}{a^{1/2}\,|X_l|}\,
B(\frac{1}{2}\,,\,\alpha-\frac{1}{2}),
\eqa
which is valid for $|\arg(\lambda/a)| < \pi$ and $\Reb\,\alpha > 1/2$.
Here $B$ denotes the Euler beta-function. Collecting the two pieces we obtain
\bqa
B_0(\alpha)
&=&
\frac{1}{2}\,\sum_{l=1}^{2}\,\frac{(-1)^l\,X_l}{|X_l|}\,
\left[
\lambda^{1/2-\alpha}\,a^{-1/2}\,B(\frac{1}{2},\alpha-\frac{1}{2})
- |X_l|\,\intfx{x}\,x^{\alpha-3/2}\,(a\,X_l^2 + \lambda\,x)^{-\alpha}
\right]
\nl
&=&
\lambda^{1/2-\alpha}\,a^{-1/2}\,B(\frac{1}{2},\alpha-\frac{1}{2})
- \frac{1}{2}\,\sum_{l=1}^{2}\,|X_l|\,
  \intfx{x}\,x^{\alpha-3/2}\,(a\,X_l^2 + \lambda\,x)^{-\alpha},
\eqa
where we have used the relation $|X_l|=(-1)^l\,X_l$, consequence of
$0 \le X \le 1$.
If instead $X < 0$ or $X > 1$ the first term vanishes and there is no 
divergence. In general we can write
\bqa
\intfx{x}\,x^{\alpha-3/2}\,(a\,X_l^2 + \lambda\,x)^{-\alpha} &=&
\frac{1}{\alpha - 1/2}\,(a\,X^2_l)^{-\alpha}\,
\hyper{\alpha}{\alpha-\frac{1}{2}}{\alpha+\frac{1}{2}}{-\,
\frac{\lambda}{a\,X^2_l}},
\eqa
where ${}_2F_1$ is the standard hypergeometric function~\cite{ellip}.
Let us define $z_l = - \lambda/(a\,X^2_l)$; then, for $\alpha = 1$ we obtain 
\bq
\hyper{1}{\frac{1}{2}}{\frac{3}{2}}{z_l} = \frac{1}{2}\,z^{-1/2}_l\,
\ln \frac{1+z^{1/2}_l}{1-z^{1/2}_l},
\eq
while for $\alpha = n$ we will use
\bqa
\hyper{n}{n-\frac{1}{2}}{n+\frac{1}{2}}{z_l} &=& 
\frac{(3/2)_{n-1}}{(1)_{n-1}\,(1/2)_{n-1}}\,
\frac{d^{n-1}}{dz^{n-1}_l}\,
\hyper{1}{\frac{1}{2}}{\frac{3}{2}}{z_l},
\eqa
where $(a)_n = \egam{a+n}/\egam{a}$ is the Pochhammer symbol and $\Gamma$ is
the Euler gamma-function.  Note that both integrals diverge for $\alpha \to
1/2$ and that the divergent parts cancel in the sum. At $\alpha = 1/2$ and
with $\xi^2_l = a\,X^2_l/\lambda$ we have the following result:
\bq
B_0(\frac{1}{2}) =
\sum_{l=1}^{2}\,(-1)^l\,\sign{X_l}\,a^{-1/2}\,
\ln\Bigl[ \xi_l + (1 + \xi^2_l)^{1/2}\Bigr].
\eq
In the same way we find that for the general case the result is
\bqa
B_n(\alpha)
&=&
\intfx{x}\,x^n\,\Bigl[ \chi_{\ssB}(x) - i\,\delta\Bigr]^{-\alpha} =
\frac{1}{2}\,\sum_{l=1}^{2}\,\sum_{k=0}^{n}\,
\left(
\ba{c}
n \nl k
\ea
\right)\,
(-1)^l\,X^{n-k}\,
\left[\frac{X_l}{|X_l|}\right]^{k+1}\,
\nl
&{}&
\left[
\lambda^{k/2+1/2-\alpha}\,a^{-k/2-1/2}\,
B(\frac{k+1}{2},\alpha-\frac{k+1}{2})
- |X_l|^{k+1}\,
  \intfx{x}\,x^{\alpha-\frac{3+k}{2}}\,(a\,X_l^2 + \lambda\,x)^{-\alpha}
\right]. \label{af1}
\nl
\eqa
The integral in the second term of \eqn{af1} can also be expressed through
an hypergeometric function:
\bqa
B_n(\alpha)
&=&
\intfx{x}\,x^n\,\Bigl[ \chi_{\ssB}(x) - i\,\delta\Bigr]^{-\alpha} =
\frac{1}{2}\,\sum_{l=1}^{2}\,\sum_{k=0}^{n}\,
\left(
\ba{c}
n \nl k
\ea
\right)\,
(-1)^l\,X^{n-k}\,
\left[\frac{X_l}{|X_l|}\right]^{k+1}\,
\nl
&{}&
\left[
\lambda^{k/2+1/2-\alpha}\,a^{-k/2-1/2}\,
B(\frac{k+1}{2},\alpha-\frac{k+1}{2})
- \right.\nl
 & & \left.|X_l|^{k+1}\,\frac{(a X_l)^{-\alpha}}{\alpha-(k+1)/2}\,
\hyper{\alpha}{\alpha-\frac{k+1}{2}}{\alpha-\frac{k-1}{2}}{-\,
\frac{\lambda}{a\,X^2_l}}
\right]. \label{af2}
\eqa
Finally, a special case will be needed in dealing with multi-leg functions:
let $\alpha = 1 +\ep/2$ and consider the expansion around $\ep = 0$;
for $n = 0$, we obtain
\bqa
B_0(1+\ep/2) &=& B_0(1) + \frac{\ep}{2}\,B'_0(1) + \ord{\ep^2},
\eqa
\bqa
B'_0(1) &=& \pi\,\Bigl[ \psi(\frac{1}{2}) - \psi(1) - \ln(\lambda - i\,\delta)
\Bigr]\,(\lambda a)^{-1/2} - \frac{1}{2}\,\sum_{i=1}^{2}\,\mid X_i\mid
\nl
{}&\times& \sum_{n=0}^{\infty}\,
\frac{(a\,X^2_i)^{-n-1}}{n+1/2}\,\Bigl[ \psi(n+1) - \psi(1) - \frac{1}{n+1/2} -
\ln(a\,X^2_i - i\,\delta)\Bigr]\,(-\lambda)^n,
\label{gtpfe}
\eqa
where $\psi$ is the Euler psi-function.
More generally, for a non-negative integer value of $n$, we have
\bqa
B_n(1+\ep/2) &=& B_n(1) + \frac{\ep}{2}\,B'_n(1) + \ord{\ep^2},
\eqa
\bq
B'_n(1) = \frac{1}{2}\,\sum_{l=1}^{2}\,\sum_{k=0}^{n}\,
\left(
\ba{c}
n \nl k
\ea
\right)\,
(-1)^{l}\,X^{n-k}\,
\left[\frac{X_l}{|X_l|}\right]^{k+1}\, \left(b'_{1; k} + b'_{2; k} 
\right) ,
\eq
where $b'_{1; k}$ and $b'_{2; k}$ are the terms proportional to $\ep/2$ that 
are obtained by expanding around $\ep =0$ the first and second term
in the square brackets in \eqn{af2}, respectively:
\bqa
b'_{1; k} &=& \lambda^{k/2-1/2}\,a^{-k/2-1/2}\,\mbox{B}\left(
\frac{1-k}{2},\frac{1+k}{2}\right)\,\left[\psi\left(\frac{1-k}{2}\right)
-\psi(1) - \ln{(\lambda  - i \delta)}\right] \, , \nl 
b'_{2; k} & = & |X_l|^{k+1}\,\sum_{m =0}^{\infty}\frac{2}{a X_l^2 
(k -1 -2m)^2}\,
\left(-\frac{\lambda}{a X_l} \right)^m\,
\left\{ 2 + (k - 1 - 2\,m)\,\left[\psi(m+1) \right. \right. \nl
&-& \left. \left.\psi(1)-\ln{(a X_l^2 - i \delta)} 
\right]
\right\} .
\eqa
\section{Three-point functions (C-family)\label{C0}}
\begin{figure}[h]
\vspace*{-8mm}
\[
  \vcenter{\hbox{
  \begin{picture}(140,130)(-15,-15)
  \SetWidth{2.}
    \Line(30,50)(73,75)        \Text(50,72)[cb]{$\mone$}
    \Line(73,75)(73,25)        \Text(50,23)[cb]{$\mtwo$}
    \Line(73,25)(30,50)        \Text(85,50)[cb]{$\mtre$}
    \Line(0,50)(30,50)    
    \Line(100,100)(73,75)   
    \Line(100,0)(73,25)
  \SetWidth{0.5}
    \LongArrow(10,57)(20,57)   \Text(13,65)[cb]{$\pone$}
    \LongArrow(89,98)(82,91)   \Text(79,103)[cb]{$\ptre$}
    \LongArrow(89,2)(82,9)     \Text(79,-3)[cb]{$\ptwo$}
  \end{picture}
\qquad
  \begin{picture}(140,130)(-15,-15)
  \SetWidth{2.}
    \Line(30,50)(73,75)  
    \Line(73,75)(73,25) 
    \Line(73,25)(30,50) 
    \Line(0,75)(30,50)    
    \Line(0,25)(30,50)    
    \Line(100,100)(73,75)   
    \Line(100,0)(73,25)
  \end{picture}}}
\]
\vspace{-0.75cm}
\caption[]{The one-loop, three-point Green function. The second diagram,
although having $4$ external lines, is included in the $C$-family of
$3$ internal lines.\label{fig:threepoint}}
\end{figure}
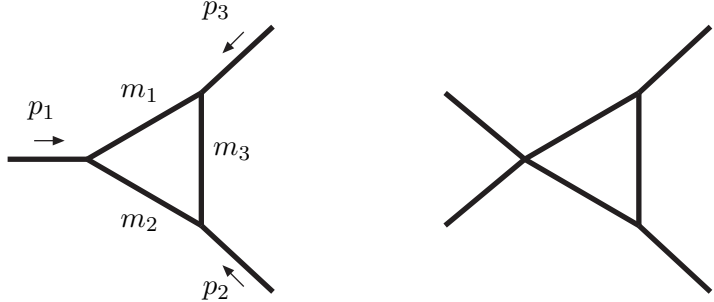

We introduce scalar one-loop $N$-point functions, with $N \ge 3$, according
to the following definition:
\bqa
S_{\ssN} &=& \frac{1}{i\,\pi^2\,\egam{N-2}}\,\int\,d^dq\,
\frac{1}{[q^2+m^2_1]\,\cdots\,[(q+p_1+\,\cdots\,
+p_{\ssN-\scriptscriptstyle{1}})^2+m^2_{\ssN}]}
\nl
{}&=& \dsimp{\scriptscriptstyle{N-1}}\,V^{2-N-\ep/2}_{\ssN}(x),
\eqa
where $d= 4 - \ep$ is the space-time dimensionality and $V_{\ssN}(x)$ is a
quadratic polynomial in the Feynman parameters
$x_{\scriptscriptstyle{1}},\ldots,x_{\scriptscriptstyle{N-1}}$.  In this
section we analyze an arbitrary scalar three-point function $C_0 \equiv S_3$
(see Fig.~\ref{fig:threepoint}), defined by
\bq
C_0 = \dsimp{2}\,V^{-1-\ep/2}(x_1,x_2),
\label{eqdefC0}
\eq
where $V$ is a quadratic polynomial
\bq
V(x_1,x_2) = a\,x_1^2 + b\,x_2^2 + c\,x_1\,x_2 + d\,x_1 +
e\,x_2 + f - i\,\delta \equiv x^t\,H\,x + 2\,K^t\,x + L ,
\eq
whose coefficients are related to the internal masses and the external
momenta by the relations $H_{ij} = -\,\spro{p_i}{p_j}, L = m^2_1$ and
\bqa
K_1 &=& \frac{1}{2} \, ( \spro{p_1}{p_1} + m_2^2 - m_1^2 ),
\quad
K_2 = \frac{1}{2} \, ( \spro{P}{P} - \spro{p_1}{p_1} + m_3^2 - m_2^2 ),
\label{Konetwo}
\eqa
with $P = p_1 + p_2$. 
\vskip 5pt
Let us define the usual BT factors\footnote{Note, however, the different sign
in the definition of $X$ with respect to~\cite{Tkachov:1997wh}.} as
$B_3 = L - K^t\,H^{-1}\,K$ and $X = -\,H^{-1}\,K$.
In the following we will present integral representations for $C_0$ which
are well suited for numerical integration and cover all options for the choice
of external momenta.
\boldmath \subsection{Evaluation of $C_0$ for $B_3 \ne 0$} \unboldmath
If $B_3 \ne 0$ we can rewrite $C_0$ according to the standard BT procedure
of \eqn{functr}. It is convenient to introduce special notations, $X_0 = 1, 
\, X_3 = 0$, and $V(\widehat {i\;i+1})$ to denote contractions, i.e.
\bq
V(\widehat{0\;1}) = V(1,x_1), \quad
V(\widehat{1\;2}) = V(x_1,x_1), \quad 
V(\widehat{2\;3}) = V(x_1,0).
\eq
In this way we obtain a simple integral representation
\bq
C_0 = \frac{1}{B_3}\,\Bigl\{ \frac{1}{2} +
\int_0^1\,dx_1\,\Bigl[ \int_0^{x_1}\,dx_2\,
\ln\,V(x_1,x_2) - 
\frac{1}{2}\,\sum_{i=0}^{2}\,(X_i - X_{i+1})\,\ln V(\widehat{i\;i+1})
\Bigr]\Bigr\}.
\label{sintr}
\eq
Note that a zero of the corresponding Gram's determinant is not a problem,
actually it even decreases the number of terms. Indeed let
$G_3 = {\rm det}\,H$ and denote by $\Delta_{ij}$ the co-determinant of the 
element $H_{ij}$. If $G_3 = 0$ we obtain
\bq
C_0(G_3 = 0) = -\,\frac{1}{2\,b_3}\,\intfx{x}\,\sum_{i=0}^{2}\,({\cal X}_i - 
{\cal X}_{i+1})\,\ln V(\widehat{i\;i+1}),
\label{zeroG}
\eq
where $B_3 = b_3/G_3$ and ${\cal X} = -\,\Delta\,K$, showing that, in this 
case, $C_0$ is a combination of three $B_0$ functions~\cite{Devaraj:1997es}.
\subsection{Form factors in the C-family}
The most general $C$-family integral that we consider in any field theory is
\bq
C[Q(x_1,x_2)] = \dsimp{2}\,Q(x_1,x_2)\,V^{-1-\ep/2}(x_1,x_2),
\quad
Q(x_1,x_2) = \sum_{n=1}^{\bar{n}}\,\sum_{m=1}^{\bar{m}}\,a_{nm}\,x^n_1\,x^m_2.
\eq
The integral representation corresponding to \eqn{sintr} will be
\bq
C[Q(x_1,x_2)] = \sum_{n=1}^{\bar{n}}\,\sum_{m=1}^{\bar{m}}\,
\frac{a_{nm}}{B_3}\,
\Bigl[
C_{m,n}^0 - \frac{1}{2}\,\dsimp{1}\,C_{m,n}^1 -
\frac{1}{2}\,\dsimp{2}\,x_1^{n-1}\,x_2^{m-1}C_{m,n}^2
\Bigr] ,
\eq
where the $C$-coefficients are
\bqa
C_{m,n}^0 &=& \frac{1}{(2+n+m)\,(1+m)},
\nl
\nl
C_{m,n}^1 &=&
(X_1-X_2)\,x^{n+m}_1\,\ln\,V(x_1,x_1) +
\delta_{m,0}\,X_2\,x_1^n\,\ln\,V(x_1,0) +
(1-X_1)\,x^m_1\,\ln\,V(1,x_1),
\nl
\nl
C_{m,n}^2 &=&
( m\,X_2\,x_1 - (2 + n + m)\,x_1\,x_2 +
n\,X_1\,x_2)\,\ln\,V(x_1,x_2).
\eqa
Examples of standard form factors of the $C$-family
(see~\cite{Passarino:1979jh} for the general definition) are
\bqa
B_3\,C_{ij} &=& \dsimp{2}\,c^k_{ij}\,\ln V(x_1,x_2) +
\dsimp{1}\,\sum_{l=0}^{2}\,c^l_{ij}\,\ln V(\widehat{i\;i+1}) + K_{ij},
\eqa
with coefficients
\bqa
c^k_{11} &=& \frac{1}{2}\,( X_1 - 3\,x_1),
\quad
c^{1}_{11} = \frac{1}{2}\,X_{12}\,x_1,
\quad
c^{2}_{11} = \frac{1}{2}\,X_2\,x_1,
\quad
c^{0}_{11} = \frac{1}{2}\,X_{01},
\quad
K_{11} = - \frac{1}{3},
\nl
c^k_{12} &=& \frac{1}{2}\,( X_2 - 3\,x_2),
\quad
c^{1}_{12} = \frac{1}{2}\,X_{12}\,x_1,
\quad
c^{2}_{12} = 0,
\quad
c^{0}_{12} = \frac{1}{2}\,X_{01}\,x_1,
\quad
K_{12} = - \frac{1}{6},
\nl
c^k_{21} &=& x_1\,( 2\,x_1 - X_1 ),
\quad
c^{1}_{21} = -\,\frac{1}{2}\,X_{12}\,x^2_1,
\quad
c^{2}_{21} = - \frac{1}{2}\,X_2\,x^2_1,
\quad
c^{0}_{21} = - \frac{1}{2}\,X_{01},
\quad
K_{21} = \frac{1}{4},
\nl
c^k_{22} &=& x_2\,( 2\,x_2 - X_2 ),
\quad
c^{1}_{22} = -\,\frac{1}{2}\,x^2_1\,X_{12},
\quad
c^{2}_{22} = 0,
\quad
c^{0}_{22} = - \frac{1}{2}\,X_{01}\,x^2_1,
\quad
K_{22} = \frac{1}{12},
\nl
c^k_{23} &=& 2\,x_1\,x_2 -\frac{1}{2}\,( X_1\,x_2 + X_2\,x_1),
\nl
c^{1}_{23} &=& - \frac{1}{2}\,x^2_1\,X_{12},
\quad
c^{2}_{23} = 0,
\quad
c^{0}_{23} = -\,\frac{1}{2}\,X_{01}\,x_1,
\quad
K_{23} = \frac{1}{8},
\eqa
where we introduced $X_{ij} = X_i - X_j$ and $X_0 = 1$. For $C_{24}$ we
directly obtain
\bq
C_{24} = \frac{1}{4}\,\Ddr - \frac{1}{2}\,\dsimp{2}\,\ln V(x_1,x_2),
\qquad
\Ddr = \frac{2}{\dre} - \eilc - \lpi,
\eq
where $\eilc = 0.577216\cdots$ is the Euler constant and $d= 4 - \ep$ is the
space-time dimensionality.

It is perhaps the right moment to summarize our strategy for tensor integrals.
Clearly a decomposition in form factors is always possible and, within our
approach, the form factors are not plagued by the appearance of inverse powers
of Gram's determinants. However, this is not exactly the way a calculation
should be organized since one has to profit from the strategy of the BT
procedure by mapping all integrals, for all topologies, into just one
integral. Therefore, there is no need to perform a pre-reduction in momentum
space. Furthermore, in any realistic calculation, the denominators in
Feynman diagrams contain scalar products, e.g. $\spro{q}{p_1}$ etc, and one
has to benefit as much as possible from algebraic simplification. However, this
is not always the case because each of the external lines, say in our original
diagram, can split into other two (three) lines (we have in mind vector bosons
that should be considered as unstable particles and, therefore, always be
attached to fermionic currents). In this case momenta will appear that are
not present in the propagators of the diagram and no simplification will occur.
For this reason, and for the reader's convenience, we list all the form factors
in the one-loop families.
\boldmath \subsection{$C_0$ at $\ord{\ep}$} \unboldmath
A two-point function with the insertion of a counterterm, which is needed in
a two-loop calculation, is formally identical to a one-loop three-point
function with one zero external momentum. Its expression at $\ord{\ep}$ is
needed and this follows from the BT algorithm quite nicely,
\bq
C_0(d) = C_0(4) + \ep\,C^{(1)}_0 + \ord{\ep^2},
\eq
\bqa
C^{(1)}_0 &=& -\,\frac{1}{2}\,\dsimp{2}\,\Bigl[ \ln V(x_1,x_2) + 
\frac{1}{2}\,\ln^2 V(x_1,x_2)\Bigr]
\nl
{}&+& \frac{1}{8}\,\dsimp{1}\,\sum_{i=0}^{2}\,( X_i - X_{i+1} )\,
\ln^2 V(\widehat{i\;i+1}),
\eqa
showing that it contains at most squares of logarithms of quadratic forms.
\boldmath \subsection{Evaluation of $C_0$ when $B_3 \approx 0$} \unboldmath
When we are around the region where $B_3 = 0$,
\eqn{sintr} cannot be used and an alternative derivation is 
needed.  Actually there are two sub-cases to be discussed starting from the 
observation that $C_0$ can be written as
\bq
C_0 = \dsimp{2}\,\Bigl[ (x - X)^t\,H\,(x - X) + B_3 \Bigr]^{-1-\ep/2}.
\label{analC0}
\eq
If $B_3 = 0$ the integral of \eqn{analC0} is singular or regular depending
on the values of $X_{1,2}$.
\subsubsection{The regular case: method I\label{rcmI}}
If $B_3 = 0$, but the condition
\bq
0 \le X_2 \le X_1 \le 1
\label{C0reg}
\eq
is not fulfilled, there is no real singularity inside the integration domain
${\cal I}_3$; still we cannot apply the standard BT techniques of \eqn{functr}.
The integral in \eqn{eqdefC0} is well defined through distortion of the 
contour integration and there is no pinch inside ${\cal I}_3$. We use this
regularity to write the following BT relations, corresponding to $B_3 \neq 0$ 
and $B_3 = 0$:
\bqa
\Big[ 1 - \frac{1}{2\,(\mu+1)}\,(x-X)\,\partial_x \Big]\,
V^{\mu+1}(x_1,x_2)
&=& B_3\,V^{\mu}(x_1,x_2),
\nl
\Big[ 1 - \frac{1}{2\,(\mu+1)}\,(x-X)\,\partial_x \Big]\,V_0^{\mu+1}(x_1,x_2)
&=& 0,
\eqa
with $V = V_0 + B_3$. Subtracting the two equations, we obtain:
\bq
V^{\mu}(x_1,x_2) =
\Big[ 1 - \frac{1}{2\,(\mu+1)}\,(x-X)\,\partial_x \Big]\,
\frac{V^{\mu+1}(x_1,x_2) - V^{\mu+1}_0(x_1,x_2)}{B_3}.
\eq
Note that this relation holds even in the case $B_3 \sim 0$.
Since the point of coordinates $x_i = X_i$ is outside ${\cal I}_3$, we can
perform an integration by parts on both $V$ and $V_0$. The result is the
following:
\bq
C_0 =
\frac{1}{B_3}\,\Bigg[
\dsimp{2}\,\ln\frac{V}{V_0}
- \frac{1}{2}\,\dsimp{1}\,\sum_{i=0}^2\,(X_i-X_{i+1})\,
   \ln\frac{V(\widehat{i\,\,i+1})}{V_0(\widehat{i\,\,i+1})}
\Bigg].
\label{C0regular:gr1}
\eq
All tensor integrals can be treated according to the same procedure.
Note that the logarithms appearing in \eqn{C0regular:gr1} are of the form
$\ln(1+x)$ with small $|x|$ and the corresponding values must retain full
relative precision, e.g. they must be computed with routines based on Chebyshev
expansion.
\subsubsection{The regular case: method II\label{rcmII}}
Even if the method of \sect{rcmI} gives a compact result for the regular
case, we derive another expression for numerical checking of our results.
Since $C_0$ is not singular in the limit $B_3 \to 0$ we can use a Taylor
expansion of the integrand of \eqn{eqdefC0} followed by a BT-algorithm 
corresponding to $B_3=0$. Therefore we have:
\bq
C_0 = \sum_{n=0}^{\infty}\,(-\,B_3)^n\,\dsimp{2}(X)\,
(x^t\,H\,x)^{-n-1-\ep/2} =
\sum_{n=0}^{\infty}\,{\cal C}(n+1)\,(-\,B_3)^n,
\label{eqrcmII}
\eq
where we kept $\ep\neq 0$ only where strictly needed and where we used 
\eqn{somedef}. By using:
\bq
\dsimp{2}(X)\,
\Bigl[ 1 + \frac{1}{2\,n+\ep}\,x\,\partial_x\Bigr]\,V^{-n-\ep/2}(x_1,x_2) = 0.
\eq
We observe that integration-by-parts is allowed since $x_i = 0$ is outside
the integration domain and derive, after changing $x$ into either $x - X_1$
or $x - X_2$,
\bq
{\cal C}(n) = \dsimp{2}(X)\,V^{-n-\ep/2}(x_1,x_2) =
- \,\frac{1}{2\,n - 2 + \ep}\,\dsimp{1}\,
\sum_{i=0}^2\,(X_i-X_{i+1})\,V^{-n-\ep/2}(T\,\widehat{i\,\,i+1}),
\label{deftaylorC}
\eq
where we employed $X_0 = 1$ and $X_3 = 0$, and we
introduced shifted arguments, i.e.~the three relevant quadratic
forms in $x$ are
\bqa
V(T\,\widehat{0\,\,1}) &\equiv& V_c = V(1-X_1,x-X_2), \quad
V(T\,\widehat{1\,\,2}) \equiv V_a = V(x-X_1,x-X_2),
\nl
V(T\,\widehat{2\,\,3}) &\equiv& V_b = V(x-X_1,-X_2),
\eqa
for which we write $V_i = h_i\,x^2 + 2\,k_i\,x + l_i$
and introduce the standard BT factor, $b_i, i=a,b,c$, and
also the BT co-factor, $X_i, i= a,b,c$. As long as all the $b_i$ are different
from zero, the corresponding BT relations hold.
For the first term in the expansion we obtain:
\bq
{\cal C}(1) =
\frac{1}{2}\,\sum_{i=a,b,c}\,\frac{X_i^{\ssL}\,}{b_i}\,
\Bigl[ {\cal C}_{0;i}(1) + \intfx{x}\,{\cal C}_{1;i}(1,x) \Bigr],
\eq
\bq
{\cal C}_{0;i}(1) =
- \frac{1}{4}\,\Big[ X_i\,\ln^2 V_i(0) + (1-X_i)\,\ln^2 V_i(1) \Big],
\quad
{\cal C}_{1;i}(1,x) = \ln V_i(x) + \frac{1}{4}\,\ln^2 V_i(x),
\eq
where the leading coefficients are:
\bq
X_a^{\ssL}= X_1 - X_2,
\qquad
X_b^{\ssL}= X_2,
\qquad
X_c^{\ssL}= 1 - X_1,
\eq
whereas for $n>1$ we get
\bq
{\cal C}(n) =
- \frac{1}{4\,(n-1)^2}\,\sum_{i=a,b,c}\,\frac{X_i^{\ssL}}{b_i}\,
\Bigl[ {\cal C}_{0;i}(n) + \intfx{x}\,{\cal C}_{1;i}(n,x) \Bigr],
\eq
\bq
{\cal C}_{0;i}(n) =
(1-X_i)\,V_i^{-n+1}(1) + X_i\,V_i^{-n+1}(0),
\qquad
{\cal C}_{1;i}(1,x) = (2\,n-3)\,V_i^{-n+1}(x).
\eq
Now we can sum the series according to the well-known results:
\bq
\sum_{n=1}^{\infty}\,\frac{x^n}{n} = -\ln(1-x),
\qquad
\sum_{n=1}^{\infty}\,\frac{x^n}{n^2} = \li{2}{x},
\eq
where $\li{2}{x}$ is the standard dilogarithm~\cite{polyl}, and obtain a 
novel representation for $C_0$:
\bq
C_0 =
\frac{1}{2}\,\sum_{i=a,b,c}\,\frac{X_i^{\ssL}}{b_i}\,
\Bigl[ C_{0;i} + \intfx{x}\,C_{1;i} \Bigr],
\label{C0regular:gr2}
\eq
\bqa
C_{0;i} &=&
- \frac{1}{4}\,X_i\,
\Big[ 2\,\li{2}{-\frac{B_3}{V_i(0)}} + \ln^2 V_i(0) \Big]
- \frac{1}{4}\,(1-X_i)\,
\Big[ 2\,\li{2}{-\frac{B_3}{V_i(1)}} + \ln^2 V_i(1) \Big],
\nl
C_{1;i} &=&
\ln \Big( V_i(x) + B_3 \Big)
+ \frac{1}{2}\,\li{2}{-\frac{B_3}{V_i(x)}}\,
+ \frac{1}{4}\,\ln^2 V_i(x) .
\eqa
The above derivation of the coefficients in the Taylor expansion,
\eqn{deftaylorC}, assumes that all sub-leading BT factors are non zero.
${\cal C}(n)$ of \eqn{deftaylorC} is a combination of
generalized two-point functions of \sect{gtpf}; if one of the $b_i \approx 0$
we still have to distinguish between two possibilities:
\begin{itemize}
\item[a)] $b_i = 0$ but $-h_i \le k_i \le 0$ ($h_i\geq\,0$) is not fulfilled.
\item[b)] $b_i = 0$ and $-h_i \le k_i \le 0$ ($h_i\geq\,0$) is fulfilled.
\end{itemize}
In both cases we can apply the results of \sect{gtpf}.
So let us denote each term of \eqn{deftaylorC} with
\bq
{\cal C}^i(n) = \frac{1}{2\,n - 2 + \ep}\,\intfx{x}\,g_i\,V^{-n-\ep/2}_i(x),
\eq
where $g_i=-X_i+X_{i+1}$. The corresponding contribution to $C_0$ will be
\bq
C^i_0 = \intfx{x}\,g_i\,V^{-1-\ep/2}_i(x)\,\sum_{n=0}^{\infty}\,
\frac{(-B_3)^n}{2\,n + \ep}\,V^{-n}_i(x).
\eq
In evaluating the sum it is more convenient to isolate the leading term
$C_0^i(n=0)$ and consider the rest separately. The latter becomes
\bqa
[C_0^i]_{\rm rest} &=&
\frac{1}{2}\,g_i\,\sum_{n=1}^{\infty}\,
\frac{(-B_3)^n}{n}\,\intfx{x}\,V^{-n-1}_i(x) =
\frac{1}{2}\,g_i\,\sum_{l=1}^{2}\,\frac{(-1)^l\,X_i^l}{|X_i^l|}\,
\sum_{n=1}^{\infty}\,\frac{(-B_3)^n}{n}\,
\nl
&{}&
\Bigg[
\frac{1}{2}\,b_i^{-1/2-n}\,h_i^{-1/2}\,B(\frac{1}{2},n+\frac{1}{2})
- |X_i^l|\,\intfx{x}\,x^{2\,n}\,(h_i\,{X_i^l}^2 + b_i\,x^2)^{-n-1}
\Bigg] ,
\eqa
where $X_{1i}= -X_i$ and $X_{2i}= 1-X_i$.
The second series can be easily summed up, while for the first one we
have:
\bqa
\sum_{n=1}^{\infty}\,\frac{(-B_3/b_i)^n}{n}\,
B(\frac{1}{2},n+\frac{1}{2}) &=&
\pi^{1/2}\,\sum_{n=0}^{\infty}\,
\frac{(n+1/2)\,\egam{n+1/2}}{(n+1)^2\,\egam{n+1}}\,(-\kappa_i)^{n+1}
\nl
&=&
- \pi\,\kappa_i\,\intfx{x}\,(1 + \frac{1}{2}\,\ln x )\,(1+\kappa_i\,x)^{-1/2} ,
\eqa
where we introduced $\kappa_i=B_3/b_i$.
Finally we obtain:
\bqa
[C_0^i]_{\rm rest} &=&
- \frac{1}{2}\,g_i\,\sum_{l=1}^{2}\,\frac{(-1)^l\,X_{li}}{|X_{li}|}\,\intfx{x}
\Bigg[
\frac{\pi\,\kappa_i}{(b_i\,h_i)^{1/2}}\,
\frac{x+x\,\ln x}{(1+\kappa_i\,x^2)^{1/2}}
\nl
&{}&
- |X_{li}|\,(h_i\,X_{li}^2 + b_i\,x^2)^{-1}\,
  \ln\frac{h_i\,X_{li}^2 + (b_i+B_3)\,x^2}{h_i\,X_{li}^2 + b_i\,x^2}
\Bigg] .
\eqa
As we argued for the generalized two-point function, the divergent
term for $b_i \approx 0$ cancels out if the corresponding $X_i$ is outside
the interval $[0,1]$.
For the leading term, after an $\ep$ expansion, we have:
\bq
C^i_0(n=0) = \intfx{x}\,g_i\,V^{-1}_i(x)\,
\Big[ \frac{1}{\ep} - \frac{1}{2}\,\ln V_i \Big],
\quad
\sum_{i=0}^2\,\intfx{x}\,g_i\,V_i^{-1} = 0.
\eq
Note that the divergent term does not contribute to $C_0$.
In order  to calculate the term $C^i_0(n=0)$,
 we need the relevant part of ${\cal C}(1)$, i.e.\ the one
corresponding to those values of $i$ for which $b_i \approx 0$.
The latter contribution can be obtained by observing that, symbolically,
\bq
\ep\,{\cal C}(1) = \sum\,B_0(1+\ep/2);
\eq
using \eqn{gtpfe} we expand the relevant generalized two-point function and
derive an expression for ${\cal C}(1)$ by equating the terms proportional to
$\ep$.

Also for the form factors we can derive a similar expansion.
Consider the auxiliary function
\bqa
C(n_1,n_2) &=& \sum_{n=0}^{\infty}\,(-\,B_3)^n\,
\dsimp{2}(X)\,\prod_{i=1}^{2}\,(x_i+X_i)^{n_i}\,(x^t\,H\,x)^{-n-1-\ep/2}
\nl
{}&=& \sum_{n=0}^{\infty}\,\sum_{l_1=0}^{n_1}\,\sum_{l_2=0}^{n_2}\,
\left(
\ba{c}
n_1 \nl l_1
\ea
\right)\,
\left(
\ba{c}
n_2 \nl l_2
\ea
\right)\,
X^{n_1-l_1}_1\,X^{n_2-l_1}_2\,
{\cal C}(n+1,l_1,l_2)\,(-\,B_3)^n,
\eqa
and the expansion is fully specified by
\bqa
{\cal C}(n,n_1,n_2) &=& \frac{1}{2-2\,n+n_1+n_2}\,\intfx{x}\,\Bigl\{
\Bigl[ (x-X_1)^{n_1}\,(x-X_2)^{n_2}\,(X_1-X_2)\,V^{-n}_a
\nl
{}&-& (x-X_1)^{n_1}\,(-\,X_2)^{n_2+1}\,V^{-n}_b +
(x-X_2)^{n_2}\,(1-X_1)^{n_1+1}\,V^{-n}_c\Bigr\}.
\eqa
Combinations of these functions account for all form factors.
\subsubsection{The singular case}
If condition \eqn{C0reg} is fulfilled then there is a pinch inside ${\cal I}_3$
and the integral in \eqn{eqdefC0} is singular. In order to minimize the
number of terms in the solution we write
\bqa
C_0 &=& \dsimp{2}\,V^{-1-\ep/2}(x_1,x_2) = \dcub{2}\,V^{-1-\ep/2}(x_1,x_2) -
\int_0^1\,dx_1\,\int_{x_1}^1\,dx_2\,V^{-1-\ep/2}(x_1,x_2)
\nl
{}&=& \dcub{2}\,V^{-1-\ep/2}(x_1,x_2) -
\dsimp{2}\,V^{-1-\ep/2}(x_2,x_1) = C^{\rm square}_0 - C^{\rm comp}_0.
\eqa
Since the point $x_i = X_i$ is now internal to the integration domain the
complementary $C_0$ function will be regular and can be computed according to
the results of \sects{rcmI}{rcmII}.
A special case is represented by $0 \le X_2 = X_1 \le 1$ where both the 
original $C_0$ and the complementary one are singular. However, due to the
symmetry of $H$, we easily obtain
\bq
C_0 = C^{\rm comp}_0 = \frac{1}{2}\,C^{\rm square}_0, \qquad
\mbox{for} \quad 0 \le X_2 = X_1 \le 1.
\eq
$C^{\rm square}_0$ is the integral over $[0,1]^2$ which we rewrite 
by changing variables, $x'_i = x_i - X_i$, as
\bqa
C^{\rm square}_0 &=& \sum_{i=1}^4\,\alpha_i\,\beta_i\,\dcub{2}
\Bigl[ Q_i(x_1,x_2) + B_3\Bigr]^{-1-\ep/2},
\eqa
where
\bq
\alpha_1 = \alpha_2 = 1 - X_1, \quad \alpha_3 = \alpha_4 = X_1,
\quad
\beta_1 = \beta_3 = 1 - X_2, \quad \beta_2 = \beta_4 = X_2,
\label{abDef}
\eq
and where the new quadrics are defined by
\bqa
Q_1 &=& Q((1-X_1) x_1\,,\,(1-X_2) x_2),
\quad
Q_2 = Q((1-X_1) x_1\,,\,-X_2 x_2),
\nl
Q_3 &=& Q(-X_1 x_1\,,\,(1-X_2) x_2),
\quad
Q_4 = Q(-X_1 x_1\,,\,-X_2 x_2),
\eqa
with $Q = x^t\,H\,x$. In general we define
\bq
Q_i(x_1,x_2) = A_i\,x^2_1 + B_i\,x^2_2 + C_i\,x_1\,x_2.
\label{refQi}
\eq
In order the derive a Laurent expansion around $B_3 = 0$ we introduce
$\rho_3 = 1/B_3$ and perform a Mellin-Barnes splitting as described in
\sect{btt},
\bq
C^{\rm square}_0 = \sum_{i=1}^4\,\frac{\alpha_i\,\beta_i}{2\,\pi\,i}\,
\int_{-i\,\infty}^{+i\,\infty}\,ds\,B(s,1-s)\,\rho^{1-s}_3\,
{\cal C}_i(s),
\quad
{\cal C}_i(s) = \dcub{2}\,Q^{-s}_i(x_1,x_2).
\label{adone}
\eq
Let us consider in detail the ${\cal C}_i$-functions. We use a simple sector
decomposition~\cite{Binoth:2000ps} to obtain
\bqa
{\cal C}_i &=& \Bigl[ \dsimp{2} + \int\,d\bar{S}_2\Bigr]\,Q^{-s}_i(x_1,x_2) =
\dcub{2}\, x^{1-2\,s}_1\,( A_i + C_i\,x_2 + B_i\,x^2_2)^{-s}
\nl
{}&+&
\dcub{2}\, x^{1-2\,s}_2\,( B_i + C_i\,x_1 + A_i\,x^2_1)^{-s} =
\frac{1}{2\,(1-s)}\,\sum_{j=1}^2\,{\cal C}_{i,j}(s).
\eqa
For each of the ${\cal C}_{ij}$-functions we have a reduced quadratic form
in one variable. Let us postpone for a moment the problem of their
evaluation. From \eqn{adone} we obtain
\bqa
C^{\rm square}_0 &=& \frac{1}{2}\,\sum_{i=1}^4\,\sum_{j=1}^2\,
\alpha_i\,\beta_i\frac{1}{2\,\pi\,i}\,
\int_{-i\,\infty}^{+i\,\infty}\,ds\,\frac{\egam{s}\,\egam{1-s}}{1-s}\,
\rho^{1-s}_3\,{\cal C}_{ij}(s),
\nl
{\cal C}_{ij}(s) &=& \intfx{x}\,(h_{ij}\,x^2 + 2\,k_{ij}\,x + l_{ij})^{-s},
\label{fconc}
\eqa
and $0 < \Reb\,s < 1$. Suppose that the corresponding BT factors, 
\bq
b_{ij} = l_{ij} - k^2_{ij}/h_{ij} = 
\alpha^2_i\,\beta^2_i\,{\rm det}(H)/h_{ij}
\label{secondT}
\eq
are not zero and that we are interested in the region of large $|\rho_3|$. 
Then we close the integration contour over the right-hand complex half-plane 
at infinity. The poles are at $s = 1\,$(double) and at $s = k + 1\,$(single) 
where $k\ge 1\,$ is an integer. For ${\cal C}_{ij}$ we use
\bqa
{\cal C}_{ij}(s) &=& \frac{1}{b_{ij}}\,\intfx{x}\,\Bigl[ 1 + 
\frac{1}{2}\,\frac{x-X_{ij}}{s-1}\,
\frac{d}{dx}\Bigr]\,Q^{1-s}_{ij}(x),
\nl
Q_{ij}(x) &=&  h_{ij}\,x^2 + 2\,k_{ij}\,x + l_{ij},
\eqa
where $X_{ij}$ is the BT co-factor for $Q_{ij}$. In the limit $s \to 1$ we 
obtain
\bqa
{\cal C}_{ij}(s) &=& {\cal C}_{ij}(1) + (s - 1)\,{\cal C}'_{ij}(1) +
\ord{(s-1)^2},
\nl
{\cal C}_{ij}(1) &=& \frac{1}{b_{ij}}\,\intfx{x}\,\Bigl[ 1 - \frac{1}{2}\,
(x - X_{ij})\,\frac{d}{dx}\,\ln\,Q_{ij}(x)\Bigl],
\nl
{\cal C}'_{ij}(1) &=& \frac{1}{b_{ij}}\,\intfx{x}\,\Bigl[ -\,\ln\,Q_{ij}(x) +
\frac{1}{4}\,(x - X_{ij})\,\frac{d}{dx}\,\ln^2\,Q_{ij}(x)\Bigl].
\eqa
Therefore the residue of the double pole at $s = 1$ is
\bq
{\cal R}_{ij}\mid_{s=1} = {\cal C}_{ij}(1)\,\ln\rho_3 -
{\cal C'}_{ij}(1).
\eq
For the single poles at $s = k + 1, k \ge 1$ we find that the residues are
\bq
{\cal R}_{ij}\mid_{s=k+1} = - \frac{(-1)^k}{k}\,B_0(k+1)\,\rho^{-k}_3,
\label{lconc}
\eq
where $B_0(k+1)$ is a generalized two-point function discussed in \sect{gtpf}.
With the help of \eqns{fconc}{lconc} we obtain the full Laurent expansion.

Also for the form factors we can use the same strategy and write the auxiliary
functions $C(n_1,n_2)$ as a combinations of Mellin-Barnes integrals
\bqa
{}&{}&
\int_{-i\,\infty}^{+i\,\infty}\,ds\,\frac{\egam{s}\,
\egam{1-s}}{1-s+(n_1+n_2)/2}\,
\rho^{1-s}_3\,{\cal C}_{ij}(n_j,s),
\nl
{\cal C}_{ij}(n_j,s) &=& \intfx{x}\,x^{n_j}\,
(h_{ij}\,x^2 + 2\,k_{ij}\,x + l_{ij})^{-s}.
\label{seeeq}
\eqa
The relevant poles are at $s = k + 1, \,k \ge 0$ and there are two
possibilities: if $n_1 + n_2 = 2\,n$ then there is a double pole for
$s = n + 1$, otherwise, if $n_1 + n_2 = 2\,n + 1$, there is one additional, 
simple, pole at $s = n + 3/2$.

However, the secondary BT factors, $b_{ij}$, could be near to zero. 
Note that the $b_{ij}$ coefficients are sub-leading of a second type since
$b_{ij} = 0$ should not be confused, in general, with the condition for the 
occurence of a Landau singularity associated with a reduced diagram.
Indeed, following \eqn{abDef} and \eqn{secondT}, they correspond to having
$0 = X_2 \le X_1 \le 1$ or $0 \le X_2 \le X_1 = 1$ in the pinch singularity 
and it is enough to select $m_i = m, \, \forall i$ in \eqn{Konetwo} to see 
that there is no correspondence with any of the two-particle cuts. From 
\eqn{rol} and \eqn{secondT} it is easily seen that for 
$G_3 = {\rm det}(H) \sim 0$ we hace $b_{ij} \sim G^{-3}_3$ so that $G_3 = 0$ 
does not imply $b_{ij} = 0$.

$b_{ij} = 0$ alone is not yet a sign of instability because
${\cal C}_{ij}$ will show a pinch singularity in $[0,1]$ only if
$h_{ij} > 0$ and $-\,h_{ij} \le k_{ij} \le 0$.
If we are in this region for some pair $ij$ then it is better to use another
expansion algorithm. From \eqn{adone} we see that the total result is a sum 
of terms with the following structure:
\bqa
\rho\,F(\rho) &=& \frac{1}{2\,\pi\,i}\,\int_{-i\,\infty}^{+i\,\infty}\,ds\,
\frac{\egam{s}\,\egam{1-s}}{s-1}\,f(s)\,\rho^{1-s},
\nl
f(s) &=& \intfx{x}\,\Bigg[ h\,\Bigl(x + \frac{k}{h}\Bigr)^2 + b \Bigg]^{-s} =
\int_{\alpha}^{1+\alpha}\,dx (h\,x^2 + b )^{-s},
\label{efour}
\eqa
where $\alpha = k/h$. With $\rho_{\ssS\ssL} = 1/b$ we write
\bq
f(s) = \rho^s_{\ssS\ssL}\,\Bigl\{ \int_0^{1+\alpha} - 
\int_0^{\alpha}\Bigr\}\,dx\,
( 1 + \rho_{\ssS\ssL}\,h\,x^2 )^{-s}.
\eq
Introducing a second Mellin-Barnes splitting gives
\bqa
{\cal F}(\rho,\rho_{\ssS\ssL}) &\equiv& \rho\,F(\rho) =
- \sum_{n=1}^{2}\,(-1)^n\,
\frac{\rho\,\beta_n}{2}\,\frac{1}{(2\,\pi\,i)^2}\,
\int_{-i\,\infty}^{+i\,\infty}\,ds\,dt
\nl
{}&\times& \frac{\egam{1-s}\,\egam{t}\,\egam{s-t}}{(s - 1)\,(t - 1/2)}\,
\Bigl(\frac{\rho_{\ssS\ssL}}{\rho}\Bigr)^s\,(\rho_{\ssS\ssL}\,h\,
\beta^2_n)^{-t},
\eqa
with $\beta_1 = \alpha$ and $\beta_2 = 1 + \alpha$. This representation is 
valid in the vertical strip $0 < \Reb\,t < \Reb\,s < 1$ and $\Reb\,t < 1/2$. 
Since we are interested in the region where both the leading $B$ and the 
sub-leading $b$ are approaching zero, we close the $t$-contour over the 
right-hand complex half-plane at infinity with poles at $t = 1/2$ and 
$t = s + k, k \ge 0$. We choose $\Reb\,s > 1/2$ and obtain
\bqa
{\cal F}(\rho,\rho_{\ssS\ssL}) &\equiv& \rho\,F(\rho) =
\sum_{n=1}^{2}\,(-1)^n\,
\frac{\rho\,\beta_n}{2}\,\frac{1}{2\,\pi\,i}\,
\int_{-i\,\infty}^{+i\,\infty}\,ds\,\frac{\egam{1-s}}{s-1}\,
\Bigl(\frac{\rho_{\ssS\ssL}}{\rho}\Bigr)^s
\nl
{}&\times& \Bigl[ \pi^{1/2}\,\egam{s-\frac{1}{2}}\,
(h\,\beta^2_n\,\rho_{\ssS\ssL})^{-1/2} + \sum_{k=0}^{\infty}\,
\frac{(-1)^{k+1}}{k\,!}\,
\frac{\egam{s+k}}{s+k-1/2}\,(h\,\beta^2_n\,\rho_{\ssS\ssL})^{-s-k}\Bigr].
\eqa
${\cal F}$ is naturally split into two parts:
\bqa
{\cal F}(\rho,\rho_{\ssS\ssL}) &=& \sum_{n=1}^2\,(-1)^n\,
\Bigl[ {\cal F}^{(n)}_1 +
\sum_{k=0}^{\infty}\,\frac{(-1)^k}{k\,!}\,{\cal F}^{(n,k)}_2\Bigr]
\nl
{}&=& \sum_{n=1}^2\,(-1)^n\,
\frac{\rho\,\beta_n}{2}\,\Bigl[
\pi^{1/2}\,(h\,\beta^2_n\,\rho_{\ssS\ssL})^{-1/2}\,{\cal F}_{1,n} +
\sum_{k=0}^{\infty}\,\frac{(-1)^k}{k\,!}\,{\cal F}_{2,n,k}\Bigr].
\label{nats}
\eqa
For the first part in \eqn{nats} we introduce the new variable
$\kappa = \rho_{\ssS\ssL}/\rho$ and derive the following coefficient:
\bqa
{\cal F}_{1,n} &=& \frac{1}{2\,\pi\,i}\,\int_{-i\,\infty}^{+i\,\infty}\,ds\,
\frac{\egam{1-s}\,\egam{s-1/2}}{s-1}\,\kappa^s.
\eqa
We close the contour over the left-hand complex half-plane at infinity.
with poles at $s = 1/2 - k, k \ge 0$, obtaining
\bqa
{\cal F}_{1,n} &=& -\,\sum_{k=0}^{\infty}\,\frac{(-1)^k}{k\,!}\,
\frac{\egam{k+1/2}}{k+1/2}\,\kappa^{1/2-k} =
- \kappa^{1/2}\,\sum_{k=0}^{\infty}\,
\frac{\egams{k+1/2}}{\egam{k+3/2}}\,\frac{(-\kappa)^{-k}}{k\,!}
\nl
{}&=& - 2\,{\pi\,\kappa}^{1/2}\,
\hyper{\frac{1}{2}}{\frac{1}{2}}{\frac{3}{2}}{-\,\frac{1}{\kappa}} =
- 2\,\pi^{1/2}\,\kappa\,
\ln\frac{1 + (1+\kappa)^{1/2}}{\kappa^{1/2}}.
\eqa
As a consequence we have
\bq
{\cal F}^{(n)}_1 = -\,\sign{\beta_n}\,\pi\,
(\frac{\rho_{\ssS\ssL}}{h})^{1/2}\,
\ln\frac{\rho^{1/2}+(\rho+\rho_{\ssS\ssL})^{1/2}}{\rho^{1/2}_{\ssS\ssL}}.
\label{resI}
\eq
For the second part in \eqn{nats} the poles are at $s = l + 1, l \ge 0$ where 
$l = 0$ gives a double pole. We obtain
\bqa
{\cal F}_{2,n,k} &=& -\,\frac{\egam{k+1}}{k+1/2}\,
(h\,\beta^2_n)^{-k-1}\,\rho^{-1}\,\rho^{-k}_{\ssS\ssL}\,
\Bigl[ -\,\frac{1}{k+1/2} + \gamma + \psi(k+1) -
\ln(h\,\beta^2_n\,\rho)\Bigr]
\nl
{}&-& \sum_{l=1}^{\infty}\,\frac{(-1)^l}{l\,!}\,
\frac{\egam{k+l+1}}{l\,(k+l+1/2)}\,(h\,\beta^2_n)^{-k-l-1}\,\rho^{-l-1}\,
\rho^{-k}_{\ssS\ssL}.
\label{resII}
\eqa
Substituting back into \eqn{adone} gives the requested result.

There is another situation that requires particular care: consider a quadratic
form $h\,x^2 + 2\,k\,x + l$ and suppose that $b \ne 0$ but that $l \approx 0$
or that, equivalently $h+ 2\,k + l \approx 0$. The solution that amounts to
iterate the BT procedure for $f(k)$ is unstable because of the surface terms
in the integration by parts. In this case we take the Mellin-Barnes 
transform of the integrand and  write
\bqa
f(s) &=& \frac{1}{2\,\pi\,i}\,\int_{-i\,\infty}^{+i\,\infty}\,dt\,
B(t\,,\,s-t)\,L^{s-t}\,\intfx{x}\,x^{-t}\,(2\,k + h\,x)^{-t}
\nl
{}&=&
\frac{1}{2\,\pi\,i}\,\int_{-i\,\infty}^{+i\,\infty}\,dt\,
B(t\,,\,s-t)\,L^{s-t}\,\frac{(2\,k)^{-t}}{1-t}\,
\hyper{t}{1-t}{2-t}{-\,\frac{h}{2\,k}},
\eqa
where $L = 1/l$ and $0 < \Reb\,t < \Reb\,s < 1$.
Therefore we obtain
\bqa
{\cal F}(\rho,L) &\equiv& \rho\,F(\rho) = -\,\frac{\rho}{(2\,\pi\,i)^2}\,
\int_{-i\,\infty}^{+i\,\infty}\,ds\,dt\,\frac{\egam{1-s}\,\egam{t}\,
\egam{s-t}}{(s - 1)\,(t - 1)}
\nl
{}&\times& \Bigl(\frac{L}{\rho}\Bigr)^s\,(2\,k\,L)^{-t}\,
\hyper{t}{1-t}{2-t}{-\,\frac{h}{2\,k}}.
\eqa
For large $\mid L\mid$ we close the $t-$integration contour over the
right-hand complex half-plane at infinity with poles at $t = 1$ and 
$t = s + n$. We get
\bqa
{\cal F}(\rho,L) &=& -\,\frac{\rho}{(2\,\pi\,i)^2}\,
\int_{-i\,\infty}^{+i\,\infty}\,ds\,\frac{\egam{1-s}}{s-1}\,
\Bigl(\frac{L}{\rho}\Bigr)^s\,\Bigl[ -\,\egam{s-1}\,(2\,k\,L)^{-1} +
\sum_{n=0}^{\infty}\,\frac{(-1)^n}{n\,!}\,\egam{s+n-1}
\nl
{}&\times& (2\,k\,L)^{-s-n}\,
\hyper{s+n}{1-s-n}{2-s-n}{-\,\frac{h}{2\,k}}\Bigr].
\eqa
For the first term we write
\bqa
\frac{1}{2\,\pi\,i}\,
\int_{-i\,\infty}^{+i\,\infty}\,ds\,\frac{\egam{1-s}\,\egam{s-1}}{s-1}\,
\Bigl(\frac{L}{\rho}\Bigr)^s &=& \sum_{n=1}^{\infty}\,\frac{1}{n^2}\,
\Bigl(-\,\frac{L}{\rho}\Bigr)^{1-n} =
-\,\frac{L}{\rho}\,\li{2}{-\,\frac{\rho}{L}},
\eqa
where we closed the $s-$integration contour over the left-hand complex 
half-plane at infinity, with poles at $s = 1 - n, n\ge 1$. 
For the second term we need special cases of the hypergeometric 
function~\cite{ellip}:
\bqa
\hyper{-n}{1+n}{2+n}{z} &=& \frac{1}{(2+n)_n}\,\Bigl(\frac{1-z}{z}\Bigr)^{n+1}\,
\frac{d^n}{dz^n}\,\Bigl[ \frac{z^{2\,n+1}}{1-z}\Bigr].
\eqa
\boldmath \subsection{A new integral representation for $C_0$} \unboldmath
It is very important to derive our results, for all kinematical regions,
in more than one form in order to be able to provide reliable cross-checks.
For this reason we derived another representation for $C_0$, to be used
when the BT coefficient $B_3$ vanishes and the integral is singular.
Starting from \eqn{refQi} a sector decomposition of the square $[0,1]^2$ gives
\bqa
C^{\rm square}_0 &=&
\sum_{i=1}^4\,\alpha_i\,\beta_i\,
\Big( \dsimp{2} + \int\,d\bar{S}_2 \Big)\,
\Big[ Q_i(x_1,x_2) + B_3 \Big]^{-1-\ep/2} 
\nl
{}&=& \sum_{i=1}^4\,\sum_{j=1}^2\,\alpha_i\,\beta_i\,\dcub{2}\,x_2\,
\Big[ x_2^2\,Q_{ij}(x_1) + B_3 \Big]^{-1-\ep/2} , 
\eqa
where we introduced $Q_{i1}(x_1) = Q_i(x_1,1)$ and
$Q_{i2}(x_1) = Q_i(1,x_1)$. They  have the following form:
$Q_{ij}(x_1)= h_{ij}\,x_1^2 + 2\,k_{ij}\,x_1 + l_{ij}$.
Note that, according to the $i\,\delta$ prescription, the following replacement
must always be understood: $Q_{ij},B_3 \to Q_{ij},B_3 - i\,\delta$.
Now we use the results of Appendix A: the coefficients in the corresponding 
master integral are:
\bq
a = Q_{ij}(x_1),
\qquad
b = B_3,
\qquad
\alpha = 1 + \ep/2,
\qquad
\beta = 1.
\eq
Therefore we obtain:
\bqa
C^{\rm square}_0 &=&
\sum_{i=1}^4\,\sum_{j=1}^2\,\alpha_i\,\beta_i\,\intfx{x_1}\,
\Bigg\{
\frac{1}{\ep}\,B_3^{-\ep/2}\,Q^{-1}_{ij}(x_1) - \intfx{x_2}\,x_2^{-1+\ep}\,
Q^{-1-\ep/2}_{ij}(x_1,x_2) \Bigg\},
\nl
Q_{ij}(x_1,x_2) &=& Q_{ij}(x_1) + B_3\,x_2^2 .
\label{c0square:gr}
\eqa
If the factors $b_{ij}=l_{ij}-k_{ij}^2/h_{ij} = 
\alpha^2_i\beta^2_i{\rm det}(H)/h_{ij}$, corresponding to the quadratic
forms $Q_{ij}(x_1)$, are different from zero we apply the following BT
relations:
\bq
Q_{ij}^{-1}(x)=
\frac{1}{b_{ij}}\,
\left[ 1 - \frac{1}{2}\,\Big( x + \frac{k_{ij}}{h_{ij}} \Big)\,
  \partial_{x}\,\ln Q_{ij}(x)
\right] ,
\eq
\bq
Q^{\mu-1}_{ij}(x_1,x_2) = \frac{1}{b_{ij}}\,\left\{
1 - \frac{1}{2\,\mu}\,  \left[
    \Big( x_1+\frac{k_{ij}}{h_{ij}} \Big)\,\partial_{x_1}
  + x_2\,\partial_{x_2}  \right]\right\}\,
Q^{\mu}_{ij}(x_1,x_2).
\eq
Let us introduce new auxiliary variables:
$\zeta_{ij1} = - k_{ij}/h_{ij}$ and $\zeta_{ij2} = 1 + k_{ij}/h_{ij}$
to obtain a compact result:
\bq
C^{\rm square}_0 =
\sum_{i=1}^4\,\sum_{j=1}^2\,\frac{\alpha_i\,\beta_i}{8\,b_{ij}}\,C_{ij} ,
\eq
\bqa
C_{ij} &=&
2\,\ln B_3\,
\Big[
- \intfx{x_1}\,\ln Q_{ij}(x_1)
+ \zeta_{ij2}\,\ln Q_{ij}(1)
+ \zeta_{ij1}\,\ln Q_{ij}(0)
- 2\Big]
\nl
&{}&
- \,4\,\intfxx{x_1}{x_2}\,
  \left( \frac{\ln Q_{ij}(x_1,x_2)}{x_2} \right)_+
+ \intfx{x_1}\,
  \Big[ \ln^2 Q_{ij}(x_1) + 4\,\ln Q_{ij}(x_1,1) \Big]
\nl
&{}&
+ \,4\,\intfx{x_2}\,
  \Big[  \zeta_{ij2}\,
    \left( \frac{\ln Q_{ij}(1,x_2)}{x_2} \right)_+
  + \zeta_{ij1}\, \left( \frac{\ln Q_{ij}(0,x_2)}{x_2} \right)_+
  \Big] - \,\zeta_{ij2}\,\ln^2 Q_{ij}(1) - \zeta_{ij1}\,\ln^2 Q_{ij}(0) .
\nl
\eqa
All the $Q_{ij}$ and $B_3$ are given a $-i\delta$.
The `+'-distribution is defined, as usual, by its action on a generic test
function $f(x)$:
\bq
\intfx{x}\,\Bigl(\frac{f(x)}{x}\Bigr)_+ = \intfx{x}\,
\frac{f(x) - f(0)}{x}.
\label{plusd}
\eq
If instead some $b_{ij}$ approaches zero we must repeat the procedure.
This can be achieved if we start from \eqn{c0square:gr} and change variables 
according to $x_1 \to x_1 - k_{ij}/h_{ij}$.
Then we decompose the $x_1$ integration interval and remap each piece into
the unit square; the result is
\bqa
C^{\rm square}_0 &=&
\sum_{i=1}^4\,\sum_{j=1}^2\,\sum_{l=1}^2\,
\alpha_i\,\beta_i\,\zeta_{ijl}\,\intfx{x_1}\,
\Bigg\{
\frac{1}{\ep}\,B_3^{-\ep/2}\,
\Big[ h_{ij}\,\zeta^2_{ijl}\,x_1^2 + b_{ij} - i\delta \Big]^{-1}
\nl
&{}&
- \intfx{x_2}\,x_2^{-1+\ep}\,
  \Big[
  h_{ij}\,\zeta^2_{ijl}\,x_1^2 + B_3\,x_2^2 + b_{ij} \Big]^{-1-\ep/2}
\Bigg\},
\eqa
where the replacement $b_{ij},h_{ij} \to b_{ij},h_{ij} - i\,\delta$ is also 
understood. Now we apply the result of Appendix A to both integrals, deriving
\bqa
C^{\rm square}_0 &=&
\sum_{i=1}^4\,\sum_{j=1}^2\,\sum_{l=1}^2\,
\alpha_i\,\beta_i\,\zeta_{ijl}\,
\Bigg\{
  \frac{\pi}{2\,\ep}\,(h_{ij}\,\zeta^2_{ijl})^{-1/2}\,
  B_3^{-\ep/2}\,b_{ij}^{-1/2}\,
\nl
&{}&
- \frac{1}{2}\,B\left( \frac{1}{2},\frac{1}{2} + \frac{\ep}{2} \right)\,
  (h_{ij}\,\zeta^2_{ijl})^{-1/2}\,\intfx{x_2}\,x_2^{-1+\ep}\,
  \Big[ B_3\,x_2^2 + b_{ij} \Big]^{-1/2-\ep/2}\,
\nl
&{}&
- \frac{1}{\ep}\,B_3^{-\ep/2}\,\intfx{x_1}\,\Big[
h_{ij}\,\zeta^2_{ijl} + b_{ij}\,x_1^2\Big]^{-1}
\nl
&{}&
+ \intfxx{x_1}{x_2}\,x_1^{\ep}\,x_2^{-1+\ep}\,
  \Big[
    h_{ij}\,\zeta^2_{ijl} 
  + \Big( B_3\,x_2^2 + b_{ij} \Big)\,x_1^2
  \Big]^{-1-\ep/2}
\Bigg\} .
\label{C0:B3/bij}
\eqa
The first three terms in \eqn{C0:B3/bij} are divergent for $B_3 \sim b_{ij} 
\sim 0$. The second integral depends only on the ratio
\bq
\kappa_{ij}= \frac{B_3-i\delta}{b_{ij}-i\delta},
\eq
so that, when we apply the master formula of Appendix A, the result is
\bq
\intfx{x_2}\,x_2^{-1+\ep}\,
\Big[ B_3\,x_2^2 + b_{ij} \Big]^{-1/2-\ep/2} =
\frac{1}{2}\,b_{ij}^{-1/2}\,B_3^{-\ep/2}\,
B\left( \frac{\ep}{2},\frac{1}{2} \right)
- \intfx{x_2}\,
\Big[ b_{ij}\,x_2^2 + B_3 \Big]^{-1/2-\ep/2}.
\label{preveq}
\eq
Note that the first term in \eqn{preveq} cancels the first term of 
\eqn{C0:B3/bij}:
\bqa
C^{\rm square}_0 &=&
\sum_{i=1}^4\,\sum_{j=1}^2\,\sum_{l=1}^2\,
\alpha_i\,\beta_i\,\zeta_{ijl}\,
\Bigg\{
- \frac{1}{\ep}\,B_3^{-\ep/2}\,\intfx{x_1}\,\Big[
h_{ij}\,\zeta^2_{ijl}  + b_{ij}\,x_1^2 \Big]^{-1}
\nl
&{}&
+\frac{1}{2}\,B\left( \frac{1}{2},\frac{1}{2} + \frac{\ep}{2} \right)\,
(h_{ij}\,\zeta^2_{ijl})^{-1/2}\,\intfx{x_2}\,
\Big[ b_{ij}\,x_2^2 + B_3 \Big]^{-1/2-\ep/2}
\nl
&{}&
+ \intfxx{x_1}{x_2}\,x_1^{\ep}\,x_2^{-1+\ep}\,
  \Big[
    h_{ij}\,\zeta^2_{ijl} 
  + \Big( B_3\,x_2^2 + b_{ij}  \Big)\,x_1^2
  \Big]^{-1-\ep/2}
\Bigg\}.
\eqa
Using this intermediate result we can apply the BT procedure to increment the
power of the quadratic forms in the first and in the last integrals while
the second one can be calculated analytically. Finally a Laurent expansion
around $\ep = 0$ is performed giving:
\bq
C^{\rm square}_0 =
\frac{1}{2}\,\sum_{i=1}^4\,\sum_{j=1}^2\,\sum_{l=1}^2\,
\alpha_i\,\beta_i\,\zeta_{ijl}\,
\Big[
C^{\rm square}_{0,ijl} + \dsimp{1}\,C^{\rm square}_{1,ijl}
+ \dsimp{2}\,C^{\rm square}_{2,ijl}
\Big] ,
\label{polescanc}
\eq
\bqa
C^{\rm square}_{0,ijl} &=&
\pi\,(h_{ij}\,\zeta^2_{ijl})^{-1/2}\,
b_{ij}^{-1/2}\,
\ln\frac{ 1 + (1+\kappa_{ij})^{1/2} }{ \kappa_{ij}^{1/2} }
\nl
&{}&
+ \frac{1}{h_{ij}\,\zeta^2_{ijl}}\,
\Bigg[
\ln B_3\,
\Big(
1 - \frac{1}{2}\,\ln P_{ijl}(1,0)
\Big)
+ \frac{1}{4}\,\ln^2 P_{ijl}(1,0)
\Bigg],
\nl
C^{\rm square}_{1,ijl} &=&
\frac{1}{h_{ij}\,\zeta^2_{ijl}}\,
\Bigg[
\frac{1}{2}\,\ln B_3\,\ln P_{ijl}(x,0)
- \frac{1}{4}\,\ln^2\frac{P_{ijl}(x,0)}{x^2}
- \left( \frac{\ln P_{ijl}(1,x)}{x} \right)_+
\Bigg],
\nl
C^{\rm square}_{2,ijl} &=&
\frac{1}{h_{ij}\,\zeta^2_{ijl}}\,
\left( \frac{\ln P_{ijl}(x_1,x_2)}{x_2} \right)_+,
\eqa
where we introduced the new quadratic form:
\bq
P_{ijl}(x_1,x_2)=
h_{ij}\,\zeta^2_{ijl} + b_{ij}\,x_1^2 + B_3\,x_2^2 -i\delta.
\eq
Note that the poles in $\ep$ cancel out in \eqn{polescanc}, as expected.
\section{\boldmath Infrared divergent $C_0$\label{IRC0} \unboldmath}
Before discussing the general classification of IR divergent configurations
for the three-point function, we consider a specific example, namely 
$p^2_1= - m^2_1, p^2_2 = - m^2_3$ and $m_2 = 0$ (see \fig{fig:threepoint}). 
We obtain
\bq
C^{\ssI\ssR}_0 \equiv C_0(Q^2,-m^2_1,-m^2_3\,;\,m_1,0,m_3) = 
\dcub{2}\,\,x^{-1-\ep}_1\,\chi^{-1-\ep/2}(x_2) =
-\,\frac{1}{\ep}\,\intfx{x}\,\chi^{-1-\ep/2}(x),
\label{varcas}
\eq
where we introduced
\bq
\chi(x) = -Q^2\,x^2 + (Q^2 + m^2_3 - m^2_1)\,x + m^2_1,
\qquad
Q^2 = (p_1+p_2)^2.
\label{irquad}
\eq
In this case the sector decomposition is trivial. The BT factor and co-factor
associated to the quadratic form of \eqn{irquad} are
\bq
B_{\rm th} = \frac{1}{4\,Q^2}\,\lambda(-Q^2,m^2_1,m^2_3), \quad
X = \frac{1}{2}\,\frac{Q^2+m^2_3-m^2_1}{Q^2}.
\label{Bth}
\eq
and $B_{\rm th} = 0$ at the normal(pseudo)-threshold of $C_0$. In the
following we discuss the various cases under which \eqn{varcas} can be
classified.
\boldmath \subsection{$B_{\rm th} \ne 0$} \unboldmath
As long as $B_{\rm th} \ne 0$ 
it is possible to apply the standard BT procedure; we derive
\bqa
C^{\ssI\ssR}_0 &=& -\,\frac{1}{B_{\rm th}\,\ep}\,\intfx{x}\,
\Bigl[ 1 + \frac{1}{\ep}\,(x - X)\,\frac{d}{dx}\Bigr]\,
\Bigl[ 1 - \frac{\ep}{2}\,\ln \chi(x) + \frac{\ep^2}{8}\,
\ln^2 \chi(x) + \ord{\ep^3}\Bigr]
\nl
{}&=& \frac{1}{B_{\rm th}}\,\Bigl( \frac{C^{\rm res}_0}{\ep} + C^{\rm fin}_0 +
\ord{\ep}\Bigr),
\eqa
where the infrared residue and the finite part are
\bqa
C^{\rm res}_0 &=& - 1 + \frac{1}{2}\,(1 - X)\,\ln\chi(1) + \frac{1}{2}\,X\,
\ln\chi(0) - \frac{1}{2}\,\intfx{x}\,\ln\chi(x),
\nl
C^{\rm fin}_0 &=& - \frac{1}{8}\,(1 - X)\,\ln^2\chi(1) + \frac{1}{8}\,X\,
\ln^2\chi(0) +  \frac{1}{2}\,\intfx{x}\,\Bigl[\ln\chi(x) + \frac{1}{4}\,
\ln^2\chi(x)\Bigr].
\eqa
\boldmath \subsection{Infrared configuration and $B_{\rm th} 
\approx 0$}\unboldmath
If $B_{\rm th} \approx 0$ then we introduce $h = - Q^2$ and write
\bq
C^{\ssI\ssR}_0 = -\,\frac{1}{\ep}\,\int_{-X}^{1-X}\,dx\,
(h\,x^2 + B_{\rm th})^{-1-\ep/2}.
\eq
and consider three sub-cases: from \eqn{Bth} we see that $B_{\rm th} = 0$
only if $Q^2 < 0$, i.e. for $-Q^2 = (m_1 \pm m_3)^2$. Note that, in this case,
we have $X = m_1/(m_1 \pm m_3)$. 
\subsubsection{$0 \le X \le 1$}
Suppose that $0 \le X \le 1$ so that the configuration is actually singular 
at $B_{\rm th} = 0$. This corresponds to the normal threshold 
$Q^2 = - (m_1 + m_3)^2$. Introducing a Mellin-Barnes splitting, as described
in \sect{btt}, and performing the $x$-integration gives
\bqa
C^{\ssI\ssR}_0 &=&  \frac{1}{2\,\ep}\,\frac{1}{2\,\pi\,i}\,
\int_{-\,i\,\infty}^{+\,i\,\infty}\,ds\,\,
\frac{\egam{-s}\,\egam{s+1+\ep/2}}{\egam{1+\ep/2}\,(s+1/2+\ep/2)}
\nl
{}&\times& (h-i\,\delta)^{-s-1-\ep/2}\,(B_{\rm th}-i\,\delta)^s\,\Bigl[
(1 - X)^{-1-2\,s-\ep} + X^{-1-2\,s-\ep}\Bigr],
\eqa
with $-1-\ep/2 \le \Reb\,s \le -1/2-\ep/2$.
Note that we explicitly indicated the infinitesimal imaginary parts.
Since we are interested in the behavior for $B_{\rm th} \to 0$ the $s$ 
contour can be closed over the right-hand complex half-plane at infinity,
with poles at $s = -1/2-\ep/2$ and at $s = k, k \ge 0$. The residue at 
$s = -1/2-\ep/2$ is
\bq
R(s = -\frac{1+\ep}{2} ) = \frac{\pi}{h^{1/2}}\,(B_{\rm th}-i\,\delta)^{-1/2}
\Bigl\{ -\,\frac{1}{\ep} + \frac{1}{2}\,\Bigl[
\ln(B_{\rm th}-i\,\delta) + \psi(1) - \psi(\frac{1}{2})\Bigr]\Bigr\}.
\eq
At $s = k$ we have instead
\bqa
R(s = k) &=& \frac{(-1)^k}{k+1/2}\,h^{-k-1}\,B^k_{\rm th}\,
\sum_{a=X,1-X}\,R(a)\,a^{-1-2\,k}\,
\nl
R(a) &=& \Bigl\{ \frac{1}{2\,\ep} - \frac{1}{4}\,\Bigl[ \frac{1}{k+1/2} +
\ln(h-i\,\delta) + 2\,\ln a + \psi(1) - \psi(k+1)\Bigr]\Bigr\}.
\label{eqdefRa}
\eqa
Finally, we present the last two sub-cases.
\subsubsection{$X < 0 $ or $X >1$}
In these cases, corresponding to the pseudo-threshold $Q^2 = -(m_1 - m_3)^2$,
we will write
\bq
C^{\ssI\ssR}_0 = -\,\frac{1}{\ep}\,\Bigl[ \theta(-X) - \theta(X-1)\Bigr]\,
\Bigl[ \int_0^{|1-X|}\,dx - \int_0^{|X|}\,dx\Bigr]\,(h\,x^2 + 
B_{\rm th})^{-1-\ep/2}.
\eq
Suppose that $X < 0$, after the Mellin-Barnes splitting we obtain
\bqa
C^{\ssI\ssR}_0 &=&  \frac{1}{2\,\ep}\,\frac{1}{2\,\pi\,i}\,
\int_{-\,i\,\infty}^{+\,i\,\infty}\,ds\,\,
\frac{\egam{-s}\,\egam{s+1+\ep/2}}{\egam{1+\ep/2}\,(s+1/2+\ep/2)}
\nl
{}&\times& (h-i\,\delta)^{-s-1-\ep/2}\,(B_{\rm th}-i\,\delta)^s\,\Bigl[
(1 - X)^{-1-2\,s-\ep} - (-X)^{-1-2\,s-\ep}\Bigr].
\eqa
There is no pole at $s = -1/2-\ep/2$ but only poles at $s = k, k \ge 0$ with
residues
\bqa
R(s = k) &=& \frac{(-1)^k}{k+1/2}\,h^{-k-1}\,B^k_{\rm th}\,
\Bigl[ (1-X)^{-1-2\,k}\,R(1-X) - \mid X\mid^{-1-2\,k}\,R(\mid X\mid),
\eqa
with a similar result if $X > 1$. $R$ is defined in \eqn{eqdefRa}.
\subsection{IR configurations and Landau equations\label{IRLEQ}}
The study of the Landau equations for a given diagram is the most elegant way
to classify its infrared divergent configurations. Again we refer to the 
one-loop vertex for illustration. The Landau equations are
\bq
\alpha_1\,(q^2 + m^2_1) = 0, \quad
\alpha_2\,((q+p_1)^2 + m^2_2) = 0, \quad
\alpha_3\,((q+p_1+p_2)^2 + m^2_3) = 0,
\label{exalanone}
\eq
\bq
\alpha_1\,q_{\mu} + \alpha_2\,(q+p_1)_{\mu} + 
\alpha_3\,(q+p_1+p_2)_{\mu} = 0.
\label{exalan}
\eq
A solution will be the leading singularity if it corresponds to $\alpha_i 
\ne 0, \forall i$. Next we multiply \eqn{exalan} by $q, p_1$ and $p_2$
respectively to obtain an homogeneous system of three equations where, 
moreover, we use $q^2 = - m^2_1$ etc, from \eqn{exalanone}.
A necessary and sufficient condition to have a proper solution, i.e. \ not all 
the $\alpha_i = 0$, requires the determinant of coefficients to be zero,
thus fixing a relation between internal and external masses.
Any configuration that satisfy this constraint is a Landau singularity for 
the diagram which, however, does not necessarily imply that the diagram itself
diverges at that configuration.

Suppose that we consider the following case: $p^2_{1,2}= - m^2, m_{1,3} = m$ 
and $m_2 = 0$. It is easily found that this configuration is a Landau 
singularity. The question is however of which kind. Let us insert the above 
values into the homogeneous system, what we obtain is
\bq
m^2\,\alpha_1 + (\frac{1}{2}\,Q^2 + m^2)\,\alpha_3 = 0,
\quad
(\frac{1}{2}\,Q^2 + m^2)\,\alpha_1 + m^2\,\alpha_3 = 0.
\eq
First of all we observe that it is not either $\alpha_1 = 0$ or $q^2 = -m^2_1$,
etc; it can be both. Secondly, our configuration, where $Q^2 = (p_1+p_2)^2$ is
unconstrained, is a singularity. Finally there is a special case of the general
configuration discussed ($Q^2$ free) which is even more singular giving, in the
infrared case, the true leading Landau singularity. To have $\alpha_{1,3} 
\neq 0$ we must require $Q^2 = - 4\,m^2$ which gives, in the annihilation
channel, the well-known threshold singularity on top of the infrared one. 
This condition emerges also from the following argument: inserting
$p^2_{1,2} = - m^2$ in the condition to have a proper solution one obtains
\bq
m^2_2\,Q^2 = 0, \quad {\mbox{and}} \quad m^2_2\,( m^2_2 - Q^2 - 4\,m^2) = 0,
\eq
In a certain sense the constraint $Q^2 = -\,4\,m^2$ is buried inside the
anomalous threshold condition. In other words, that all the propagators in a
diagram are on-shell and that the consistency relation is satisfied does not
necessarily imply that all $\alpha_i$ are different from zero: the infrared
case is a clear example.  Note the presence of a potential singularity also
at $Q^2 = 0$ which, however, is not physical, i.e.~not on the physical
Riemann sheet. The latter fact can be seen by inspecting the explicit
analytic result,
\bq
C^{\ssI\ssR}_0 = \frac{2}{\beta\,Q^2}\,\ln\frac{\beta+1}{\beta-1}\,
\frac{1}{\ep} + \,\,\mbox{IR finite}, \quad \beta^2 = 1 + 4\,\frac{m^2}{Q^2}.
\eq
Our strategy in the general classification of infrared divergent 
configurations, diagram by diagram, will be to assume a certain number of 
zero internal masses with at least one unconstrained external momentum. Then
we fix the remaining parameters to satisfy the consistency relation for the 
Landau equations. Finally, we return to the original set of Landau equations 
and look for additional constraints that are necessary in order to arrive to
the true leading singularity. The presence of a threshold-like singularity on
top of the infrared poles is the sign that, after extracting these poles, we
still have complications for the residual integrations that cannot be solved
with naive methods.

For $C_0$ and arbitrary masses we look for a solution of the homogeneous 
system of the three Landau equations in a situation where $Q^2$ is not
constrained (due to the symmetry of the problem it is sufficient to consider 
$Q^2$). The condition to have a proper solution is a quadratic equation in 
$Q^2$ with $m^2_2$ as the coefficient of $Q^4$; therefore we require $m_2 = 0$,
after which we obtain the following condition:
\bq
(p^2_1+m^2_1)\,(p^2_2 + m^2_3)\,Q^2 -
\Bigl[ (p^2_1+m^2_1) - (p^2_2 + m^2_3)\Bigr]\,\Bigl[
m^2_1\,(p^2_2+m^2_3) - m^2_3\,(p^2_1+m^2_1)\Bigr] = 0,
\eq
requiring $p^2_1 = -m^2_1$ and $p^2_2 = - m^2_3$, i.e.~the configuration that
we already discussed. If we substitute back into the Landau equations 
we obtain
\bqa
2\,m^2_1\,\alpha_1 + (Q^2 + m^2_1 + m^2_3)\,\alpha_3 &=& 0,
\nl
(Q^2 + m^2_1 + m^2_3)\,\alpha_1 + 2\,m^2_3\,\alpha_3 &=& 0,
\eqa
so that the leading singularity, i.e.~all $\alpha$'s different from zero,
is obtained for 
\bq
\lambda(-Q^2,m^2_1,m^2_3) = 0 \quad \equiv \quad B_3 = 0,
\eq
which clarifies once more the physical meaning of the BT factor. 
\section{Four-point functions (D-family)\label{D0}}
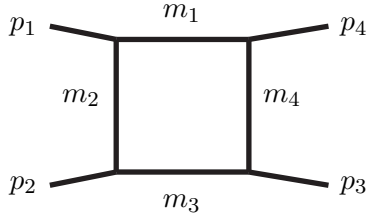
\begin{figure}[h]
\vspace*{-8mm}
\[
  \vcenter{\hbox{
  \begin{picture}(150,100)(0,0)
  \SetWidth{2.}
    \Line(25,80)(50,75)        
    \Line(50,25)(50,75)        
    \Line(25,20)(50,25)        
    \Line(130,80)(100,75)     
    \Line(100,75)(100,25)
    \Line(130,20)(100,25)     
    \Line(50,75)(100,75)
    \Line(50,25)(100,25)
    \Text(75,10)[cb]{$\mtre$}
    \Text(37,50)[cb]{$\mtwo$}
    \Text(75,82)[cb]{$\mone$}
    \Text(113,50)[cb]{$\mfor$}
    \Text(10,80)[lc]{$\pone$}
    \Text(10,20)[lc]{$\ptwo$}
    \Text(135,20)[lc]{$\ptre$}
    \Text(135,80)[lc]{$\pfor$}
  \end{picture}}}
\]
\vspace{-0.75cm}
\caption[]{The one-loop, four-point Green function of \eqn{origD0}.
All momenta are flowing inwards.}
\label{fig:fourpoint}
\end{figure}
\vskip 5pt
The generic, scalar, box (see Fig.~\ref{fig:fourpoint}) is given by
\bq
D_0 = \dsimp{3}\,V^{-2-\ep/2}(x_1,x_2,x_3),
\quad
V(x) = x^t\,H\,x + 2\,K^t\,x + L,
\label{origD0}
\eq
where $H_{ij} = -\,\spro{p_i}{p_j}$, $L = m_1^2$ and where
\bqa
K_1 &=& - \frac{1}{2} \, ( m_1^2 - m_2^2 - \spro{p_1}{p_1} ),
\quad
K_2 = - \frac{1}{2} \, ( m_2^2 - m_3^2 - 2\,\spro{p_1}{p_2} - 
\spro{p_2}{p_2} ),
\nl
K_3 &=& - \frac{1}{2} \, ( m_3^2 - m_4^2 - 2\,\spro{p_1}{p_3} - 
2\,\spro{p_2}{p_3} - \spro{p_3}{p_3} ).
\eqa
We are now in a position to write the answer for the case $B_4 \ne 0$.
\boldmath \subsection{Evaluation of $D_0$ when $B_4 \ne 0$
\label{regularD0}} \unboldmath
If we are away from $B_4 = L - K^t\,H^{-1}\,K = 0$ we can apply a BT iteration
obtaining 
\bq
B_4\,D_0 = -\,\frac{1}{2}\,
\dsimp{3}\,V^{-1-\ep/2}(x_1,x_2,x_3) + \frac{1}{2}\,\dsimp{2}\,
\sum_{i=0}^{3}\,( X_i - X_{i+1} )\,V^{-1-\ep/2}(\widehat{i\;i+1}),
\eq
where $X = -\,K^t\,H^{-1}$ and, moreover, $X_0 = 1, X_4 = 0$.
Contractions are defined as before with
\bq
\widehat{0\;1} = (1,x_1,x_2), \quad
\widehat{3\;4} = (x_1,x_2,0),
\eq
etc. Therefore $D_0(d=4)$ is a combination of a non-contracted term which can
be seen as $D_0(d=6)$ and of four pinches which can be represented as the 
triangles arising when we shrink a line of the box to a point. The latter 
are three-point functions that can be treated according
to the results of \sect{C0}. Furthermore, if $B_4 \ne 0$, 
$D_0(d=6)$ can be treated with a second BT iteration giving rise to logarithms,
\bqa
B_4\,D_0(d = 6) &=& \dsimp{3}\,V^{-1-\ep/2}(x_1,x_2,x_3) =
\dsimp{3}\,\Bigl[ \frac{3}{2}\,\ln V(x_1,x_2,x_3) + 1 \Bigr]
\nl
{}&-& \frac{1}{2}\,\dsimp{2}\,\sum_{i=0}^{3}\,
(X_i - X_{i+1})\,\ln V(\widehat{i\,i+i}).
\label{eqdefD0}
\eqa
Note that each three-point function will have, after the second iteration,
its own sub-leading BT coefficient. As long as $B_4 \ne 0$, the case where
some or all of the sub-leading $B$ are zero is fully covered by the results
of \sect{C0}. Similarly to \eqn{zeroG}, the case of $G_4 = {\rm det}(H) = 0$
reduces $D_0$ to a combination of four $C_0$ functions.
\subsection{Form factors in the D-family}
Also form factors can be written in a compact way by using the functions
\bqa
{\cal D}_{n_1,n_2,n_3} &=& \dsimp{3}\,
x^{n_1}_1\,x^{n_2}_2\,x^{n_3}_3\,V^{-2-\ep/2}(x_1,x_2,x_3),
\nl
{\cal C}_i(\widehat{j\,j+1}) &=& \dsimp{2}\,x_i\,
V^{-1-\ep/2}(\widehat{j\,j+1}) .
\eqa
Here we give one explicit example,
\bqa
B_4\,{\cal D}_{100} &=& \frac{1}{2}\,X_1\,D_0(d = 6) - {\cal D}_{100}(d=6) +
\frac{1}{2}\,\Bigl[ (1 - X_1)\,C_0(\widehat{0\,1}) + 
\sum_{i=1}^{3}\,(X_i - X_{i+1})\,{\cal C}_1(\widehat{i\,i+1})\Bigr],
\nl
\eqa
where the $d = 6$ component gives
\bqa
B_4\,{\cal D}_{100}(d=6) &=& \dsimp{3}\,\Bigl[
(2\,x_1 - \frac{1}{2}\,X_1)\,\ln V(x_1,x_2,x_3) + x_1\Bigr]
\nl
{}&-& \frac{1}{2}\,\dsimp{2}\,\Bigl[
(1 - X_1)\,\ln V(\widehat{0\,1}) + \sum_{i=1}^{3}\,(X_i - X_{i+1})\,x_1\,
\ln V(\widehat{i\,i+1})\Bigr]. 
\nl
\eqa
Suppose now that $B_4 \approx 0$: an alternative derivation for $D_0$ is 
needed. However, we first consider an extension of \eqn{eqdefD0}. 
\boldmath \subsection{$D_0$ at $\ord{\ep}$} \unboldmath
A three-point function with the insertion of a counterterm, which is needed in
a two-loop calculation, is formally identical to a one-loop four-point
function with one zero external momentum. Its expression at $\ord{\ep}$ is
needed and this follows from the BT algorithm quite nicely,
\bq
D_0(d) = D_0(4) + \ep\,D^{(1)}_0 + \ord{\ep^2}.
\eq
We obtain
\bq
16\,B_4\,D^{(1)}_0 = \frac{1}{B_4}\,\dsimp{3}\,I_3 +
\dsimp{2}\,\Bigl[ \frac{I^{\ssA}_{2}}{B_4} - 2\,
\sum_{i=w,s,e,n}\,\frac{d_i\,\,I^{\ssB}_{2i}}{B_{3i}}\Bigr] +
\sum_{i=w,s,e,n}\,\sum_{j=r,g,b}\,
\frac{d_i\,d_{ij}}{B_{3i}}\,\dsimp{1}\,I_{1ij} ,
\eq
where the integrand is given by
\bq
I_3 = 2 + 22\,\ln V(x_1,x_2,x_3) + 3\,\ln^2 V(x_1,x_2,x_3),
\eq
\bqa
I^{\ssA}_{2} &=& - \sum_{i=w,s,e,n}\,d_i\,\Bigl[ 6\,\ln V_i(x_1,x_2) +
\ln^2 V_i(x_1,x_2)\Bigr],
\eqa
\bqa
I^{\ssB}_{2i} &=& 1 + 4\,\ln V_i(x_1,x_2) + \ln^2 V_i(x_1,x_2),
\qquad
I_{1ij} = 2\,\ln V_{ij}(x_1) + \ln^2 V_{ij}(x_1),
\eqa
where we introduced the quadratics
\bq
V_w = V(x_1,x_1,x_2),
\quad
V_s = V(x_1,x_2,x_2),
\quad
V_e = V(x_1,x_2,0),
\quad
V_n = V(1,x_1,x_2), 
\label{fabove}
\eq
\bq
V_{ir} = V_i(x_1,x_1),
\quad
V_{ig} = V_i(x_1,0),
\quad
V_{ib} = V_i(1,x_1),
\label{sabove}
\eq
and the auxiliary factors
\bq
d_w = X_1 - X_2, \quad d_s = X_2 - X_3, \quad d_e =  X_3, \quad d_n = 1 - X_1,
\eq
\bq
d_{ir} = X_{i1} - X_{i2}, \quad d_{ig} = X_{i2} \quad d_{ib} = 1 - X_{i1}.
\eq
In the last equation $X_i$ and $X_{ij}$ are the BT co-factors of the
corresponding quadratics defined by \eqns{fabove}{sabove}.
\boldmath \subsection{Evaluation of $D_0$ when $B_4 \approx 0$ and $D_0$ is
regular\label{ed0reg}} \unboldmath
This is the case when $B_4 \to 0$ but the condition $0 < X_3 < X_2 < X_1 < 1$
is not fulfilled, i.e. the point of coordinates $X_i$ is outside the original
simplex. Thus $D_0$ is regular and the integration hyper-contour can be 
distorted to avoid the zeros of $V$ without pinches.
Two methods will be introduced to cover this case.
\subsubsection{Method I}
We proceed analogously to the case of $C_0$.
First of all we define $V_0(x_1,x_2,x_3) = V(x_1,x_2,x_3) - B_4$ and write 
down the BT relations corresponding to $B_4 \neq 0$ and $B_4 = 0$:
\bqa
\Big[ 1 + \frac{1}{2+\ep}\,(x-X)\,\partial_x \Big]\,
V^{-1-\ep/2}(x_1,x_2,x_3)
&=& B_4\,V^{-2-\ep/2}(x_1,x_2,x_3),
\nl
\Big[ 1 + \frac{1}{2+\ep}\,(x-X)\,\partial_x \Big]\,V_0^{-1-\ep/2}(x_1,x_2,x_3)
&=& 0 .
\eqa
Then we sum each side of the two equations and integrate by parts, obtaining:
\bq
D_0 =
- \,\frac{1}{2}\,\dsimp{3}\,B^{-1}_4\,
  V^{-1-\ep/2}(x_1,x_2,x_3)\,\mid_{\rm sub}
+ \frac{1}{2}\,\dsimp{2}\,\sum_{i=0}^3\,(X_i - X_{i+1})\,
  B^{-1}_4\,V^{-1-\ep/2}(\widehat{i\,\,i+1})\mid_{\rm sub},
\label{D0regular:gr1}
\eq
where $V^{\mu}|_{\rm sub} = V^{\mu} - V^{\mu}_0$.
The twofold integrals are nothing but differences of three-point functions
that can be easily treated according to the results obtained for $C_0$ in
\sect{C0}. For the first integral of \eqn{D0regular:gr1} we set $\ep = 0$ and 
apply again the BT algorithms:
\bqa
1 - \frac{1}{2}\,(x-X)\,\partial_x\,\ln V(x_1,x_2,x_3)
&=& B_4\,V^{-1}(x_1,x_2,x_3),
\nl
B_4\,\Big[ 1 + \frac{1}{2}\,(x-X)\,\partial_x \Big]\,
V_0^{-1}(x_1,x_2,x_3) &=& 0,
\nl
1 - \frac{1}{2}\,(x-X)\,\partial_x\,\ln V_0(x_1,x_2,x_3) &=& 0 .
\label{D0regular:BT2}
\eqa
By summing the first two equations and by subtracting the last one, we get:
\bq
B^{-1}_4\, V^{-1}(x_1,x_2,x_3)\,\mid_{\rm sub} =
- \frac{1}{2}\,(x-X)\,\partial_x\,
\left( \frac{\ln V(x_1,x_2,x_3)}{B_4^2} \right)_{++} ,
\eq
where we introduced the '++ distribution'
\bq
\left( \frac{\ln V(x_1,x_2,x_3)}{B_4^2} \right)_{++} \equiv
\frac{\ln V(x_1,x_2,x_3) - \ln V_0(x_1,x_2,x_3) - B_4\,V_0^{-1}(x_1,x_2,x_3)}
{B_4^2} .
\eq
Now another integration by parts is performed and we obtain:
\bq
\dsimp{3}\,B^{-1}_4\, V^{-1}(x_1,x_2,x_3)\,\mid_{\rm sub} =
\frac{3}{2}\,\dsimp{3}\,\left( \frac{\ln V(x_1,x_2,x_3)}{B_4^2} \right)_{++}
\eq
\bq
- \frac{1}{2}\,\dsimp{2}\,\sum_{i=0}^3\,(X_i - X_{i+1})\,
\left( \frac{\ln V(\widehat{i\,\,i+1})}{B_4^2} \right)_{++} .
\eq
In the twofold integral the '++' distribution contains terms such as
$V_0^{-1}(\widehat{i\,\,i+1})$ which are three-point functions with $\ep = 0$
and can be calculated in the usual way.
In the first integral we have $V_0^{-1}(x_1,x_2,x_3)$ 
which can be reduced again to a sum of three-point functions:
\bq
\dsimp{3}\,V_0^{-1}(x_1,x_2,x_3) =
\dsimp{2}\,\sum_{i=0}^3\,(X_i - X_{i+1})\,V_0^{-1}(\widehat{i\,\,i+1}) .
\eq
This result is obtained by using the second relation of \eqn{D0regular:BT2} 
together with an integration by parts.
An analogous procedure can be derived for tensor integrals.
\subsubsection{Method II}
Also for the four-point function another method is available.
First of all we perform a Taylor expansion around $B_4 = 0$ in
\eqn{origD0}:
\bq
D_0 = \dsimp{3}(X)\,\Bigl[ x^t\,H\,x + B_4\Bigr]^{-2-\ep/2} = 
\sum_{n=0}^{\infty}\,(n+1)\,{\cal D}(n+2)\,(-\,B_4)^n,
\label{calDcoeff}
\eq
and then we use the BT relation
\bq
\Bigl[ 1 + \frac{x\,\partial_x}{2\,n+\ep}
\Bigr]\,( x^t\,H\,x)^{-n-\ep/2} = 0.
\eq
In this way, an arbitrary coefficient in the Taylor expansion in $B_4$
can be reduced to a twofold integral to be treated according to a standard
BT algorithm. There will be several new quadratic forms to be introduced,
in two and one variables. We obtain the following form for the coefficients
\bq
{\cal D}(n) = \dsimp{3}(X)\,( x^t\,H\,x)^{-n-\ep/2} = 
\frac{1}{3 - 2\,n}\,\sum_{i=0}^{3}\,X_i^{\ssL}\,\dsimp{2}\,
Q^{-n-\ep/2}(T\,\widehat{i\,\,i+1}),
\label{TexpD0}
\eq
where $X_i^{\ssL}=X_i-X_{i+1}$ and the quadratic forms $Q(T\,\widehat{i\,\,i+1})$ are
\bqa
Q(T\,\widehat{0\,\,1}) \equiv
Q_0 &=& Q(1-X_1,x_1-X_2,x_2-X_3),
\nl
Q(T\,\widehat{1\,\,2}) \equiv
Q_1 &=& Q(x_1-X_1,x_1-X_2,x_2-X_3),
\nl
Q(T\,\widehat{2\,\,3}) \equiv
Q_2 &=& Q(x_1-X_1,x_2-X_2,x_2-X_3),
\nl
Q(T\,\widehat{3\,\,4}) \equiv
Q_3 &=& Q(x_1-X_1,x_2-X_2,-X_3).
\eqa
In general we will write $Q_i = x^t\,H_i\,x + 2\,K^t_i\,x + L_i$
and introduce corresponding sub-leading BT factors
$B_i=L_i-K^t_i\,H_i^{-1}\,K_i$ and co-factors $X_i= - K^t_i\,H_i^{-1}$ with
element $X_{ij}$ ($j=1,2$).
A second iteration will bring into the result quadratic forms in one variable,
\bq
Q_{i,0}(x) = Q_i(1,x),
\qquad
Q_{i,1}(x) = Q_i(x,x),
\qquad
Q_{i,2}(x) = Q_i(x,0),
\eq
which are written as $Q_{ij}(x) = h_{ij}\,x^2 + 2\,k_{ij}\,x + l_{ij}$ and
also have a sub-sub-leading BT factor $b_{ij}=l_{ij}-k_{ij}^2/h_{ij}$ and a
co-factor$X_{ij}$.
By applying again the BT procedure we finally obtain for ${\cal D}(2)$
\bq
{\cal D}(2) =
\frac{1}{4}\,\sum_{i=0}^{3}\,\sum_{j=0}^{2}\,
\frac{X_i^{\ssL}\,X_{ij}^{\ssL}}{B_{i}\,b_{ij}}\,
\Big[ {\cal D}_{0,ij}(2) + \intfx{x}\,{\cal D}_{1,ij}(2) \Big],
\eq
\bq
{\cal D}_{0,ij}(2) =
\Big( 1 + \frac{k_{ij}}{h_{ij}} \Big)\,\ln Q_{ij}(1)
- \frac{k_{ij}}{h_{ij}}\,\ln Q_{ij}(0) - 2,
\qquad
{\cal D}_{1,ij}(2) = - \ln Q_{ij}(x) ,
\eq
while for $n\neq 2$
\bq
{\cal D}(n) =
\frac{1}{(3-2\,n)\,(n-1)}\,\sum_{i=0}^{3}\,\frac{X_i^{\ssL}}{B_{i}}\,
\Big[ {\cal D}_{0,i}(n) + \intfx{x}\,{\cal D}_{1,i}(n)
+ \dsimp{2}\,{\cal D}_{2,i}(n) \Big],
\eq
\bq
{\cal D}_{0,i}(n) =
\frac{1}{4\,(n-2)}\,\sum_{j=0}^{2}\,\frac{X_{ij}^{\ssL}}{b_{ij}}\,
\Bigg[
  \Big( 1 + \frac{k_{ij}}{h_{ij}} \Big)\,Q_{ij}^{-n+2}(1)
- \frac{k_{ij}}{h_{ij}}\,Q_{ij}^{-n+2}(0)
\Bigg],
\eq
\bq
{\cal D}_{1,i}(n) =
\frac{1}{4\,(n-2)}\,\sum_{j=0}^{2}\,X_{ij}^{\ssL}\,
\Big( \frac{2\,(n-2)}{B_i} + \frac{2\,n-5}{b_{ij}} \Big)\,
Q_{ij}^{-n+2}(x),
\qquad
{\cal D}_{2,i}(n) = \frac{n-3}{B_i}\,Q_i^{-n+2}(x_1,x_2) .
\eq 
Now we can sum the series according to well-known results:
\bq
\sum_{n=1}^{\infty}\,x^n = \frac{x}{1-x},
\qquad
\sum_{n=1}^{\infty}\,\frac{x^n}{n} = - \ln(1-x),
\qquad
\sum_{n=1}^{\infty}\,\frac{x^n}{(2\,n+1)\,n} =
- \intfx{y}\,\ln (1-x\,y^2),
\eq
and obtain:
\bq
D_0 =
  {\cal D}(2)
+ \sum_{i=0}^{3}\,\frac{X_i^{\ssL}}{B_{i}}\,
\Big[ D_{0,i} + \intfx{x}\,D_{1,i} + \dsimp{2}\,D_{2,i} \Big],
\eq
where
\bqa
D_{0,i} &=&
\frac{1}{4}\,\sum_{j=0}^{2}\,\frac{X_{ij}^{\ssL}}{b_{ij}}\,
\intfx{y}\,
\Bigg[
  \Big( 1 + \frac{k_{ij}}{h_{ij}} \Big)\,
  \ln \frac{Q_{ij}(1)+B_4\,y^2}{Q_{ij}(1)}
- \frac{k_{ij}}{h_{ij}}\,\ln \frac{Q_{ij}(0)+B_4\,y^2}{Q_{ij}(0)}
\Bigg],
\nl
D_{1,i} &=&
\frac{1}{4}\,\sum_{j=0}^{2}\,X_{ij}^{\ssL}\,
\Bigg[
  \Big( \frac{1}{B_i} + \frac{1}{b_{ij}} \Big)\,
  \ln \frac{Q_{ij}(x)+B_4}{Q_{ij}(x)}
- \Big( \frac{1}{B_i} + \frac{2}{b_{ij}} \Big)\,
  \intfx{y}\,\ln \frac{Q_{ij}(x)+B_4\,y^2}{Q_{ij}(x)}
\Bigg],
\nl
D_{2,i}(n) &=&
\frac{1}{2\,B_i}\,
\Bigg[
  \frac{B_4}{Q_i(x_1,x_2)+B_4}
- \frac{3}{2}\,\ln \frac{Q_i(x_1,x_2)+B_4}{Q_i(x_1,x_2)}
+ \frac{3}{2}\,\intfx{y}\,\ln \frac{Q_i(x_1,x_2)+B_4\,y^2}{Q_i(x_1,x_2)}
\Bigg] .
\quad
\eqa
In the twofold integral we still have a polynomial with a negative power;
it is possible to increment the power of the latter by employing the BT 
relation:
\bq
(B_i+B_4)\,(Q_i+B_4)^{-1} =
1 - \frac{1}{2}\,(x-X_i)\,\partial_x\,\ln (Q_i+B_4) .
\eq
The usual integration by parts gives:
\bq
\dsimp{2}\,(Q_i+B_4)^{-1} =
\frac{1}{2\,(B_i+B_4)}\,
\Bigg[
1 + 2\,\dsimp{2}\,\ln(Q_i+B_4)
- \sum_{j=0}^2\,X_{ij}^{\ssL}\,\dsimp{1}\,\ln(Q_{ij}+B_4)
\Bigg].
\eq
In this way all coefficients in \eqn{calDcoeff} are explicitly computed. 
We can treat form factors with the same procedure; indeed the auxiliary 
functions
\bqa
D(\{n\}) &=& \dsimp{3}\,\prod_{i=1}^{3}\,x^{n_i}_i\,
V^{-2-\ep/2}(x_1,x_2,x_3) ,
\eqa
become linear combinations of
\bqa
{\cal D}(n,n_1,n_2,n_3) &=& \frac{1}{3-2\,n+n_1+n_2+n_3}\,\dsimp{2}
\nl
{}&\times& \Bigl\{
(x_1-X_2)^{n_2}\,(x_2-X_3)^{n_3}\,(1-X_1)^{n_1+1}\,Q^{-n}_0(x_1,x_2)
\nl
{}&+& (x_1-X_1)^{n_1}\,(x_1-X_2)^{n_2}\,(x_2-X_3)^{n_3}\,
      (X_1-X_2)\,Q^{-n}_1(x_1,x_2)
\nl
{}&+& (x_2-X_2)^{n_2}\,(x_2-X_3)^{n_3}\,(x_1-X_1)^{n_1}\,
      (X_2-X_3)\,Q^{-n}_2(x_1,x_2)
\nl
{}&-& (x_1-X_1)^{n_1}\,(x_2-X_2)^{n_2}\,(-\,X_3)^{n_3+1}\,Q^{-n}_3(x_1,x_2)
\Bigr\},
\eqa
which are form factors of the $C$-family.
\subsubsection{Sub-leading behavior}
The above derivation assumes that the sub-leading BT factors are not zero. If
this is the case we start from the fact that $D_0$ is a sum of terms of the
form
\bq
D^i_0 = -\,\sum_{n=0}^{\infty}\,\frac{n+1}{2\,n+1}\,(-\,B_4)^n\,
\dsimp{2}\,q_i\,Q^{-n-2}_i,
\label{markmu}
\eq
with $q_i$ constant.
When $B_i \approx 0$ we cannot apply the BT algorithm and it is necessary
to adopt an alternative strategy. 
If the generalized three-point function associated to the vanishing
coefficient $B_i$ is regular, we can perform another Taylor expansion
exactly as it was done for $C_0$. Otherwise we rewrite the sum as
follows:
\bqa
\sum_{n=0}^{\infty}\,\frac{n+1}{2\,n+1}\,Q^{-n-2}_i\,(-B_4)^n &=&
(B_4)^{-1/2}\,\int_0^{\scriptstyle \sqrt{B_4}}\,dz\,(Q_i + z^2)^{-2}.
\label{mark}
\eqa
To the price of introducing an extra integration we are now in the situation
where the series is replaced by a single term and the sub-leading BT factors
for the $x_1 - x_2$ integration are replaced by $B_i + z^2$ and the singular
behavior is left in the last $z$ integration. Note that, if $B_4 < 0$ we
simply have to replace everywhere $B_4$ and $Q_i$ with $-B_4, -Q_i$.
Actually the best way of dealing with the singular behavior is to change
variable and obtain
\bq
\dsimp{2}(X)\,\intfx{z}\,(x^t\,H_i\,x + B_{3i} + B_4\,z^2)^{-2},
\eq
and perform a Mellin-Barnes splitting (accompanied by a sector
decomposition) of $x^t\,H_i\,x$ and $B_{3i} + B_4\,z^2$. The leading
behavior comes from the single pole at $s = 1$ and it is, therefore,
proportional to
\bq
(B_4\,B_{3i})^{-1/2}\,\arctan\Bigl(\frac{B_4}{B_{3i}}\Bigr)^{1/2},
\eq
where $B \to B - i\,\delta$.
Using the identity of \eqn{mark} in \eqn{markmu} we obtain
\bq
D_0^i = - q_i\,\int_0^1 dz\,\dsimp{2}\,(Q_i + B_4\,z^2)^{-2}\,.
\eq
The first step toward the evaluation of the integral consists in splitting
the integral in the sum of two parts:
\bq
D_0^i \equiv I_1^i - I_2^i,
\qquad
I_{1;2}^i = - q_i\,\int_0^1 dz\,\int_0^1 dx_1\,\int_{0;x_1}^1 dx_2\,
(Q_i + B_4\,z^2)^{-2}.
\eq
Now the first integral is singular when $B_{3i} \to 0$ while the second is not.
Next, we consider  the calculation of $I_1^i$. Introducing the new variables of
integration $x'_i = x_i -X_i$, we have that
\bq
I_1^i = - q_i\,\dcub{2}(X)\,\int_0^1 dz\,(x^t\,H\,x + B_{3i} + B_4\,z^2)^{-2}\,;
\eq
then we apply the Mellin-Barnes transform and obtain
\bq
I_1^i = - \frac{q_i}{2 \pi i}\,\dcub{2}(X)\,\int_0^1 dz\,
\int_{-i \infty}^{i \infty} ds\,B(s,2-s)\,(x^t\,H\,x)^{-s}\,
\rho^{2-s}_{\rm mix}\,,
\eq
where $1/\rho_{\rm mix} = B_4\,z^2 + B_{3i}$.
It is possible to perform the integration over the variables $x_i$
and the result can be cast into the following, general, form:
\bq
\dcub{2}(X)\,(x^t\,H\,x)^{-s} = \frac{f(s)}{s-1}.
\eq
As a consequence we obtain
\bq
I_1^i = - \frac{q_i}{2 \pi i}\,\int_0^1 dz\,
\int_{-i \infty}^{i \infty} ds\,B(s,2-s)\,\rho^{2-s}_{\rm mix}\,
\frac{f(s)}{s-1}\, .
\eq
We can now perform the integration over $s$ and the leading divergent
behavior is given by the pole at $s = 1$:
\bq
I_1^{i,\rm sing} = q_i\,\Bigl( B_4\,B_{3i}\Bigr)^{-1/2}\,f(1)\,
\arctan\left(\frac{B_4}{B_{3i}} \right)^{1/2}\, ,
\eq
where the determination of the $\arctan$ function is fixed by the
$- i \delta$ prescription.
The regular part of the integral $I_1^i$ is
\bq
I_1^{i,\rm  reg}  = - q_i\,\sum_{n =0}^{\infty}\,(-1)^n\int_0^1 dz\,f(n+2)\,
(B_4 z^2 +B_{3i})^n\, ,
\eq
the integration over $z$ can then be performed:
\bq
I_1^{i, \rm reg}  =
q_i\,\sum_{n =0}^{\infty}\,(-1)^{n+1}\,f(n+2)\,(B_{3 i})^n\,
\hyper{\frac{1}{2}}{-n}{\frac{3}{2}}{-\frac{B_4}{B_{3 i}}} .
\eq
The integral $I_2$ can be calculated by performing a Taylor expansion around
$B_4 z^2 + B_{3i} = 0$:
\bqa
I_2^i &=& - q_i\,\sum_{k =0}^{\infty}\,\int_0^1 dz\,(-B_4\,z^2-B_{3i})^k
\int_{0}^{1} dx_1\,\int_{x_1}^{1} dx_2\,(k+1)\,[(x-X)^t\,H_i\,(x-X)]^{-2-k}\,, \nl
 &= &  - q_i\,\sum_{k =0}^{\infty}\,(-B_{3i})^k\,(k+1)\,
\hyper{\frac{1}{2}}{-n}{\frac{3}{2}}
{-\frac{B_4}{B_{3 i}}}\,{\cal I}_2(n+2)\, ,
\eqa
where we introduced the integral ${\cal I}_2(n)$ defined by
\bqa
{\cal I}_2(n) &=& \int_{0}^{1} dx_1\,\int_{x_1}^{1}  dx_2\,[
(x-X)^t\,H_i\,(x-X)]^{-n}\, = \int\,d\bar{S}_2(X)\,\Bigl[
x^t\,H_i\,x \Bigr]^{-n} .
\eqa
These integrals can be treated in analogy to what was done for
${\cal C}(n)$ (\eqn{deftaylorC}) and ${\cal D}(n)$ (\eqn{TexpD0}); using the
relation
\bq
\left[1+\frac{x \partial}{2 n} \right]\,
(x^t\,H\,x)^{-n} =0\,,
\eq
we obtain
\bqa
{\cal I}_2(n) &=& \frac{1}{2(n-1)}\,\int_0^1\,dx\,
\left[ (1-X_2)\,{\cal P}^{-n}_1 +
(X_2-X_1)\,{\cal P}^{-n}_2 +
X_1\,{\cal P}^{-n}_3 \right],
\nl
\eqa
where ${\cal P}(x_1,x_2) = x^t\,H\,x$ and where
\bq
{\cal P}_1 = {\cal P}(1-X_2,x-X_1), \qquad
{\cal P}_2 = {\cal P}(x-X_2,x-X_1), \qquad
{\cal P}_3 = {\cal P}(x-X_2,X_1).
\eq
The remaining one-dimensional integrals are now generalized $B_0$ functions
which can be calculated employing the methods introduced in \sect{gtpf}.
\boldmath \subsection{Evaluation of $D_0$ when $B_4 \approx 0$ and $D_0$ is 
singular\label{ed0sing}} \unboldmath
This is the case when $B_4 \to 0$ and the condition $0 < X_3 < X_2 < X_1 < 1$
is fulfilled, i.e. the point of coordinates $X_i$ is inside the original
simplex. Thus $D_0$ is singular and the integration hyper-contour cannot be
distorted due to the occurrence of a pinch.
In order to minimize the number of terms in the solution we write
\bqa
\int_0^1\,\prod_{i=1}^{3}\,dx_i &=& \sum_{\rm cycl. perm.}\,
\intfx{x_i}\,\int_0^{x_i}\,dx_j\,\int_0^{x_i}\,dx_k.
\eqa
Furthermore, for each term we may use the following identity
\bq
\int_0^{x_1}\,dx_2\,\int_0^{x_1}\,dx_3 =
\int_0^{x_1}\,dx_2\,\int_0^{x_2}\,dx_3 +
\int_0^{x_1}\,dx_3\,\int_0^{x_3}\,dx_2.
\eq
Therefore the integral over $[0,1]^3$ can be written as
\bqa
\int_0^1\,\prod_{i=1}^{3}\,dx_i\,V^{-2-\ep/2}(x_1,x_2,x_3) =
\dsimp{3}\,\sum_{\rm perm}\,V^{-2-\ep/2}(x_1,x_2,x_3),
\eqa
where the sum is over all permutations of $\{x_1,x_2,x_3\}$. The first term 
in the sum is exactly our original $D_0$ function while the rest gives the 
five complementary functions which, by construction, are regular and can be 
treated according to the strategy of \sect{ed0reg}.
As a consequence, we now have to evaluate $D^{\rm cube}_0$ when $B_4 
\approx 0$ and the point of coordinates $X_i$ is inside the unit cube.
Using standard techniques, already applied to the $C_0$-function, we
obtain
\bqa
D^{\rm cube}_0 &=& \Bigl\{ \prod_{i=1}^{3}\,\int_{-X_i}^{1-X_i}\,dx_i\,
\Bigr\}\,\Bigl( x^t\,H\,x + B_4\Bigr)^{-2-\ep/2}
\nl
{}&=& \sum_{i=1}^{8}\,\int_0^{\alpha_i}\,dx_1\,\int_0^{\beta_i}\,dx_2\,
\int_0^{\gamma_i}\,dx_3\,(Q_i + B_4)^{-2-\ep/2}.
\eqa
If $Q(x_1,x_2,x_3) = x^t\,H\,x$ the quadrics $Q_i$ are defined as
\bqa
Q_1 &=& Q(x_1,x_2,x_3), \quad
Q_2 = Q(-x_1,x_2,x_3), \quad 
Q_3 = Q(-x_1,-x_2,x_3), \nl
Q_4 &=& Q(-x_1,-x_2,-x_3), \quad
Q_5 = Q(-x_1,x_2,-x_3), \quad
Q_6 = Q(x_1,-x_2,x_3), \nl
Q_7 &=& Q(x_1,-x_2,-x_3), \quad
Q_8 = Q(x_1,x_2,-x_3), 
\eqa
and moreover the coefficients $\alpha_i, \beta_i$ and $\gamma_i$ are defined
by
\bqa
\alpha_1 &=& \alpha_6 = \alpha_7 = \alpha_8 = 1 - X_1,
\qquad
\alpha_2 = \alpha_3 = \alpha_4 = \alpha_5 = X_1,
\nl
\beta_1 &=& \beta_2 = \beta_5 = \beta_8 = 1 - X_2,
\qquad
\beta_3 = \beta_4 = \beta_6 = \beta_7 = X_2,
\nl
\gamma_1 &=& \gamma_2 = \gamma_3 = \gamma_6 = 1 - X_3,
\qquad
\gamma_4 = \gamma_5 = \gamma_7 = \gamma_8 = X_3.
\eqa
In this way the evaluation of $D^{\rm cube}_0$ is reduced to the one of
\bq
D_i = \int_0^{\alpha_i}\,dx_1\,\int_0^{\beta_i}\,dx_2\,
\int_0^{\gamma_i}\,dx_3\,(Q_i + B_4)^{-2-\ep/2},
\label{refDi}
\eq
for which we introduce a Mellin-Barnes transform
\bq
D_i = \frac{1}{2\,\pi\,i}\,\int_{-\,i\,\infty}^{+\,i\,\infty}\,
ds\,\rho^{2-s}_4\,{\cal D}_i(s),
\eq
where $\rho_4 = 1/B_4$. It follows that
\bq
{\cal D}_i(s) = \alpha_i\,\beta_i\,\gamma_i\,B(s\,,\,2-s)\,d_i(s),
\quad
d_i(s) = \dcub{3}\,
Q^{-s}_i(\alpha_i\,x_1,\beta_i\,x_2,\gamma_i\,x_3).
\eq
After a sector decomposition in $[0,1]^3$ we obtain
\bqa
d_i(s) &=& \dcub{3}\,\Bigl[ x^{2-2\,s}_1\,
Q^{-s}_i(\alpha_i,\beta_i\,x_2,\gamma_i\,x_3) +
x^{2-2\,s}_2\,Q^{-s}_i(\alpha_i\,x_1,\beta_i,\gamma_i\,x_3) +
x^{2-2\,s}_3\,Q^{-s}_i(\alpha_i\,x_1,\beta_i\,x_2,\gamma_i)\Bigr].
\nl
\eqa
Collecting all pieces we obtain
\bq
D^{\rm cube}_0 = \sum_{i=1}^{8}\,\sum_{j=1}^{3}\,\alpha_i\,\beta_i\,
\gamma_i\,D_{ij},
\quad
D_{ij} =  
\frac{1}{2\,\pi\,i}\,\int_{-\,i\,\infty}^{+\,i\,\infty}\,
ds\,\rho^{2-s}_4\,\frac{B(s,2-s)}{3 - 2\,s}\,\dcub{2}\,
Q^{-s}_{ij}(x_1,x_2) .
\eq
The evaluation of $D^{\rm cube}_0$ has been therefore reduced to the one
of an inverse Mellin-Barnes transform:
\bqa
D &=& \frac{1}{2\,\pi\,i}\,
\int_{-\,i\,\infty}^{+\,i\,\infty}\,ds\,\rho^{2-s}\,\frac{B(s,2-s)}{3-2\,s}\,
f(s).
\label{fcons}
\eqa
For large values of $|\rho|$ we close the $s$-contour over the right-hand 
complex half-plane at infinity with a single pole at $s = 3/2$ and also at 
$s = 2 + k, k \ge 0$. The result is straightforward
\bqa
D &=& \frac{\pi}{4}\,f(\frac{3}{2})\,\rho^{1/2} -
\sum_{k=0}^{\infty}\,\frac{\egam{k+2}}{2\,k+1}\,f(k+2)\,
\frac{(-\rho)^{-k}}{k\,!}.
\label{scons}
\eqa
Since $f$ stands generically for
\bq
f(s) \sim \dcub{2}\,Q^{-s}_{ij}(x_1,x_2),
\label{tcons}
\eq
we have to evaluate this integral for $s = 3/2$ as well as for $s$ equal to
an arbitrary integer $\geq 2$. We will use the following two recursion
relations:
\bqa
\dsimp{1}\,Q^{-s}(x_1) &=& \frac{1}{2\,(s-1)\,B}\,\Bigl[
(2\,s - 3)\,\dsimp{1}\,Q^{1-s}(x_1) + X_1\,Q^{1-s}(0) + 
(1 - X_1)\,Q^{1-s}(1)\Bigr],
\nl
\nl
\dcub{2}\,Q^{-s}(x_1,x_2) &=&
\frac{1}{2\,(s-1)\,B}\,\Bigl\{
2\,(s - 2)\,\dcub{2}\,Q^{1-s}(x_1,x_2) +
\dsimp{1}\,\Bigl[ X_2\,Q^{1-s}_1(x_1)
\nl
{}&+&
X_1\,Q^{1-s}_2(x_1) + (1 - X_2)\,Q^{1-s}_3(x_1) + (1 - X_1)\,Q^{1-s}_4(x_1)
\Bigr]\Bigr\},
\label{tworec}
\eqa
where $Q$ is an arbitrary quadratic form in one (two) variables, $B$ is the
associated BT factor, $X$ the associated BT co-factor  and $Q_i$ are 
secondary quadratic forms defined by
\bq    
Q_1(x) = Q(x,0), \quad
Q_2(x) = Q(0,x), \quad
Q_3(x) = Q(x,1), \quad
Q_4(x) = Q(1,x).
\eq
We give in the following the explicit solution for $s = 3/2$.
\bqa
\dcub{2}\,Q^{-3/2}(x_1,x_2) &=& \frac{1}{B}\,\Bigl[
-\,\frac{3}{B}\,\dcub{2}\,Q^{1/2}(x_1,x_2) +
\dsimp{1}\,f_1(\frac{3}{2}) - f_0(\frac{3}{2})\Bigr],
\eqa
where we have
\bqa
f_1(\frac{3}{2}) &=& X_2\,(\frac{2}{B_1} + \frac{1}{B})\,Q^{1/2}_1(x_1) +
X_1\,(\frac{2}{B_2} + \frac{1}{B})\,Q^{1/2}_2(x_1) 
\nl
{}&+&
(1 - X_2)\,(\frac{2}{B_3} + \frac{1}{B})\,Q^{1/2}_3(x_1) +
(1 - X_1)\,(\frac{2}{B_4} + \frac{1}{B})\,Q^{1/2}_4(x_1),
\nl
f_0(\frac{3}{2}) &=& \frac{X_2}{B_1}\,\Bigl[ X_{11}\,Q^{1/2}_1(0) + 
        (1 - X_{11})\,Q^{1/2}_1(1)\Bigr] +
        \frac{X_1}{B_2}\,\Bigl[ X_{12}\,Q^{1/2}_2(0) + 
        (1 - X_{12})\,Q^{1/2}_2(1)\Bigr] 
\nl
{}&+&
        \frac{1 - X_2}{B_3}\,\Bigl[ X_{13}\,Q^{1/2}_3(0) +
        (1 - X_{13})\,Q^{1/2}_3(1)\Bigr] 
\nl
{}&+&
        \frac{1 - X_1}{B_4}\,\Bigl[ X_{14}\,Q^{1/2}_4(0) + 
        (1 - X_{14})\,Q^{1/2}_4(1)\Bigr],
\eqa
where $B$ and $X_1, X_2$ are the leading BT factor and co-factors of the 
original quadric while $B_i$ and $X_{1i}$ are the sub-leading ones for the
secondary quadratic forms. Note that in this case the BT procedure allows
us to derive a result in terms of roots of quadratic forms.
For $s = 2$ we derive instead
\bqa
\dcub{2}\,Q^{-2}(x_1,x_2) &=& \frac{1}{4\,B}\,\Bigl[
\dsimp{1}\,f_1(2) - f_0(2)\Bigr],
\eqa
where we have
\bqa
f_1(2) &=& \frac{X_2}{B_1}\,\ln Q_1(x_1) +
           \frac{X_1}{B_2}\,\ln Q_2(x_1) +
           \frac{1 - X_2}{B_3}\,\ln Q_3(x_1) +
           \frac{1 - X_1}{B_4}\,\ln Q_4(x_1),
\nl
f_0(2) &=& \frac{X_2}{B_1}\,\Bigl[ X_{11}\,\ln Q_1(0) +
           (1 - X_{11})\,\ln Q_1(1) - 2 \Bigr] +
           \frac{X_1}{B_2}\,\Bigl[ X_{12}\,\ln Q_2(0) +
           (1 - X_{12})\,\ln Q_2(1) - 2 \Bigr]
\nl
{}&+&
           \frac{1 - X_2}{B_3}\,\Bigl[ X_{13}\,\ln Q_3(0) +
           (1 - X_{13})\,\ln Q_3(1) - 2 \Bigr]
\nl
{}&+&
           \frac{1 - X_1}{B_4}\,\Bigl[ X_{14}\,\ln Q_4(0) +
           (1 - X_{14})\,\ln Q_4(1) - 2 \Bigr].
\eqa
For $s \ge 3$ we derive  similar expressions, not to be reported here.
$D^{\rm cube}_0$ is, therefore, explicitly known. 

The procedure cannot work if one of the sub-leading $B$ is approaching zero.
Again, the $B$ are sub-leading of second type as explained in the discussion
before \eqn{efour}.
In the case $B \approx 0$ the solution consists in a double Mellin-Barnes 
splitting. Consider \eqns{fcons}{tcons}, the evaluation of $D^{\rm cube}_0$ 
is reduced to the one of
\bqa
D &=& \frac{1}{2\,\pi\,i}\,
\int_{-i\,\infty}^{+i\,\infty}\,ds\,\rho^{2-s}_{\ssL}\,
\frac{B(s,2-s)}{3-2\,s}\,\dcub{2}\,Q^{-s}(x_1,x_2),
\eqa
where $Q$ denotes one of the quadratic forms of indices $ij$, which we write as
$Q(x_1,x_2) = x^t\,H\,x + 2\,K^t\,x + L$.
Let $B_{\ssS\ssL} = L - K^t\,H^{-1}\,K$, $\rho_{\ssS\ssL} = 1/B_{\ssS\ssL}$
($\rho_{\ssL} \equiv \rho_4$) and $X = - H^{-1}\,K$. After transforming 
$x_1,x_2$ we obtain
\bqa
D &=& \frac{1}{2\,\pi\,i}\,
\int_{-i\,\infty}^{+i\,\infty}\,ds\,\rho^{2-s}_{\ssL}\,
\frac{B(s,2-s)}{3-2\,s}\,
\dcub{2}(X)\,( x^t\,H\,x + B_{\ssS\ssL} )^{-s}.
\eqa
With $0 < \Reb\,t < \Reb\,s < 2$ we write
\bq
D = \frac{1}{(2\,\pi\,i)^2}\,
\int_{-i\,\infty}^{+i\,\infty}\,ds\,dt\,
\rho^{2-s}_{\ssL}\,\rho^{s-t}_{\ssS\ssL}\,\frac{B(s,2-s)}{3-2\,s}\,B(t,s-t)\,
\dcub{2}(X)\,( x^t\,H\,x  )^{-t}.
\eq
The $x_1,x_2$ integral has been already considered, starting from \eqn{adone},
and we know that it develops a pole at $t = 1$ times the integral of a
quadratic form in one variable. If, for the latter, the BT factor is not around
zero we can safely proceed. If we write
\bq
\dcub{2}(X)\,( x^t\,H\,x  )^{-t} = \frac{f(t)}{t-1},
\eq
then we get
\bqa
D &=& \frac{1}{2\,\pi\,i}\,\int_{-i\,\infty}^{+i\,\infty}\,ds\,
\Bigl[ - f(1)\,\frac{\rho^2_{\ssL}}{\rho_{\ssS\ssL}}\,
\frac{\egam{2-s}\,\egam{s-1}}{3-2\,s}\,
\Bigl(\frac{\rho_{\ssS\ssL}}{\rho_{\ssL}}\Bigr)^s
\nl
{}&+& \rho^{2-s}_{\ssL}\,\sum_{k=0}^{\infty}\,
\frac{\egam{2-s}\,\egam{s+k-1}}{3-2\,s}\,f(s+k)\,
\frac{(-\rho_{\ssS\ssL})^{-k}}{k\,!}\Bigr].
\eqa
Next we use $\kappa = \rho_{\ssS\ssL}/\rho_{\ssL}$ and
\bqa
\frac{1}{2\,\pi\,i}\,\int_{-i\,\infty}^{+i\,\infty}\,ds\,
\frac{\egam{2-s}\,\egam{s-1}}{3-2\,s}\,\kappa^s &=&
\sum_{k=0}^{\infty}\,\frac{(-1)^k}{k\,!}\,\frac{\egam{k+1}}{2\,k+1}\,
\kappa^{1-k}
\nl
{}&=& \frac{\kappa}{2}\,\sum_{k=0}^{\infty}\,
\frac{\egam{k+1}\,\egam{k+1/2}}{\egam{k+3/2}}\,\frac{(-\kappa)^{-k}}{k\,!}
\nl
{}&=& \kappa\,\,\hyper{\frac{1}{2}}{1}{\frac{3}{2}}{-\,\frac{1}{\kappa}} =
\kappa^{3/2}\,\arctan(\kappa^{-1/2}).
\eqa
Therefore we get
\bqa
D &=& - (\rho_{\ssL}\,\rho_{\ssS\ssL})^{1/2}\,f(1)\,\arctan\,
\Bigl(\frac{\rho_{\ssL}}{\rho_{\ssS\ssL}}\Bigr)^{1/2} +
\frac{1}{2}\,(\pi\,\rho_\ssL)^{1/2}\,
\sum_{k=0}^{\infty}\,\egam{k+\frac{1}{2}}\,f(k+\frac{3}{2})\,
\frac{(-\rho_{\ssS\ssL})^{-k}}{k\,!}
\nl
{}&-& \sum_{k,l=0}^{\infty}\,\frac{\egam{k+l+1}\,f(k+l+2)}{2\,l+1}\,
\frac{(-\rho_{\ssS\ssL})^{-k}}{k\,!}\,\frac{(-\rho_{\ssL})^{-l}}{l\,!}.
\label{resIII}
\eqa
In the above results it is always understood that $\rho_{\ssL} \to
\rho_{\ssL} - i\,\delta$ and that $\rho_{\ssS\ssL} \to
\rho_{\ssS\ssL} - i\,\delta$. 

As illustrated in the previous section the evaluation of $f(k)$ is reduced
to computing a one dimensional integral of a quadratic form in one variable
(with a small negative imaginary part) which can be done by BT iteration,
unless the corresponding BT factor is around zero. If this is the case, we
start wondering why the world is against us or return to the starting
expression and use one more Mellin-Barnes splitting.
\boldmath \subsection{A new integral representation for $D_0$} \unboldmath
Also for $D_0$ we are able to present a new integral representation which is
suitable for numerical treatment, especially when $B_4 = 0$. Consider again
\eqn{refDi}: after a sector decomposition in $[0,1]^3$ we obtain
\bqa
D_i &=&
  \alpha_i\,\beta_i\,\gamma_i\,
  \dcub{3}\,
  \Bigg\{
  x_1^2\,\Big[
  x_1^2\,Q_i(\alpha_i,\beta_i\,x_2,\gamma_i\,x_3) + B_4 - i\delta
  \Big]^{-2-\ep/2}
\nl
&{}&
+ x_2^2\,\Big[
  x_2^2\,Q_i(\alpha_i\,x_1,\beta_i,\gamma_i\,x_3) + B_4 - i\delta
  \Big]^{-2-\ep/2}
+ x_3^2\,\Big[
  x_3^2\,Q_i(\alpha_i\,x_1,\beta_i\,x_2,\gamma_i) + B_4 - i\delta
  \Big]^{-2-\ep/2}
  \Bigg\}
\nl
&=&
\alpha_i\,\beta_i\,\gamma_i\,
\sum_{j=1}^{3}\,\dcub{3}\,x_3^2\,
\Big[ x_3^2\,Q_{ij}(x_1,x_2) + B_4 - i\delta \Big]^{-2-\ep/2} ,
\label{Di:eq1}
\eqa
where we introduced the quadratic form
\bqa
Q_{i1}(x_1,x_2) &=& Q_i(\alpha_i\,x_1,\beta_i\,x_2,\gamma_i),
\nl
Q_{i2}(x_1,x_2) &=& Q_i(\alpha_i\,x_1,\beta_i,\gamma_i\,x_2),
\nl
Q_{i3}(x_1,x_2) &=& Q_i(\alpha_i,\beta_i\,x_1,\gamma_i\,x_2).
\eqa
They are expressible as
\bq
Q_{ij}(x_1,x_2)=
x^t\,H_{ij}\,x + 2\,K_{ij}^t\,x + L_{ij}=
\bar{x}^t\,M_{ij}\,\bar{x} ,
\eq
where $x^t=(x_1,x_2)$, $\bar{x}^t= (x_1,x_2,1)$ and the matrix $M$ is given by
\bq
M_{ij}=
\left(
\ba{cc}
H_{ij}   \quad & K_{ij} \nl
K_{ij}^t \quad & L_{ij}
\ea
\right) .
\eq
Now we apply the results for the master integral of Appendix A with
coefficients
\bq
a = Q_{ij}(x_1,x_2),
\qquad
b = B_4,
\qquad
\alpha = 2 + \ep/2,
\qquad
\beta = 2.
\eq
With the usual replacement $Q_{ij},B_4 \to Q_{ij},B_4 - i\,\delta$ we obtain
\bqa
D_i &=& \alpha_i\,\beta_i\,\gamma_i\,\sum_{j=1}^{3}\dcub{2}\,
\Bigg\{
\frac{1}{2}\,B_4^{-1/2-\ep/2}\,
B\left( \frac{3}{2},\frac{1}{2}+\frac{\ep}{2} \right)\,
Q^{-3/2}_{ij}(x_1,x_2) - \intfx{x_3}\,x_3^{\ep}\,
Q^{-2-\ep/2}_{ij}(x_1,x_2,x_3)
\Bigg\}.
\nl
\eqa
In the last equation polynomials $Q_{ij}(x_1,x_2)$ and $Q_{ij}(x_1,x_2,x_3) =
Q_{ij}(x_1,x_2) + B_4\,x_3^2$ make their appearance; their powers can be 
easily incremented with the BT method, as long as the corresponding factors 
$B_{3ij}$ are different from zero. The BT relations to be used for this 
purpose are as follows:
\bq
Q_{ij}^{\mu-1}(x_1,x_2)= \frac{1}{B_{3ij}}\,\left[
1 - \frac{1}{2\,\mu}\,  
\sum_{l=1}^2\,(x_l - X_{lij})\,\partial_{x_l}\right]\,
Q_{ij}^{\mu}(x_1,x_2) ,
\eq
\bq
Q^{\mu-1}_{ij}(x_1,x_2,x_3) =
\frac{1}{B_{3ij}}\,\left[ 1 - \frac{1}{2\,\mu}\,  \left(
\sum_{l=1}^2\,(x_l - X_{lij})\,\partial_{x_l}   + x_3\,\partial_{x_3}
  \right)\right]\,
Q^{\mu}_{ij}(x_1,x_2,x_3),
\eq
where the sub-leading BT-factors are defined by
\bq
B_{3ij} =
\frac{{\rm det}(M_{ij})}{G_{3ij}}=
\frac{\alpha_i^2\,\beta_i^2\,\gamma_i^2}{G_{3ij}}\,{\rm det}(H),
\qquad
G_{3ij} = \Delta_{33}(M_{ij}),
\eq
\bq
X_{1ij} = \frac{\Delta_{13}(M_{ij})}{G_{3ij}},
\qquad
X_{2ij} = \frac{\Delta_{23}(M_{ij})}{G_{3ij}},
\eq
where $\Delta_{lk}(M_{ij})$ is the co-determinant of the 
element $lk$ of the matrix $M_{ij}$.
The BT method cannot be used when also some of the $B_{3ij}$ are near zero
and, in this case, we have to proceed along the same lines used for the 
three-point function. We rewrite the polynomials $Q_{ij}(x_1,x_2)$ in the 
following way:
\bq
Q_{ij}(x_1,x_2)= (x-X_{ij})^t\,H_{ij}\,(x-X_{ij}) + B_{3ij}.
\eq
Next we transform variables according to $x_1 = x'_1 + X_{1ij}$,
$x_2 = x'_2 + X_{2ij}$ and split the $x_1 - x_2$ integration
interval according to
\bq
\int_{-X_{lij}}^{1-X_{lij}}\,dx_l =
\Big( -\,\int_0^{-X_{lij}} + \int_0^{1-X_{lij}} \Big)\,dx_l.
\eq
Each of the integrals is mapped into the unit square and we derive
\bqa
D_i &=& \alpha_i\,\beta_i\,\gamma_i\,
\sum_{j=1}^{3}\,\sum_{l=1}^4\,\eta_{ijl}\,\xi_{ijl}\,
\dcub{2}\,
\Bigg\{
\frac{1}{2}\,B_4^{-1/2-\ep/2}\,
B\left( \frac{3}{2},\frac{1}{2}+\frac{\ep}{2} \right)\,
\Big[ Q_{ijl}(x_1,x_2) + B_{3ij} \Big]^{-3/2}
\nl
{}&-& \intfx{x_3}\,x_3^{\ep}\,\Big[
Q_{ijl}(x_1,x_2) + B_{3ij} + B_4\,x_3^2\Big]^{-2-\ep/2}
\Bigg\},
\eqa
where the replacement $B_{3ij} \to B_{3ij} - i\,\delta$ is understood and
where we introduced new auxiliary quantities:
\bqa
\eta_{ij1} &=& \eta_{ij2} = 1 - X_{1ij},
\quad
\eta_{ij3} = \eta_{ij4} = X_{1ij},
\nl
\xi_{ij1} &=& \eta_{ij3} = 1 - X_{2ij},
\quad
\xi_{ij2} = \xi_{ij4} = X_{2ij}.
\eqa
The new quadrics are defined by
\bqa
Q_{(ij)1} &=& Q_{ij}'((1-X_{1ij})\,x_1\,,\,(1-X_{2ij})\,x_2),
\quad
Q_{(ij)2} = Q_{ij}'((1-X_{1ij})\,x_1\,,\,-X_{2ij}\,x_2),
\nl
Q_{(ij)3} &=& Q_{ij}'(-X_{1ij}\,x_1\,,\,(1-X_{2ij})\,x_2),
\quad
Q_{(ij)4} = Q_{ij}'(-X_{1ij}\,x_1\,,\,-X_{2ij}\,x_2),
\eqa
with $Q_{ij}' = x^t\,H_{ij}\,x$.
At this point, after performing the usual sector decomposition of the square 
$[0,1]^2$, we apply again the master formula of Appendix A and get:
\bqa
D_i &=&
\alpha_i\,\beta_i\,\gamma_i\,
\sum_{j=1}^{3}\,\sum_{l=1}^4\,\sum_{m=1}^2\,
\eta_{ijl}\,\xi_{ijl}\,
\intfx{x_1}\,D_{ijlm},
\eqa
\bqa
D_{ijlm} &=&
\frac{1}{2}\,B_4^{-1/2-\ep/2}\,B_{3ij}^{-1/2}\,
B\left( \frac{3}{2},\frac{1}{2}+\frac{\ep}{2} \right)\,Q^{-1}_{ijlm}(x_1)
\nl
&{}&
- \frac{1}{2}\,B(1,1+\ep/2)\,Q^{-1}_{ijlm}(x_1)\,\intfx{x_3}\,x_3^{\ep}\,
\Big[ B_4\,x_3^2 + B_{3ij} \Big]^{-1-\ep/2}\,
\nl
&{}&
- \frac{1}{2}\,B_4^{-1/2-\ep/2}\,
B\left( \frac{3}{2},\frac{1}{2}+\frac{\ep}{2} \right)\,
\intfx{x_2}\, Q^{-3/2}_{ijlm}(x_1,x_2) + 
\intfxy{x_2}{x_3}\,x_3^{\ep}\,Q^{-2-\ep/2}_{ijlm}(x_1,x_2,x_3),
\nl
\label{Di:B4/B3}
\eqa
where we introduced $Q_{ijl1}(x_1)= Q_{ijl}(x_1,1)$ and
$Q_{ijl2}(x_1)= Q_{ijl}(1,x_1)$. Likewise we have
\bqa
Q_{ijlm}(x_1,x_2) &=& Q_{ijlm}(x_1) + B_{3ij}\,x_2^2,
\nl
Q_{ijlm}(x_1,x_2,x_3) &=& Q_{ijlm}(x_1) + B_{3ij}\,x_2^2 + B_4\,x_3^2.
\eqa
The divergent behavior of the first and of the third term is evident and
the second integral depends essentially on the ratio
\bq
\kappa_{ij}= \frac{B_4-i\delta}{B_{3ij}-i\delta}.
\eq
Therefore, by applying the master formula of Appendix A, we obtain
\bqa
\intfx{x_3}\,x_3^{\ep}\,\Big[ B_4\,x_3^2 + B_{3ij} \Big]^{-1-\ep/2} &=&
\frac{1}{2}\,B^{-1/2}_{3ij}\,B_4^{-1/2-\ep/2}\,
B\left( \frac{1}{2}+\frac{\ep}{2},\frac{1}{2} \right)  
- \intfx{x_3}\,\Big[ B_{3ij}\,x_3^2 + B_4 \Big]^{-1-\ep/2} 
\nl
{}&=&
\frac{\pi}{2}\,B_{3ij}^{-1/2}\,B_4^{-1/2}\,
- B_4^{-1}\,\kappa_{ij}^{1/2}\,{\rm arctan}(\kappa_{ij}^{-1/2})
+ \ord{\ep} .
\label{againpeq}
\eqa
Note that the first term in \eqn{againpeq} cancels the first term of
\eqn{Di:B4/B3} giving our integral representation for $D_i$:
\bqa
D_i &=& \alpha_i\,\beta_i\,\gamma_i\,
\sum_{j=1}^{3}\,\sum_{l=1}^4\,\sum_{m=1}^2\,\eta_{ijl}\,\xi_{ijl}\,
\intfx{x_1}\,
\Bigg\{
\frac{1}{2}\,B_4^{-1/2}\,B_{3ij}^{-1/2}\,
{\rm arctan}(\kappa_{ij}^{-1/2})\,Q^{-1}_{ijlm}(x_1)
\nl
{}&-& \frac{\pi}{4}\,B_4^{-1/2}\,
\intfx{x_2}\,
Q^{-3/2}_{ijlm}(x_1,x_2) + \intfxy{x_2}{x_3}\,
Q^{-2}_{ijlm}(x_1,x_2,x_3)\Bigg\}.
\nl
\label{tober}
\eqa
As a final step, the power of the polynomials appearing in \eqn{tober} can
be incremented by using the following set of BT relations:
\bqa
Q_{ijlm}^{-1}(x_1) &=& \frac{1}{b_{ijlm}}\,
\left[ 1 - \frac{1}{2}\,\Big( x_1+\frac{k_{ijlm}}{h_{ijlm}} \Big)\,
  \partial_{x_1}\,\ln Q_{ijlm}(x_1)
\right] ,
\eqa
\bqa
Q^{\mu-1}_{ijlm}(x_1,x_2) &=&
\frac{1}{b_{ijlm}}\,
\left\{ 1 - \frac{1}{2\,\mu}\,
  \left[
    \Big( x_1+\frac{k_{ijlm}}{h_{ijlm}} \Big)\,\partial_{x_1}
  + x_2\,\partial_{x_2}
  \right]
\right\} Q^{\mu}_{ijlm}(x_1,x_2) ,
\nl
\eqa
\bqa
Q^{\mu-1}_{ijlm}(x_1,x_2,x_3) &=&
\frac{1}{b_{ijlm}}\,
\left\{ 1 - \frac{1}{2\,\mu}\,
  \left[
    \Big( x_1+\frac{k_{ijlm}}{h_{ijlm}} \Big)\,\partial_{x_1}
  + \sum_{k=2}^{3}\,x_k\,\partial_{x_k} \right]\right\}\,
Q^{\mu}_{ijlm}(x_1,x_2,x_3) ,
\eqa
where we defined
\bq
Q_{ijlm}(x_1)= h_{ijlm}\,x_1^2 + 2\,k_{ijlm}\,x_1 + l_{ijlm}
\qquad
b_{ijlm}= l_{ijlm} - \frac{k_{ijlm}^2}{h_{ijlm}}.
\eq
If also $b_{ijlm}$ tends to 0, we must repeat the procedure for the
quadratic form $Q_{ijlm}(x_1)$.
\section{\boldmath Infrared divergent $D_0$\label{IRD0} \unboldmath}
At the one-loop level the evaluation of infrared divergent $N$-point
functions with $N \ge 4$ is greatly simplified insofar the divergent parts
can always be reduced to three-point functions. This decomposition is even
more relevant in our approach where the evaluation of the infrared finite
remainder does not require any additional work. Let us give an example for
$D_0$~\cite{Beenakker:1988jr}, where we consider the configuration with
\bq
p^2_1 = p^2_2= - m^2_2, \quad p^2_3 = p^2_4 = - m^2_4, \quad m_1 = m_3 = 0.
\label{IRD0conf}
\eq
If we introduce the following notation:
\bq
[1] = q^2, \quad [2] = (q+p_1)^2 + m^2_2, \quad
[3] = (q+P)^2, \quad [4] = (q-p_4)^2+ m^2_4,
\eq
where $P = p_1+p_2$, then the following decomposition holds
\bqa
-P^2\,D_0 &=& \frac{2}{i\,\pi^2}\,\int\,d^dq\,
\frac{\spro{q}{(q+P)}}{[1]\,[2]\,[3]\,[4]} - 
C_0((p_2+p_3)^2,p^2_2,p^2_3\,;\,m_2,0,m_4) 
\nl
{}&-& C_0(p^2_4,p^2_1,(p_2+p_3)^2\,;\,0,m_2,m_4),
\eqa
showing two IR divergent $C_0$ functions and an IR finite integral. For the
latter we obtain
\bq
D^{\rm rem}_0 = \frac{1}{i\,\pi^2}\,
\int\,d^dq\,\frac{\spro{q}{(q+P)}}{[1]\,[2]\,[3]\,[4]} =
\dsimp{3}\,V^{-2-\ep/2}\,( k^2 - \spro{k}{P} + 2\,V),
\label{defrem}
\eq
where
\bq
P = p_1 + p_2, \quad Q= p_2 + p_3, \quad k = \sum_{i=1}^{3}\,x_i\,p_i,
\eq
and the quadric
\bqa
V &=& m^2_2\,(x^2_1 + x^2_2) + m^2_4\,x^2_3 - (P^2 + 2\,m^2_2)\,x_1\,x_2 +
(Q^2 + P^2 + m^2_2 + m^2_4)\,x_1\,x_3 
\nl
{}&-& (Q^2 + m^2_2 + m^2_4)\,x_2\,x_3 + P^2\,( x_2 - x_3),
\eqa
shows a zero for $x_1 = x_2 = x_3 = 0$. The numerator in \eqn{defrem}
becomes
\bq
k^2 - \spro{k}{P} + 2\,V = V - \frac{1}{2}\,P^2\,(x_1 - x_2 + x_3),
\eq
so that we obtain
\bqa
D^{\rm rem}_0 &=& \dsimp{3}\,\Bigl[
V^{-1-\ep/2} - \frac{1}{2}\,P^2\,(x_1 - x_2 + x_3)\,V^{-2-\ep/2}\Bigr].
\eqa
By infrared power counting the expression for $D^{\rm rem}_0$ is manifestly
finite and can be computed according to standard BT techniques when
$B_4 \ne 0$. The explicit expression of the coefficient $B_4$ is
\bq
B_4 = \frac{1}{{\rm det}(H)}\,\frac{P^4}{16}\,\Bigl[ 4\,m_2^2\,m_4^2 -
(Q^2 + m_2^2 + m_4^2)^2 \Bigr] =
\frac{1}{{\rm det}(H)}\,\frac{P^4}{16}\,\lambda(-Q^2, m_2^2, m_4^2),
\eq
where ${\rm det}(H)$ is the determinant of the matrix $H$ in the BT relation:
\bqa
{\rm det}(H) = \frac{P^2}{4}\,
           \Bigl[ \lambda(-Q^2, m_2^2, m_4^2) + Q^2\,P^2 \Bigr].
\eqa
As a consequence $B_4 = 0$ only for $Q^2 < 0$, i.e. for 
normal(pseudo)-threshold $Q^2 = - (m_2 \pm m_4)^2$. Around the thresholds
other techniques are used since the BT procedure overestimates the singularity,
see \sect{ed0reg} for the pseudo-threshold (where $D^{\rm rem}_0$ is regular)
and \sect{ed0sing} for normal threshold (where $D^{\rm rem}_0$ is indeed
singular).
\subsection{Classification of IR divergent cases for the 
D-family\label{classD0IR}}
A short classification of IR divergent four-point functions can be based on
IR power counting which, for one-loop diagrams, requires $\int_0 d^4q/q^4$.
The configurations of interest are those where some of the internal masses
vanish and the Mandelstam invariants of the process are not constrained.

Without loss of generality we may assume that $m_1 = 0$. For $q = \Delta$
and $\Delta$ vanishing the requested behavior of $\ord{\Delta^4}$ is obtained 
in three different cases~\footnote{This is the well-known eikonal 
approximation~\cite{Abarbanel:1969ek}, whose validity we assume without 
further justifications~\cite{Sterman:ce}.}:
\begin{enumerate}
\item $p^2_1 = - m^2_2 \qquad p^2_4 = - m^2_4$;
\item $p^2_1 = - m^2_2 \qquad P^2 = - m^2_3$;
\item $p^2_4 = - m^2_4 \qquad P^2 = - m^2_3$.
\end{enumerate}
Note that only the first one is of interest, so the IR configuration is
\bq
m_1 = 0, \quad p^2_1 = - m^2_2, \quad  p^2_4 = - m^2_4.
\eq
If also $m_3 = 0$ then there is an additional pole for $q = - P$ and we 
introduce $q = -P + \Delta$. The four denominators in $D_0$ behave as
\bq
[1] \sim P^2, \quad
[2] \sim p^2_2 + m^2_2 - 2\,\spro{p_2}{\Delta},
\quad
[3] \sim \Delta^2,
\quad
[4] \sim p^2_3 + m^2_4 + 2\,\spro{p_3}{\Delta},
\eq
and the requested behavior of $\ord{\Delta^4}$ is obtained if $p^2_2 = - m^2_2$
and $p^2_3 = - m^2_4$, as expected. Therefore another IR configuration is
\bq
m_1 = m_3 = 0, \quad 
\{p^2_1 = - m^2_2 \; p^2_4 = - m^2_4\} 
\quad \mbox{and/or} \quad
\{p^2_2 = - m^2_2 \; p^2_3 = - m^2_4\}.
\eq
If instead we have $m_1 =0$ and $m_2 = 0$ there will be another pole
at $q = - p_1$ whose IR nature depends on $p^2_1 = 0$ and $p^2_2= - m^2_3$.
The new IR configuration is
\bq
m_1 = m_2 = 0, \; p^2_1= 0, \quad
\{p^2_4 = - m^2_4\}
\quad \mbox{and/or} \quad
\{p^2_2 = - m^2_3\}.
\eq 
Furthermore, if $m_1 = m_2 = m_3 = 0$ there are the following IR 
configurations:
\bq
m_1 = m_2 = m_3 = 0, \quad
\{p^2_1 = 0 \; p^2_4 = - m^2_4\}
\quad \mbox{or} \quad
\{p^2_1 = 0 \; p^2_2 = 0\}
\quad \mbox{or} \quad
\{p^2_2 = 0 \; p^2_3 = - m^2_4\}.
\eq
In all cases the IR box can be reduced to a combination of (possibly) IR 
$C_0$ functions and a IR finite reminder.

To further clarify the connection between IR configurations and Landau 
singularities we consider the $D_0$ specified by \eqn{IRD0conf}.
The corresponding Landau equations are 
\bqa
\alpha_1\,q^2 = 0\,, &\alpha_2\,[ (q +p_1)^2 + m_2^2] = 0 \,,\quad 
\alpha_3\,(q + P)^2 = 0\,,& \alpha_4\,[ (q-p_4)^2 + m_4^2] = 0 \nl
&\alpha_1\,q_{\mu}  + \alpha_2\,(q + p_1)_{\mu} + \alpha_3\,(q + P)_{\mu} 
+ \alpha_4\,(q - p_4)_{\mu} = 0\,.
\label{Q5}
\eqa
The equations in the first row of \eqn{Q5} are satisfied, for $\alpha_i 
\neq 0$, if
\bq
q^2 = 0 \, , \quad 
\spro{q}{p_1} = 0 \,,  \quad 
\spro{q}{P} = -\frac{1}{2}\,P^2 \,, 
\quad
\spro{q}{p_4} = 0\,. 
\label{Q6}
\eq
By contracting the free Lorentz index in the equation in the second row of 
\eqn{Q5} by $q_{\mu}$, $(p_1)_{\mu}$, $P_{\mu}$, and ${p_4}_{\mu}$, and by
using \eqn{Q6} we obtain the system of equations
\bqa
\alpha_3\,P^2 = 0 \,, &\quad& 
-2\,\alpha_2\,m_2^2 + \alpha_3\,P^2 -
2\,\alpha_{4}\,\spro{p_1}{p_4} = 0\, , 
\nl
-\alpha_1\,P^2 + \alpha_3\,P^2 = 0\,, &\quad& 
 2\,\alpha_2\,\spro{p_1}{p_4} - \alpha_3\,P^2 +
2\,\alpha_{4}\,m_4^2 = 0\, . 
\label{Q7}
\eqa
If the system of \eqn{Q7} admits a proper solution (i.e. a solution in which 
not all the $\alpha_i$ vanish), the determinant of the system must be equal 
to zero. One can readily verify that the determinant is
\bq
\Delta = -\,\frac{P^4}{16}\,\lambda(-Q^2,m^2_2,m^2_4).
\eq
Therefore the leading Landau singularity requires $Q^2 < 0$ and corresponds 
to the usual normal threshold $Q^2 = - (m_2 + m_4)^2$ on top of the infrared 
pole.
\section{\boldmath Five-point functions (E-family)\label{E0} \unboldmath}
Several results are known in the literature about the reduction of
dimensionally regulated one-loop $N$-point integrals with $N\geq5$
(\cite{vanNeerven:1983vr}, \cite{Fleischer:1999hq}
and~\cite{Binoth:1999sp}). The generic, scalar, pentagon is given by
\bq
E_0 = \dsimp{4}\,V^{-3-\ep/2}(x_1,x_2,x_3,x_4),
\qquad
V(x) = x^t\,H\,x + 2\,K^t\,x + L,
\label{EqDefE}
\eq
where $H_{ij} = -\,\spro{p_i}{p_j}$ with $i,j = 1,\dots,4$,
$L = m_1^2$, and where
\bqa
K_1 &=& - \frac{1}{2} \, ( m_1^2 - m_2^2 - \spro{p_1}{p_1} ),
\nl
K_2 &=& - \frac{1}{2} \, ( m_2^2 - m_3^2 - 2\,\spro{p_1}{p_2} -
\spro{p_2}{p_2} ),
\nl
K_3 &=& - \frac{1}{2} \, ( m_3^2 - m_4^2 - 2\,\spro{p_1}{p_3} -
2\,\spro{p_2}{p_3} - \spro{p_3}{p_3} ),
\nl
K_4 &=& - \frac{1}{2} \, ( m_4^2 - m_5^2 - 2\,\spro{p_1}{p_4} -
2\,\spro{p_2}{p_4} - 2\,\spro{p_3}{p_4} - \spro{p_4}{p_4} ).
\eqa
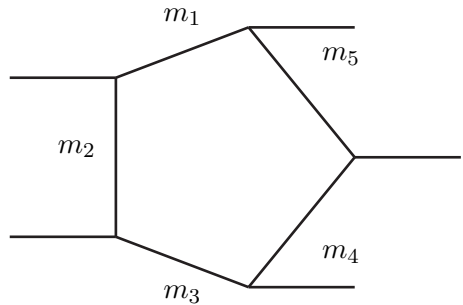
\begin{figure}[h]
\vspace*{0.3cm}
\[
\hspace{0.5cm}
  \vcenter{\hbox{
  \begin{picture}(150,100)(-40,-40)
  \SetScale{0.5}
  \SetWidth{2.}
  \Line(100,0)(20,97.98)
  \Line(20,97.98)(-80,60)
  \Line(-80,60)(-80,-60)
  \Line(-80,-60)(20,-97.98)
  \Line(20,-97.98)(100,0)
  \Line(-80,60)(-160,60)
  \Line(-80,-60)(-160,-60)
  \Line(20,97.98)(100,97.98)
  \Line(20,-97.98)(100,-97.98)
  \Line(100,0)(180,0)
  \Text(-15,50)[cb]{$\mone$}
  \Text(-55,0)[cb]{$\mtwo$}
  \Text(-15,-55)[cb]{$\mtre$}
  \Text(45,-40)[cb]{$\mfor$}
  \Text(45,35)[cb]{$\mfiv$}
  \end{picture}}}
\]
\vspace{0.5cm}
\caption[]{The one-loop, five-point Green function of \eqn{EqDefE}.
Propagators are $q^2+m^2_1\cdots\,$ $(q+p_1+\cdots + p_4)^2+m^2_5$.}
\label{fig:fivepoint}
\end{figure}
\vskip 5pt
\boldmath \subsection{Evaluation of $E_0$ when $B_5 \ne 0$\label{E0BT}} 
 \unboldmath
It is a virtue of the BT algorithm that we can show the decomposition
of $E_0$ in five boxes (in $d = 4$) with just one iteration, as depicted
in \fig{fig:pentagon}. We obtain
\bq
E_0 = \frac{1}{4\,B_5}\,\sum_{i=0}^{4}\,w_i\,D^{(i)}_0,
\label{pdec}
\eq
where the weights are
\bq
w_i = X_i - X_{i+1}, \qquad X_0 = 1, \quad X_5 = 0,
\eq
and where $B_5 = L - K^t\,H^{-1}\,K$ and $X = - K^t\,H^{-1}$. The further 
advantage in this derivation is that the nature of the weights is  
transparent since $B_5 = 0$ corresponds to a Landau singularity of the 
pentagon. Furthermore the boxes are specified by
\bq
D^{(i)}_0 = \dsimp{3}\,V^{-2-\ep/2}(\widehat{i\;i+1}), 
\eq
where the contractions are
\bq
\widehat{0\;1} = (1,x_1,x_2,x_3), \quad
\widehat{4\;5} = (x_1,x_2,x_3,0),
\eq
etc. As long as $B_5$ is not around zero the derivation for the pentagon is 
completed since we know how to deal with boxes even if their (sub-leading)
BT factors are around zero.
\subsection{Form factors in the E-family}
The nice feature of the scalar pentagon is that fourfold integrals disappear
in the final answer, \eqn{pdec}. The same is not immediately true for form 
factors, $E_{11}$ etc, which contain powers of the Feynman parameters in the 
numerator. We can use a new identity,
\bq
(x_i - X_i)\,V^{\mu}(x_1,x_2,x_3,x_4) = \frac{1}{2\,(\mu+1)}\,H^{-1}_{ij}\,
\partial_j\,V^{\mu+1}(x_1,x_2,x_3,x_4).
\eq
With the help of the identity and after integration-by-parts we are able to
remove again all fourfold integrals in the form factors of the $E_{1i}$-series.
To give an example we consider
\bq
E_{11} = \dsimp{4}\,x_1\,V^{-3-\ep/2}(x_1,x_2,x_3,x_4) =
\frac{1}{4\,B_5}\,\dsimp{3}\,( E^{-2}_{11} -
\frac{1}{2}\,E^{-1}_{11}).
\eq
Secondary quadrics are defined as:
\bqa
V_a(x_1,x_2,x_3) &=& V(x_1,x_1,x_2,x_3),
\quad
V_b(x_1,x_2,x_3) = V(x_1,x_2,x_2,x_3),
\nl
V_c(x_1,x_2,x_3) &=& V(x_1,x_2,x_3,x_3),
\quad
V_d(x_1,x_2,x_3) = V(x_1,x_2,x_3,0),
\nl
V_e(x_1,x_2,x_3) &=& V(1,x_1,x_2,x_3).
\eqa
The $E_{11}$ form factor is fully specified by
\bqa
E^{-2}_{11} &=& (X_1 - X_2)\,x_1\,V^{-2-\ep/2}_a +
(X_2 - X_3)\,x_1\,V^{-2-\ep/2}_b +
(X_3 - X_4)\,x_1\,V^{-2-\ep/2}_c
\nl
{}&+& X_4\,x_1\,V^{-2-\ep/2}_d +
(1 - X_1)\,V^{-2-\ep/2}_e,
\eqa
\bqa
E^{-1}_{11} &=& (H^{-1}_{11} - H^{-1}_{12})\,V^{-1-\ep/2}_a +
(H^{-1}_{12} - H^{-1}_{13})\,V^{-1-\ep/2}_b +
(H^{-1}_{13} - H^{-1}_{14})\,V^{-1-\ep/2}_c
\nl
{}&+&
H^{-1}_{14}\,V^{-1-\ep/2}_d -
H^{-1}_{11}\,V^{-1-\ep/2}_e.
\eqa
Similarly we obtain
\bqa
E^{-1}_{i1} &=& (H^{-1}_{i1} - H^{-1}_{i2})\,V^{-1-\ep/2}_a +
(H^{-1}_{i2} - H^{-1}_{i3})\,V^{-1-\ep/2}_b +
(H^{-1}_{i3} - H^{-1}_{i4})\,V^{-1-\ep/2}_c
\nl
{}&+&
H^{-1}_{i4}\,V^{-1-\ep/2}_d -
H^{-1}_{i1}\,V^{-1-\ep/2}_e,
\eqa
for $i=1,\dots,4$ and
\bqa
E^{-2}_{21} &=& (X_1 - X_2)\,x_1\,V^{-2-\ep/2}_a +
(X_2 - X_3)\,x_2\,V^{-2-\ep/2}_b +
(X_3 - X_4)\,x_2\,V^{-2-\ep/2}_c
\nl
{}&+& X_4\,x_2\,V^{-2-\ep/2}_d +
(1 - X_1)\,x_1\,V^{-2-\ep/2}_e,
\nl
E^{-2}_{31} &=& (X_1 - X_2)\,x_2\,V^{-2-\ep/2}_a +
(X_2 - X_3)\,x_2\,V^{-2-\ep/2}_b +
(X_3 - X_4)\,x_3\,V^{-2-\ep/2}_c
\nl
{}&+& X_4\,x_3\,V^{-2-\ep/2}_d +
(1 - X_1)\,x_2\,V^{-2-\ep/2}_e,
\nl
E^{-2}_{41} &=& (X_1 - X_2)\,x_3\,V^{-2-\ep/2}_a +
(X_2 - X_3)\,x_3\,V^{-2-\ep/2}_b +
(X_3 - X_4)\,x_3\,V^{-2-\ep/2}_c
\nl
{}&+& (1 - X_1)\,x_3\,V^{-2-\ep/2}_e .
\eqa
As it has been done before and for the reader's convenience we list all the
form factors in the $E$-family.
Note that there are also delicate points connected with the
decomposition into objects belonging to the $E$-family since the decomposition
itself is strictly defined in $4$ dimensions. Furthermore, starting from six 
powers of momenta in the numerator, we will encounter UV divergent terms so 
that some care is needed. Clearly we can have such a term while working in a 
general $R_{\xi}$-gauge.

With two momenta in the numerator we define form factors as
\bqa
E_{2;i} &=& E_{2}(ii), \qquad i,j= 1,\cdots,4,
\nl
E_{2;5} &=& E_2(12), \quad
E_{2;6} = E_2(13), \quad
E_{2;7} = E_2(14), \quad
E_{2;8} = E_2(23),
\nl
E_{2;9} &=& E_2(24), \quad
E_{2;10} = E_2(34),
\eqa
where the auxiliary function $E_2(ij)$ is
\bq
E_2(ij) = \dsimp{4}\,x_i\,x_j\,V^{-3-\ep/2}(x_1,x_2,x_3,x_4).
\eq
However, a new form factor arises, the one proportional to $\delta_{\mu\nu}$,
\bq
E_{2;11} = \frac{1}{4}\,\dsimp{4}\,
V^{-2-\ep/2}(x_1,x_2,x_3,x_4).
\eq
$E_{2;11}$ can still be reduced to form factors of the $E$-family by writing
\bqa
E_{2;11} &=& \frac{1}{4}\,\Bigl[ \sum_{i,j=1}^{4}\,H_{ij}\,E_2(ij) + 2\,
\sum_{i=1}^{4}\,K_i\,E_{1i} + L\,E_0\Bigr].
\eqa
Let us give the complete expressions for the form factors of the $E_2$-series.
\bq
4\,B_5\,E_2(ij) = - \dsimp{4}\,H^{-1}_{ij}\,V^{-1-\ep/2}(x_1,x_2,x_3,x_4) +
\dsimp{3}\,( E^{-2}_{2;ij} - E^{-1}_{2;ij}),
\eq
where the various coefficients are:
\bqa
E^{-2}_{2;11} &=&  (X_1 - X_2)\,x^2_1\,V^{-2-\ep/2}_a +
(X_2 - X_3)\,x^2_1\,V^{-2-\ep/2}_b
\nl
{}&+&
(X_3 - X_4)\,x^2_1\,V^{-2-\ep/2}_c +
X_4\,x^2_1\,V^{-2-\ep/2}_d +
(1 - X_1)\,V^{-2-\ep/2}_e,
\nl
E^{-2}_{2;22} &=&  (X_1 - X_2)\,x^2_1\,V^{-2-\ep/2}_a +
(X_2 - X_3)\,x^2_2\,V^{-2-\ep/2}_b
\nl
{}&+&
(X_3 - X_4)\,x^2_2\,V^{-2-\ep/2}_c +
X_4\,x^2_2\,V^{-2-\ep/2}_d +
(1 - X_1)\,x^2_1\,V^{-2-\ep/2}_e,
\nl
E^{-2}_{2;33} &=&  (X_1 - X_2)\,x^2_2\,V^{-2-\ep/2}_a +
(X_2 - X_3)\,x^2_2\,V^{-2-\ep/2}_b
\nl
{}&+&
(X_3 - X_4)\,x^2_3\,V^{-2-\ep/2}_c +
X_4\,x^2_3\,V^{-2-\ep/2}_d +
(1 - X_1)\,x^2_2\,V^{-2-\ep/2}_e,
\nl
E^{-2}_{2;44} &=&  (X_1 - X_2)\,x^2_3\,V^{-2-\ep/2}_a +
(X_2 - X_3)\,x^2_3\,V^{-2-\ep/2}_b
\nl
{}&+&
(X_3 - X_4)\,x^2_3\,V^{-2-\ep/2}_c +
(1 - X_1)\,x^2_3\,V^{-2-\ep/2}_e,
\nl
E^{-2}_{2;12} &=&  (X_1 - X_2)\,x^2_1\,V^{-2-\ep/2}_a +
(X_2 - X_3)\,x_1\,x_2\,V^{-2-\ep/2}_b
\nl
{}&+&
(X_3 - X_4)\,x_1\,x_2\,V^{-2-\ep/2}_c +
X_4\,x_1\,x_2\,V^{-2-\ep/2}_d +
(1 - X_1)\,x_1\,V^{-2-\ep/2}_e,
\eqa
\bqa
E^{-1}_{2;11} &=& (H^{-1}_{11} - H^{-1}_{12})\,x_1\,V^{-1-\ep/2}_a +
(H^{-1}_{12} - H^{-1}_{13})\,x_1\,V^{-1-\ep/2}_b
\nl
{}&+& (H^{-1}_{13} - H^{-1}_{14})\,x_1\,V^{-1-\ep/2}_c +
H^{-1}_{14}\,x_1\,V^{-1-\ep/2}_d -
H^{-1}_{11}\,V^{-1-\ep/2}_e,
\nl
E^{-1}_{2;22} &=& (H^{-1}_{21} - H^{-1}_{22})\,x_1\,V^{-1-\ep/2}_a +
(H^{-1}_{22} - H^{-1}_{23})\,x_2\,V^{-1-\ep/2}_b
\nl
{}&+& (H^{-1}_{23} - H^{-1}_{24})\,x_2\,V^{-1-\ep/2}_c +
H^{-1}_{24}\,x_2\,V^{-1-\ep/2}_d -
H^{-1}_{21}\,x_1\,V^{-1-\ep/2}_e,
\nl
E^{-1}_{2;33} &=& (H^{-1}_{31} - H^{-1}_{32})\,x_2\,V^{-1-\ep/2}_a +
(H^{-1}_{32} - H^{-1}_{33})\,x_2\,V^{-1-\ep/2}_b
\nl
{}&+& (H^{-1}_{33} - H^{-1}_{34})\,x_3\,V^{-1-\ep/2}_c +
H^{-1}_{34}\,x_3\,V^{-1-\ep/2}_d -
H^{-1}_{31}\,x_2\,V^{-1-\ep/2}_e,
\nl
E^{-1}_{2;44} &=& (H^{-1}_{41} - H^{-1}_{42})\,x_3\,V^{-1-\ep/2}_a +
(H^{-1}_{42} - H^{-1}_{43})\,x_3\,V^{-1-\ep/2}_b
\nl
{}&+& (H^{-1}_{43} - H^{-1}_{44})\,x_3\,V^{-1-\ep/2}_c -
H^{-1}_{41}\,x_3\,V^{-1-\ep/2}_e,
\nl
2\,E^{-1}_{2;12} &=&
(H^{-1}_{11} - H^{-1}_{22})\,x_1\,V^{-1-\ep/2}_a
+
\Bigl[
(H^{-1}_{12} - H^{-1}_{13})\,x_2 +
(H^{-1}_{22} - H^{-1}_{23})\,x_1
\Bigr]\,V^{-1-\ep/2}_b
\nl
{}&+&
\Bigl[
(H^{-1}_{13} - H^{-1}_{14})\,x_2 +
(H^{-1}_{23} - H^{-1}_{24})\,x_1
\Bigr]\,V^{-1-\ep/2}_c
\nl
{}&+&
  (H^{-1}_{14}\,x_2 + H^{-1}_{24}\,x_1)\,V^{-1-\ep/2}_d
- (H^{-1}_{11}\,x_1 + H^{-1}_{21} )\,V^{-1-\ep/2}_e .
\eqa
All functions $E_{ij}$ contain a term which is proportional to
$E_0(d = 8)$. Consider again $E_{2;11}$ which becomes
\bqa
4\,B_5\,E_{2;11} &=& - E_0(d = 8) + \frac{1}{2}\,\dsimp{3}\,\Bigl[
(X_1 - X_2)\,V^{-1-\ep/2}_a + (X_2 - X_3)\,V^{-1-\ep/2}_b
\nl
{}&+& (X_3 - X_4)\,V^{-1-\ep/2}_c +
X_4\,V^{-1-\ep/2}_d + (1 - X_1)\,V^{-1-\ep/2}_e
\Bigr].
\eqa
When we consider
\bq
E_{\mu\nu} =
\frac{1}{i\pi^2\,\egam{3}}\,
\int\,d^dq\,
\frac{q_{\mu}\,q_{\nu}}{(q^2+m^2_1)\,\cdots\,((q+p_1+\,
\cdots\,+p_{4})^2+m^2_{5})},
\eq
the contribution proportional to $E_0(d = 8)$ is
\bq
E_{\mu\nu} = -\,\frac{E_0(d = 8)}{4\,B_5}\,\Bigl[ \delta_{\mu\nu} +
\sum_{i,j=1}^{4}\,H^{-1}_{ij}\,p_{i\mu}\,p_{j\nu}\Bigr] \quad +
\quad \mbox{form factors} \quad N < 5.
\eq
Since the four-vectors $p_{i\mu}$ span $d = 4$ space-time the term
proportional to $E_0(d = 8)$ disappears and, therefore, there is no need to
compute it.
Indeed let us contract with $p_l, l=1,\dots,4$ to obtain
\bqa
p_{l\nu}\,( \delta_{\mu\nu} +
\sum_{i,j=1}^{4}\,H^{-1}_{ij}\,p_{i\mu}\,p_{j\nu} ) =
p_{l\mu} - \sum_{i,j=1}^{4}\,H^{-1}_{ij}\,p_{i\mu}\,H_{il} = 0.
\eqa
When we go to three powers of $q$ in the numerator the following result is
valid:
\bq
E_{\mu\nu\alpha} = -\,\frac{1}{2\,B_5}\,
\dsimp{4}\,V^{-1-\ep/2}\,
\Bigl[ \sum_{i}
x_i\,\{\delta p_i\}_{\mu\nu\alpha} +
\sum_{i \le j \le l}\,e^3_{ijl}\,\{p_i p_j p_l\}_{\mu\nu\alpha}\Bigr]
\; + \; \mbox{form factors} \; N < 5.
\eq
where the coefficients $e^3$ are
\bq
e^3_{ijl} = H^{-1}_{ij}\,x_l \quad + \quad \mbox{cyclic},
\eq
and where we introduced
\bq
\{\delta p\}_{\mu\nu\alpha} = \delta_{\mu\nu}\,p_{\alpha} +
\delta_{\mu\alpha}\,p_{\nu} +
\delta_{\nu\alpha}\,p_{\mu},
\quad
\{p q k\}_{\mu\nu\alpha} = p_{\mu}\,q_{\nu}\,k_{\alpha} \quad + \quad
\mbox{cyclic},
\eq
showing that $E_0(d = 8)$ disappears again. It is straightforward to extend
the demonstration up to five powers of loop momentum in the numerator. This
is the maximum of non-contracted powers that we can have in a renormalizable
theory and starting from six powers we will have numerators of the following
structure,
\bq
\{\spro{q}{p_i}\,,\,q^2\}\,q_{\mu_1}\cdots q_{\mu_5},
\eq
where scalar products can be simplified according to $q^2 = [1] - m^2_1$ etc.
There is only one case where the argument fails: suppose that one 
external line of momentum $p_i$ splits into two lines of momenta $p_{ij},
j= 1,2$. Then scalar products $\spro{q}{p_{ij}}$ do not occur in any of the
propagators and in the final answer we would end up with evanescent operators
like
\bq
\frac{1}{\ep}\,\delta^{\mu\nu}_{[-\ep]}.
\eq
We now describe the solution around $B_5 = 0$.
\boldmath \subsection{Evaluation of $E_0$ when $B_5 \approx 0$ and $E_0$ is 
regular\label{E0reg}} \unboldmath
If $B_5 \approx 0$ and the condition $0 < X_i < X_{i+1} < 1, i=1,\cdots,4$ 
is not satisfied $E_0$ is regular but the decomposition into a sum of boxes 
fails. We write
\bqa
E_0 &=& \dsimp{4}(X)\,(x^t\,H\,x + B_5)^{-3-\ep/2}.
\eqa
Since $E_0$ is regular at $B_5 = 0$ we perform a Taylor expansion in $B_5$
with coefficients ${\cal E}(n)$,
\bq
E_0 = \frac{1}{2}\,\sum_{n=0}^{\infty}\,(n+1)\,(n+2)\,{\cal E}(n+3)\,
(-\,B_5)^n,
\eq
\bq
{\cal E}(n) = \dsimp{4}(X)\,(x^t\,H\,x)^{-n-\ep/2},
\eq
with $n \ge 3$. Using
\bq
\dsimp{4}(X)\,
\Bigl[ 1 + \frac{x\,\partial_x}{2\,n + \ep}\Bigr]\,(x^t\,H\,x)^{-n-\ep/2} = 0,
\eq
we easily obtain
\bqa
{\cal E}(n) &=& \frac{1}{4 - 2\,n}\,\sum_{i=0}^{4}\,
\dsimp{3}\,( X_i - X_{i+1} )\,Q^{-n-\ep/2}(T\,\widehat{i\;i+1}),
\eqa
where, as usual, we introduced shifted arguments.
If we denote $x^t\,H\,x$ with $Q(x_1,x_2,x_3,x_4)$ the secondary quadrics and
the corresponding coefficients are
\bqa
Q(T\,\widehat{0\;1}) &\equiv& Q_1 = Q(1-X_1,x_1-X_2,x_2-X_3,x_3-X_4),
\nl
Q(T\,\widehat{1\;2}) &\equiv& Q_2 = Q(x_1-X_1,x_1-X_2,x_2-X_3,x_3-X_4),
\nl
Q(T\,\widehat{2\;3}) &\equiv& Q_3 = Q(x_1-X_1,x_2-X_2,x_2-X_3,x_3-X_4),
\nl
Q(T\,\widehat{3\;4}) &\equiv& Q_4 = Q(x_1-X_1,x_2-X_2,x_3-X_3,x_3-X_4),
\nl 
Q(T\,\widehat{4\;5}) &\equiv& Q_5 = Q(x_1-X_1,x_2-X_2,x_3-X_3,-X_4),
\eqa
and $X_0 = 1, X_5 = 0$.
Again, each coefficient in the Taylor expansion is written as a combination
of threefold integrals which can be evaluated with standard BT techniques.
This will introduce sub-leading quadrics, i.e.
\bq
Q_{i1} = Q_i(\widehat{0\;1}) = Q_i(1,x_1,x_2),
\eq
etc, and sub-subleading quadrics, i.e.
\bq
Q_{ij1} = Q_{ij}(\widehat{0\;1}) = Q_{ij}(1,x_1),
\eq
etc, and also constant terms, e.g. \ $Q_{ij1}(1,1)$ etc. At each step 
non-leading BT factors are introduced and the procedure fails when one of 
the sub-leading BT factors is zero. In this case, since $E_0$ is the sum of
$5$ terms of the form
\bq
E^i_0= -\,\frac{1}{4}\,\sum_{n=0}^{\infty}\,(n+2)\,\dsimp{2}\,q_i\,
Q^{-n-3}_i\,(-\,B_5)^n,
\eq
with $q_i$ constant, we rewrite the sum as
\bq
\frac{1}{4}\,\sum_{n=0}^{\infty}\,(n+2)\,Q^{-n-3}_i\,(-\,B_5)^n =
\frac{1}{2}\,\intfx{z}\,(Q_i + B_5\,z)^{-3}.
\eq
Likewise, by changing variables we obtain
\bq
\intfx{z}\,\dsimp{3}\,q_i\,(Q_i + B_5\,z)^{-3} =
\dsimp{3}(X)\,\intfx{z}\,q_i\,(x^t\,H_i\,x + B_{4i} + B_5\,z)^{-3},
\eq
where $B_{4i}$ is the relevant sub-leading BT factor and we can apply a 
Mellin-Barnes splitting, followed by a sector decomposition, to 
$x^t\,H_i\,x$ and $B_{4i} + B_5\,z^2$. For the Mellin-Barnes anti-transform
the leading contribution comes from the pole at $s = 3/2$ giving
\bq
\intfx{z}\,(B_{4i} + B_5\,z)^{-3/2} = -\,\frac{2}{B_5}\,\,\Bigl[
(B_{4i} + B_5)^{-1/2} - B^{-1/2}_{4i}\Bigr].
\eq
Alternatively we define $V_0(x_1,\dots,x_4)= V(x_1,\dots,x_4)-B_5$ and write
down the BT relations corresponding to $B_5 \neq 0$ and $B_5 = 0$:
\bqa
\Big[ 1 + \frac{1}{4+\ep}\,(x-X)\,\partial_x \Big]\,
[V_0(x_1,\dots,x_4)+B_5]^{-2-\ep/2}
&=& B_5\,[V_0(x_1,\dots,x_4)+B_5]^{-3-\ep/2},
\nl
\Big[ 1 + \frac{1}{4+\ep}\,(x-X)\,\partial_x \Big]\,
V_0^{-2-\ep/2}(x_1,\dots,x_4) &=& 0 .
\eqa
Then, after having summed each side of the two equations, we integrate by parts
and set $\ep = 0$, obtaining:
\bq
E_0 =
\frac{1}{4}\,\dsimp{2}\,\sum_{i=0}^4\,(X_i - X_{i+1})\,B^{-1}_5\,
V^{-2}(\widehat{i\,\,i+1})\mid_{\rm sub}.
\label{E0regular:gr}
\eq
Hence the study of the scalar pentagon reduces again to the study of
four-point functions.
The tensor pentagons too can be treated analogously to the case $B_5 \neq 0$.
\boldmath \subsection{Evaluation of $E_0$ when $B_5 \approx 0$ and $E_0$ is
singular\label{E0sing}} \unboldmath
For the pentagon we are in a special situation even if $B_5 \approx 0$
and the condition $X_i < X_{i-1}, i=1,\dots,5$ (with $X_0 = 1$ and
$X_5 = 0$) is satisfied. Indeed, in this case, the BT algorithm already
gives the correct leading behavior, $E_0 \sim 1/B_5$. This is easily seen
by the Mellin-Barnes technique combined with a sector decomposition that
would require to anti-transform the product
\bq
\frac{\egam{s}\,\egam{3-s}}{s-2}\,\rho^{3-s}_5, \qquad \rho_5 = \frac{1}{B_5}.
\eq
Closing the $s$ contour over the right-hand complex half-plane at infinity,
thus selecting the poles at $s = 2$ and $s = 3+k, k \ge 0$, gives $\rho_5$ as 
the leading term.
Therefore, in this case, we simply go on in treating the four-point functions
according to their (sub-leading) BT factors.
\section{\boldmath Infrared divergent $E_0$\label{IRE0} \unboldmath}
For the $E$-family, as long as $B_5 \ne 0$, we can use \eqn{pdec} even for
IR divergent configurations, therefore reducing the problem to the one of 
boxes. If instead $B_5 \approx 0$ another procedure is more convenient. 
The physical relevance of the $E$-family in photonic virtual corrections
to four-fermion production in $e^+e^-$ annihilation, as well as examples of
the corresponding $E_0$ reduction, have been illustrated
in~\cite{Denner:1997ia}.

Let us give an example for $E_0$ where we consider the configuration with
\bq
p^2_1 = p^2_5= - m^2, \quad
m_1 = 0, \quad
m_2 = m_5 = m, \quad
m_3 = m_4 = M.
\label{IRE0conf}
\eq
If we introduce the notation,
\bq
[1] = q^2, \quad
[2] = (q+p_1)^2 + m^2, \quad
[3] = (q+p_1+p_2)^2 + M^2, \quad
\eq
\bq
[4] = (q+p_1+p_2+p_3)^2 + M^2, \quad
[5] = (q-p_5)^2 + m^2,
\eq
then the following decomposition holds
\bq
E^{\rm rem}_0 = -\,\frac{1}{2}\,\Bigl[ (p_1+p_2)^2 + M^2\Bigr]\,
E^{\ssI\ssR}_0 + \frac{1}{2}\,( D^{15}_0 + D^{23}_0 ),
\label{IRdecE0}
\eq
where the IR finite reminder is
\bq
E^{\rm rem}_0 = \frac{1}{i\,\pi^2}\,\int d^dq\,
\frac{\spro{q}{(q+p_1+p_2)}}{[1]\,\cdots\,[5]},
\eq
and where the two $D_0$ functions are specified by their list of arguments:
\bqa
D^{15}_0 &\equiv& D_0(p^2_2\,,\,p^2_3\,,\, p^2_4\,,\,(p_1+p_5)^2\,,\,
(p_2+p_3)^2\,,\,(p_3+p_4)^2\,,\,m\,,\,M\,,\,M\,,\,m),
\nl
D^{23}_0 &\equiv& D_0(-m^2\,,\,(p_2+p_3)^2\,,\,p^2_4\,,\,-m^2\,,\,
(p_4+p_5)^2\,,\,(p_1+p_5)^2\,,\,0\,,\,m\,,\,M\,,\,m).
\eqa
As done in \sect{IRD0} the reminder is computed without simplifications
between numerator and denominator. In particular, if $B_5 \ne 0$ we use the
method of \sect{E0BT}; otherwise, for $B_5 = 0$ and $E^{\rm rem}_0$ regular
(singular) we use the results of \sect{E0reg} (\sect{E0sing}).  The
classification of arbitrary IR divergent $E_0$ functions follows from IR
power counting similarly to the discussion of \sect{classD0IR}. In all cases
the IR divergent part is expressed through $C_0$ functions that are
contained in some of the $D_0$ which appear in decompositions of the type
expressed by \eqn{IRdecE0}.
\section{Six-point functions (F-family)\label{F0}}
To compute the generic, scalar, six-point function~\cite{Binoth:2001vm}
we introduce more compact notations:
\bqa
H^{-1}_{\ssN,ij} &=& \frac{\Delta_{\ssN,ij}}{G_{\ssN}},
\eqa
where $H_{\ssN}$ is the $(N-1)\,\times\,(N-1)$ matrix with elements
$-\,\spro{p_i}{p_j}$, $G_{\ssN} = {\rm det}\,H_{\ssN}$ is its determinant
and $\Delta_{\ssN,ij}$ is the co-determinant of the element $H_{ij}$. The
quadric $V_{\ssN}$ will be written as $V_{\ssN} = x^t\,H_{\ssN}\,x +
2\,K^t_{\ssN}\,x + L_{\ssN}$. It follows that
\bq
B_{\ssN} = \frac{b_{\ssN}}{G_{\ssN}}, 
\quad 
b_{\ssN} = G_{\ssN}\,L_{\ssN} - K^t_{\ssN}\,\Delta_{\ssN}\,K_{\ssN},
\quad
X_{\ssN} = \frac{{\cal X}_{\ssN}}{G_{\ssN}}, 
\quad
{\cal X}_{\ssN} = -\,\Delta_{\ssN}\,K_{\ssN}.
\eq
Furthermore, $L_{\ssN} = m^2_1$ and $K_{\ssN}$ has elements
\bqa
K_{\ssN i} &=& \frac{1}{2}\,\Bigl( [i+1] - [i] \Bigr), \quad
i = 1,\ldots,N-1,
\eqa
where $[1] = m^2_1$ and $[j] = (p_1 + \,\cdots\,+\, p_{j-1})^2 + m^2_j$,
with $j = 2,\ldots,N$.  Since the scalar hexagon is UV finite we can work in
$d = 4$ dimensions where $G_6 = 0$. Therefore, one BT iteration gives
\bqa
b_6\,F_0 = \frac{1}{6}\,\dsimp{4}\,\sum_{i=0}^{5}\,
({\cal X}_i - {\cal X}_{i+1})\,V^{-3-\ep/2}_{6i},
\label{standarhexagon}
\eqa
where ${\cal X}_0 = {\cal X}_6 = 0$ and where
\bqa
V_{60} &=& V(\widehat{0\;1}) = V(1,x_1,\,\cdots\,,x_4),
\nl
V_{61} &=& V(\widehat{1\;2}) = V(x_1,x_1,\,\cdots\,,x_4),
\nl
V_{62} &=& V(\widehat{2\;3}) = V(x_1,x_2,x_2,\,\cdots\,,x_4),
\nl
{}&\cdots&
\nl
V_{65} &=& V(\widehat{5\;6}) = V(x_1,\,\cdots\,,x_4,0),
\eqa
and the scalar hexagon is the sum of six pentagons, i.e.~of $30$ 
boxes~\cite{Fleischer:1999hq}.

In computing the hexagon we have to decide about the input parameter set 
(hereafter IPS). Clearly internal masses, $m_1,\,\cdots\,,m_6$ belong to the 
IPS and for the rest we prefer to use invariants. For a total of $N$ momenta
we have $3\,N - 10$ independent invariants which is considerably less than 
the number of inner products. The energy-momentum conservation leads to a
number of linear relations among the inner products, of which a typical example
is that connecting $s,t,u$ for the box. Secondly, the dimensionality 
constraints lead to quadratic relations. For the hexagon there is just one,
expressed by $G_6 = 0$. The complete set of invariants will be
(remember that in our conventions all momenta are flowing inwards)
$s_{ijk\cdots} = -\,(p_i + p_j + p_k + \,\cdots )^2$,
and we will select $8$ of them for our IPS.
\subsection{Form factors in the F-family}
Consider the integral
\bq
F_{\mu} = \frac{1}{i\pi^2\,\egam{4}}\,\int\,d^dq\,
\frac{q_{\mu}}{(q^2+m^2_1)\,\cdots\,((q+p_1+\,\cdots\,+p_5)^2+m^2_6)} =
 \sum_{i=1}^{5}\,F_{1i}\,p_{i\mu}.
\eq
We immediately obtain
\bq
F_{1i} = \frac{{\cal X}_i}{6\,b_6}\,\dsimp{5}\,V^{-3-\ep/2}_6(\{x\}) +
           F^{4}_{1i},
\eq
where $F^{4}_{1i}$ is a fourfold integral. Therefore there is a piece in
$F_{\mu}$ proportional to $F_0(d = 6)$, where the coefficient is
\bq
\sum_{i=1}^{5}\,{\cal X}_i\,p_{i\mu}.
\eq
In order to prove that this coefficient is zero it is enough to contract it
with a set of vectors $p_l$ that span $d = 4$ space-time.
In this way we obtain
\bq
-\,\sum_{i=1}^{5}\,\Delta_{ij}\,K_j\,\spro{p_i}{p_l} =
\sum_{i=1}^{5}\,\Delta_{ij}\,K_j\,H_{il}.
\eq
Since $\Delta_{ij}$ is the co-determinant of the element $H_{ij}$ we can use
Laplace's theorem, $\Delta_{ij}\,H_{il} = \delta_{jl}\,G$, to obtain
\bq
\sum_{i=1}^{5}\,\Delta_{ij}\,K_j\,H_{il} = G\,K_l = 0,
\eq
due to the vanishing of the Gram's determinant.
For two or more powers of loop momenta in the numerator we need a new relation;
let $d = 4 -\ep$ and consider
\bq
\Bigl( x - \frac{\cal X}{G}\Bigr)\,V^{\mu-1} = \frac{1}{2\,\mu}\,
\frac{\Delta}{G}\,\partial\,V^{\mu},
\qquad
G = \sum_{n=1}^{\infty}\,G^{(n)}\,\ep^n,
\eq
where $\mu$ is arbitrary. We obtain
\bq
{\cal X}\,V^{\mu-1} = - \frac{1}{2\,\mu}\,\Delta\,\partial\,V^{\mu},
\quad
(G^{(1)})^2\,x\,V^{\mu-1} + G^{(2)}\,(
{\cal X}\,V^{\mu-1} + \frac{1}{2\,\mu}\,\Delta\,\partial\,V^{\mu}) = 0,
\label{reduced}
\eq
etc. With the help of \eqn{reduced} we can show that the part proportional to
$F_0(d = 8)$ in $F_{\mu\nu}$ has a coefficient
\bq
\sum_{i,j=1}^{5}\,\Delta_{ij}\,p_{i\mu}\,p_{j\nu},
\eq
which contracted with $p_l,\,l=1,\cdots,5$ gives
\bq
\sum_{i,j=1}^{5}\,\Delta_{ij}\,p_{i\mu}\,\spro{p_j}{p_l} =
-\,\sum_{i,j=1}^{5}\,\Delta_{ij}\,H_{jl}\,p_{i\mu} = -\,G\,p_{l\mu} = 0.
\eq
In computing form factors of the $F$-family we will use results already
obtained for the $E$-family showing that, once more, everything is reducible
to form factors of the $D$-family.  For instance for $F_{\mu}$
we use
\bq
\sum_{i=1}^{5}\,{\cal X}_{ij}\,p_{j\mu} = 0,
\quad
\frac{X_i}{B_6} = \frac{{\cal X}_i}{b_6},
\quad
G_6 = 0,
\eq
and also
\bqa
V^{-4-\ep/2} &=& \frac{1}{B_6}\,
\Bigl[ 1 + \frac{(x-X)\,\partial}{6+\ep}\Bigr]\,V^{-3-\ep/2},
\quad \mbox{or} \quad
V^{-4-\ep/2} = -\,\frac{1}{b_6}\,
\frac{{\cal X}\,\partial}{6+\ep}\,V^{-3-\ep/2}.
\eqa
Furthermore, secondary quadrics are
\bqa
V(\widehat{0\;1}) &=& V(1,x_1,x_2,x_3,x_4)  =
V_1(x_1,x_2,x_3,x_4),
\nl
V(\widehat{1\;2}) &=& V(x_1,x_1,x_2,x_3,x_4) =
V_2(x_1,x_2,x_3,x_4),
\nl
V(\widehat{2\;3}) &=& V(x_1,x_2,x_2,x_3,x_4) =
V_3(x_1,x_2,x_3,x_4),
\nl
V(\widehat{3\;4}) &=& V(x_1,x_2,x_3,x_3,x_4) =
V_4(x_1,x_2,x_3,x_4),
\nl
V(\widehat{4\;5}) &=& V(x_1,x_2,x_3,x_4,x_4) =
V_5(x_1,x_2,x_3,x_4),
\nl
V(\widehat{5\;6}) &=& V(x_1,x_2,x_3,x_4,0)  =
V_6(x_1,x_2,x_3,x_4),
\eqa
to which we apply a standard BT algorithm
\bqa
V^{-3-\ep/2}_i &=& \frac{1}{B_{5i}}\,\Bigl[
1 + \frac{1}{4+\ep}\,(x_j - X_{ij})\,\partial_j\Bigr]\,V^{-2-\ep/2}_i,
\quad i=1,\cdots,6,
\eqa
or a special BT algorithm,
\bqa
x_j\,V^{-2-\ep/2}_i &=& X_{ij}\,V^{-2-\ep/2}_i -
\frac{1}{2+\ep}\,(H_i)^{-1}_{jl}\,\partial_l\,V^{-1-\ep/2}_i,
\eqa
to show that $F_{\mu}$ is given by a combination of $813$ threefold integrals,
i.e. form factors of the $D$-family.
\boldmath \subsection{Special configurations for $F_0$} \unboldmath
If we have $b_6 = -\,K^t\,\Delta\,K = 0$ on top of $G_6 = 0$ then
\eqn{standarhexagon} cannot be applied. In this case we can proceed as follows:
since the vectors $p_i$ are linearly dependent, we substitute
\bq
p_{5\mu} = -\,\sum_{i=1}^{4}\,\frac{{\cal X}_i}{{\cal X}_5}\,p_{i\mu},
\eq
into the expression of $V$ and show that
\bq
\sum_{i,j=1}^{5}\,H_{ij}\,x_i\,x_j = \sum_{i,j=1}^{4}\,H^{(4)}_{ij}\,
X_i\,X_j, \qquad
X_i = x_i - \frac{{\cal X}_i}{{\cal X}_5}\,x_5,
\eq
where $H^{(4)}$ is a $4\,\times\,4$ matrix with elements $-\,\spro{p_i}{p_j}$.
Likewise we have
\bq
\sum_{i=1}^{5}\,K_i\,x_i =
\sum_{i=1}^{4}\,K_i\,X_i + \frac{b_6}{{\cal X}_5}\,x_5.
\eq
If we define
\bq
L_i = ( 1- r_i )\,x_5 = \Bigl( 1 - \frac{{\cal X}_i}{{\cal X}_5}\Bigr)\,x_5,
\qquad
U_i = x_{i-1} + ( r_{i-1} - r_i )\,x_5,
\eq
with $x_0 = 1$ and $r_0 = 0$, the expression for $F_0$ can be cast into the
following form:
\bq
F_0 = \intfx{x_5}\,{\cal F}_0(x_5),
\quad
{\cal F}_0(x_5) = \prod_{i=1}^{4}\,\int_{\ssL_i}^{\ssU_i}\,dx_i\,
\Bigl( Q_4 + 2\,\frac{b_6}{{\cal X}_5}\,x_5\Bigr)^{-4-\ep/2},
\label{BTfnads}
\eq
where $Q_4$ is a quadric in four variables, i.e.
\bq
Q_4 = \sum_{i,j=1}^{4}\,H^{(4)}_{ij}\,x_i\,x_j + 2\,\sum_{i=1}^{4}\,
K_i\,x_i + L.
\eq
To this quadric is associated a BT factor $B_4$ so that we can apply the BT
algorithm to \eqn{BTfnads} with a total BT factor $B_4 + 2\,b_6/{\cal
X}_5\,x_5$.
\section{Conclusions}
In this paper we presented a detailed investigation of the algorithms, based
on the Bernstein-Tkachov theorem~\cite{Tkachov:1997wh}, which form the basis
for a fast and reliable numerical integration of one-loop multi-leg (up to
six in this paper) diagrams. The rationale for this work is represented by 
the need of encompassing a number of problems that one encounters in 
assembling a calculation of some complicated process, e.g. full one-loop 
corrections to $e^+e^- \to 4\,$fermions. Furthermore, in any attempt to 
compute physical observables at the two-loop level, we will have to include 
the one-loop part, and it is rather obvious that the two pieces should be 
treated on equal footing. 

Finally, our method represents a new strategy for the
so-called problem of reduction of tensor integrals (for additional
alternatives we refer to~\cite{Fleischer:1999hq},\cite{Binoth:1999sp} and
\cite{Bauer:2001ig}).

All algorithms that aim to compute Feynman diagrams numerically are based on
some manipulation of the original integrands that brings the final answer
into something smooth. This has the consequence of bringing the original
(Landau) singularity of the diagram into some overall denominator and,
usually, the method overestimates the singular behavior around some
threshold. In these regions an alternative derivation is needed. Instead of
using the method of asymptotic expansions~\cite{aexp}, we introduced a novel
algorithm based on a Mellin-Barnes decomposition of the singular integrand,
followed by a sector decomposition that allows us to write the Laurent
expansion around threshold. 

Particular care has been devoted to analyze those situations where a 
sub-leading singularity may occur, and to properly account for those cases 
where the algorithm cannot be applied because the corresponding BT factor 
is zero although the singular point in parametric space does not belong to 
the integration domain.

Clearly, no numerical evaluation of Feynman diagrams should be attempted unless
we have a detailed knowledge of their analytic structure; in other words, we
must know beforehand where the real singularities are sitting, which
apparent singularity can be the origin of numerical instabilities, and what
to do when one of these cases is met.

One of the main by-products of the BT approach, \eqn{functr}, is that any 
diagram $G$ has an associated factor $B_{\ssG}$ which is immediately computed 
for any one-loop diagram and gives information about the leading solution of 
the corresponding set of Landau equations (these solutions are notoriously hard
to derive with standard methods~\cite{Wu:fr}). Technically speaking, 
$B_{\ssG} = 0$ guarantees that the Landau equations admit a proper solution.  
Repeated applications of the algorithm, as explained for instance in
\sect{regularD0}, will also introduce sub-leading
$B_{\ssG|\ssS\ssL}$-factors whose zeros correspond to sub-leading Landau
singularities. Likewise, our solution around $B_{\ssG} = 0$ will introduce
sub-leading $B$-coefficients of a second kind (as explained, for instance, in
the discussion after \eqn{seeeq}), not associated with Landau
singularities of the reduced diagrams.

When these pieces of knowledge are at hand, we can automatize
our calculations, even without having to worry in advance about the physical
nature~\cite{Kershaw:1971rc} of the singular points (a much more difficult 
assignment): it will be enough to be able to handle all cases and let the 
program decide when $B_{\ssG} = 0$ corresponds to a singular behavior or
when a coincidence occurs. 
Furthermore, we have a conjecture stating that when $B_{\ssG} = 0$, then the 
simultaneous occurrence of $B_{\ssG|\ssS\ssL} = 0$ is not physical, but we 
have not been able to prove it. Instead we decided to cover also this case, 
so that our expressions for the one-loop diagrams are quite general and not 
limited to the physical region of a specific process.

In general it is known that when a given Landau singularity curve touches one 
of its sub-leading singularity curves the determination of the nature of the 
singularity breaks down completely (because the $\alpha_i$'s are no longer
uniquely determined) and changes suddenly and drastically. Similarly, our 
analysis via the multiple Mellin - Barnes techniques (cf. 
\eqns{resI}{resII} and \eqn{resIII}), dealing with the coincidence 
$B_{\ssG} = B_{\ssG|\ssS\ssL} = 0$ (where $B_{\ssG|\ssS\ssL}$ is of 
second kind), represents a novel result.
Alternatively, we have been able to derive new integral representations that
bypass the use of Laurent expansions.

Finally we have given a description of infrared divergent one-loop virtual
configurations in the framework of dimensional regularization: here both the 
residue of the infrared pole and the infrared finite remainder are cast into 
a form that can be safely computed numerically.

The collection of formulas that cover all corners of phase space have been
translated into a set of FORM codes~\cite{Vermaseren:2000nd} and the output 
has been used to create a FORTRAN code whose technical description will be 
given elsewhere. 
\Acknowledgments
For the idea of presenting a comprehensive discussion of all algorithms
needed in the numerical evaluation of one-loop diagrams we gratefully
acknowledge the incitation of Ansgar Denner, Stefan Dittmaier and
Stanislaw Jadach. G.~P.~would like to express his gratitude to Fyodor Tkachov 
for important discussions on the general idea of evaluating multi-loop 
Feynman diagrams numerically, and to recognize the critical spur of Dima 
Bardin. M.~P.~would like to thank J.~Gasser for enlightening discussions
on the analytic properties of Feynman diagrams.
\clearpage
\appendix
\section{A master integral}
In this appendix we calculate the master integral 
\bq
I_{(\alpha,\beta)}(a,b) =
\intfx{x}\,x^{\beta}\,(a\,x^2 + b - i\delta)^{-\alpha},
\eq
where $a$ and $b$ are real numbers with $b$ approaching $0$ and $a$ far from 
$0$. If $\beta> 2\,\alpha-1$ the integral is not divergent, otherwise we 
split the integral into two pieces: the first one contains the divergent 
part as $b \to 0$, while the second is finite:
\bq
I_{(\alpha,\beta)}(a,b) =
\int_0^\infty\,dx\,x^\beta\,(a\,x^2 + b - i\delta)^{-\alpha}
- \int_1^\infty\,dx\,x^\beta\,(a\,x^2 + b - i\delta)^{-\alpha}.
\eq
Since we are in the limit of $\delta \to 0$, we may write
$ a\,x^2 + b - i\delta = (a-i\delta)\,x^2 + b - i\delta$.
Next we perform the following transformations in the two integrals 
respectively: $x \to x^{1/2}$ and $x \to x^{-1}$ and obtain
\bq
I_{(\alpha,\beta)}(a,b) =
\frac{1}{2}\,\int_0^\infty\,dx\,x^{\frac{\beta-1}{2}}\,
\Big[ (a-i\delta)\,x + b - i\delta \Big]^{-\alpha}
- \intfx{x}\,x^{2\,\alpha-\beta-2}\,
\Big[ a - i\delta + (b-i\delta)\,x^2 \Big]^{-\alpha}.
\eq
In the first integral an additional change of variable is performed:
\bq
x \to \frac{b-i\delta}{a-i\delta}\,\frac{1-x}{x},
\eq
and we make use of the following property:
\bq
\left( \frac{b-i\delta}{a-i\delta} \right)^{\mu}=
\frac{(b-i\delta)^{\mu}}{(a-i\delta)^{\mu}} ~.
\eq
Hence the final result is:
\bq
I_{(\alpha,\beta)}(a,b) =
\frac{(b-i\delta)^{-\alpha+\frac{\beta+1}{2}}}
     {2\,(a-i\delta)^{\frac{\beta+1}{2}}}\,
B\left( \frac{\beta+1}{2},\alpha-\frac{\beta+1}{2} \right)
- \intfx{x}\,x^{2\,\alpha-\beta-2}\,
\Big[ a - i\delta + (b-i\delta)\,x^2 \Big]^{-\alpha}.
\eq
The second term is well defined only for $\beta < 2\,\alpha-1$ and
we perform an analytic continuation to the region $\beta < 2\,\alpha+1$
(with $a,b \to a,b - i\,\delta$):
\bqa
I_{(\alpha,\beta)}(a,b) &=&
\frac{b^{-\alpha+\frac{\beta+1}{2}}}
     {2\,a^{\frac{\beta+1}{2}}}\,
B\left( \frac{\beta+1}{2},\alpha-\frac{\beta+1}{2} \right)\,
- \,\frac{a^{-\alpha}}{2\,\alpha-\beta-1} - \,\intfx{x}\,x^{2\,\alpha-\beta}\,
\Bigg\{
\frac{[ a + b\,x^2 ]^{-\alpha}}{x^2}
\Bigg\}_{++}.
\label{eqapp}
\eqa From \eqn{eqapp} we see that the two divergent terms cancel out
also when $\beta = 2\,\alpha -1$. For $\alpha = n$ and $\beta = 2\,n-1$ 
we get
\bq
I_{(\alpha,\beta)}(a,b) =
- \frac{a^{-n}}{2}\,
\Big( \ln \frac{b}{a} + \sum_{j=1}^{n-1}\,\frac{1}{j} \Big)
- \,\intfx{x}\,
\Bigg\{ \frac{[ a + b\,x^2 ]^{-n}}{x} \Bigg\}_+.
\eq
\clearpage
\begin{figure}[th]
\bqas
  \vcenter{\hbox{
  \begin{picture}(80,50)(70,50)
  \SetScale{0.4}
  \SetWidth{2.}
  \Line(100,0)(20,97.98)
  \Line(20,97.98)(-80,60)
  \Line(-80,60)(-80,-60)
  \Line(-80,-60)(20,-97.98)
  \Line(20,-97.98)(100,0)
  \Line(-80,60)(-160,60)
  \Line(-80,-60)(-160,-60)
  \Line(20,97.98)(100,97.98)
  \Line(20,-97.98)(100,-97.98)
  \Line(100,0)(180,0)
  \Text(110,-4)[cb]{=}
  \end{picture}}}
&{}&
  \vcenter{\hbox{
  \begin{picture}(-30,50)(-50,50)
  \SetScale{0.4}
  \SetWidth{2.}
  \Line(100,0)(-80,60)
  \Line(-80,60)(-80,-60)
  \Line(-80,-60)(20,-97.98)
  \Line(20,-97.98)(100,0)
  \Line(100,0)(180,0)
  \Line(20,-97.98)(100,-97.98)
  \Line(-80,-60)(-160,-60)
  \Line(-80,60)(-160,60)
  \Line(-80,60)(100,97.98)
  \Text(0,-4)[cb]{[1]}
  \Text(110,-4)[cb]{+}
  \end{picture}}}
\eqas
\end{figure}
\vspace{0.8cm}
\begin{figure}[th]
\bqas
  \vcenter{\hbox{
  \begin{picture}(80,50)(70,50)
  \SetScale{0.4}
  \SetWidth{2.}
  \Line(100,0)(20,97.98)
  \Line(20,97.98)(-80,0)
  \Line(-80,-0)(20,-97.98)
  \Line(20,-97.98)(100,0)
  \Line(-80,0)(-160,60)
  \Line(-80,0)(-160,-60)
  \Line(20,97.98)(100,97.98)
  \Line(20,-97.98)(100,-97.98)
  \Line(100,0)(180,0)
  \Text(110,-4)[cb]{+}
  \Text(10,-4)[cb]{[2]}
  \end{picture}}}
&{}&
  \vcenter{\hbox{
  \begin{picture}(-30,50)(-50,50)
  \SetScale{0.4}
  \SetWidth{2.}
  \Line(100,0)(20,97.98)
  \Line(20,97.98)(-80,60)
  \Line(-80,60)(-80,-60)
  \Line(-80,-60)(100,0)
  \Line(-80,60)(-160,60)
  \Line(-80,-60)(-160,-60)
  \Line(100,0)(180,0)
  \Line(20,97.98)(100,97.98)
  \Line(-80,-60)(100,-97.98)
  \Text(0,-4)[cb]{[3]}
  \Text(110,-4)[cb]{+}
  \end{picture}}}
\eqas
\end{figure}
\vspace{0.8cm}
\begin{figure}[th]
\bqas
  \vcenter{\hbox{
  \begin{picture}(80,50)(70,50)
  \SetScale{0.4}
  \SetWidth{2.}
  \Line(20,-97.98)(20,97.98)
  \Line(20,97.98)(-80,60)
  \Line(-80,60)(-80,-60)
  \Line(-80,-60)(20,-97.98)
  \Line(-80,60)(-160,60)
  \Line(-80,-60)(-160,-60)
  \Line(20,97.98)(100,97.98)
  \Line(20,-97.98)(100,-97.98)
  \Line(20,-97.98)(180,0)
  \Text(110,-4)[cb]{+}
  \Text(-10,-4)[cb]{[4]}
  \end{picture}}}
&{}&
  \vcenter{\hbox{
  \begin{picture}(-30,50)(-50,50)
  \SetScale{0.4}
  \SetWidth{2.}
  \Line(20,-97.98)(20,97.98)
  \Line(20,97.98)(-80,60)
  \Line(-80,60)(-80,-60)
  \Line(-80,-60)(20,-97.98)
  \Line(-80,60)(-160,60)
  \Line(-80,-60)(-160,-60)
  \Line(20,97.98)(100,97.98)
  \Line(20,-97.98)(100,-97.98)
  \Line(20,+97.98)(180,0)
  \Text(-10,-4)[cb]{[5]}
  \end{picture}}}
\eqas
\vspace{4cm}
\caption[]{Diagrammatical representation of the BT algorithm of \eqn{pdec}
for the pentagon. The symbol $[i]$ denotes multiplication of the corresponding
box by a factor $w_i/(4\,B_5)$.\label{fig:pentagon}}
\end{figure}
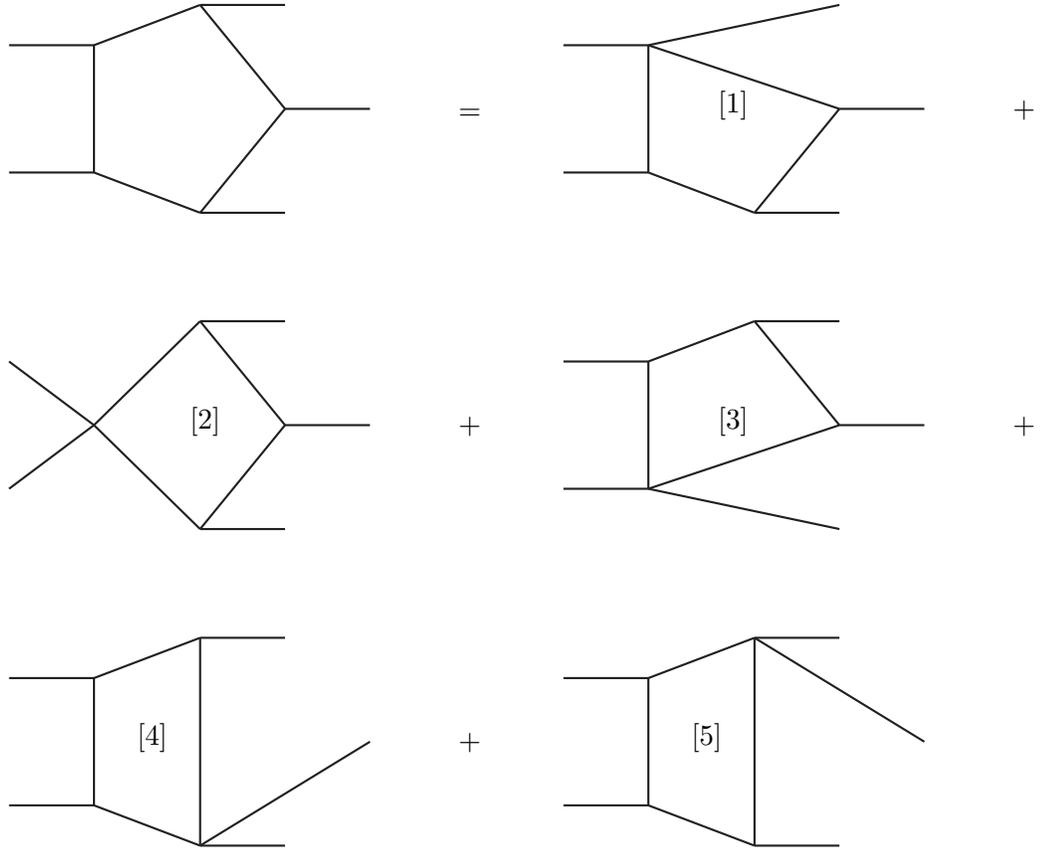


\clearpage

\begin{thebibliography}{99}

\bibitem{'tHooft:1979xw}
G.~'t Hooft and M.~Veltman,
Nucl.\ Phys.\ B {\bf 153} (1979) 365.

\bibitem{Kotikov:1990kg}
A.V.~Kotikov,
Phys.\ Lett.\ B {\bf 254} (1991) 158;\\
A.V.~Kotikov,
Phys.\ Lett.\ B {\bf 267} (1991) 123.

\bibitem{Landau:1959fi}
L.D.~Landau,
Nucl.\ Phys.\  {\bf 13} (1959) 181.

\bibitem{elop}
R.J.~Eden, P.V.~Landshoff, D.I.~Olive, and J.C.~Polkinghorne,
{\it The Analytic S-Matrix}, Cambridge Univ. Press, 1966.

\bibitem{vanNeerven:1983vr}
W.L.~van Neerven and J.A.~Vermaseren,
Phys.\ Lett.\ B {\bf 137} (1984) 241;\\
Z.~Bern, L.J.~Dixon and D.A.~Kosower,
Nucl.\ Phys.\ B {\bf 412} (1994) 751
[arXiv:hep-ph/9306240];\\
A.I.~Davydychev,
J.\ Math.\ Phys.\  {\bf 33} (1992) 358.

\bibitem{cac}
A.~Vicini, talk given at Loopfest, Brookhaven National Laboratory, Upton, 
NY, May 9-10, 2002 (http://www.pas.rochester.edu/~dow/loopfest/parallel.html).

\bibitem{Beenakker:1996kn}
W.~Beenakker {\it et al.},
Nucl.\ Phys.\ B {\bf 500} (1997) 255
[arXiv:hep-ph/9612260].

\bibitem{Passarino:1979jh}
G.~Passarino and M.~Veltman,
Nucl.\ Phys.\ B {\bf 160} (1979) 151;\\
D.~Bardin and G.~Passarino,
{\it The standard model in the making: Precision study of the electroweak
interactions},
{\it  Oxford, UK: Clarendon (1999)}.

\bibitem{Passarino:2001wv}
G.~Passarino,
Nucl.\ Phys.\ B {\bf 619} (2001) 257
[arXiv:hep-ph/0108252];\\
G.~Passarino and S.~Uccirati,
Nucl.\ Phys.\ B {\bf 629} (2002) 97
[arXiv:hep-ph/0112004].

\bibitem{Bardin:2000cf}
D.Y.~Bardin, L.V.~Kalinovskaya and F.V.~Tkachov,
hep-ph/0012209;\\
G.~Passarino,
hep-ph/0101299.

\bibitem{Tkachov:1997wh}
F.V.~Tkachov,
Nucl.\ Instrum.\ Meth.\ A {\bf 389} (1997) 309
[hep-ph/9609429];\\
L.N.~Bertstein, Functional Analysis and its Applications, {\bf 6}(1972)66.

\bibitem{ellip} A.~Erdelyi et al., {\it Higher Transcendental Functions}, 
vol.~2, Bateman Manuscript Project, McGraw-Hill, 1953;\\
L.J.~Slater, {\it Generalized Hypergeometric Functions},
Cambridge Univ. Press, 1966.

\bibitem{Binoth:2000ps}
T.~Binoth and G.~Heinrich,
Nucl.\ Phys.\ B {\bf 585} (2000) 741
[arXiv:hep-ph/0004013];\\
M.~Roth and A.~Denner,
Nucl.\ Phys.\ B {\bf 479} (1996) 495
[arXiv:hep-ph/9605420];\\
K.~Hepp,
Commun.\ Math.\ Phys.\  {\bf 2} (1966) 301.

\bibitem{Fujimoto:1991bm}
J.~Fujimoto, Y.~Shimizu, K.~Kato and Y.~Oyanagi,
Prog.\ Theor.\ Phys.\  {\bf 87} (1992) 1233;\\
J.~Fujimoto, Y.~Shimizu, K.~Kato, N.~Nakazawa and T.~Kaneko, prepared for  
{\it 10th International Workshop on High-energy Physics and 
Quantum Field Theory (NPI MSU 95)}, Zvenigorod, Russia, 19--26 Sep 1995;\\
F.~Caravaglios,
Nucl.\ Phys.\ B {\bf 589} (2000) 475
[arXiv:hep-ph/0004030];\\
R.~Easther, G.~Guralnik and S.~Hahn,
Phys.\ Rev.\ D {\bf 61} (2000) 125001
[arXiv:hep-ph/9903255];\
M.~Steinhauser,
Phys.\ Rept.\  {\bf 364} (2002) 247
[arXiv:hep-ph/0201075].

\bibitem{Montagna:1993ai}
G.~Montagna, F.~Piccinini, O.~Nicrosini, G.~Passarino and R.~Pittau,
Comput.\ Phys.\ Commun.\  {\bf 76} (1993) 328;\\
G.~Montagna, F.~Piccinini, O.~Nicrosini, G.~Passarino and R.~Pittau,
Nucl.\ Phys.\ B {\bf 401} (1993) 3.

\bibitem{Boos:1990rg}
E.E.~Boos and A.I.~Davydychev,
Theor.\ Math.\ Phys.\  {\bf 89} (1991) 1052
[Teor.\ Mat.\ Fiz.\  {\bf 89} (1991) 56];\\
A.I.~Davydychev and J.B.~Tausk,
Nucl.\ Phys.\ B {\bf 397} (1993) 123;\\
S.~Bauberger, F.A.~Berends, M.~Bohm and M.~Buza,
Nucl.\ Phys.\ B {\bf 434} (1995) 383
[arXiv:hep-ph/9409388];\\
A.I.~Davydychev and J.B.~Tausk,
Phys.\ Rev.\ D {\bf 53} (1996) 7381
[arXiv:hep-ph/9504431].

\bibitem{oldies}
R.~Karplus, C.~M.~Sommerfield and E.~H.~Wichmann, 
Phys.\ Rev. {\bf 111} (1958) 1187;\\
G.~K\"allen and A.~S.~Wightman,
K. Dan. Vidensk. Selsk. Mat. - Fys. Skr. {\bf 1} N0. 6 (1958).

\bibitem{Cutkosky:1960sp}
R.~E.~Cutkosky,
J.\ Math.\ Phys.\  {\bf 1} (1960) 429.

\bibitem{Devaraj:1997es}
G.~Devaraj and R.G.~Stuart,
Nucl.\ Phys.\ B {\bf 519} (1998) 483
[arXiv:hep-ph/9704308].

\bibitem{polyl} L.~Lewin, {\it Poly-logarithms and Associated Functions},
North Holland (New York 1981).

\bibitem{Beenakker:1988jr}
W.~Beenakker and A.~Denner,
Nucl.\ Phys.\ B {\bf 338} (1990) 349.

\bibitem{Abarbanel:1969ek}
H.D.~Abarbanel and C.~Itzykson,
Phys.\ Rev.\ Lett.\  {\bf 23} (1969) 53;\\
M.~Levy and J.~Sucher,
Phys.\ Rev.\  {\bf 186} (1969) 1656.

\bibitem{Sterman:ce}
G.~Sterman, {\it An Introduction to Quantum Field Theory},
Cambridge Univ. Press, 1993.

\bibitem{Fleischer:1999hq}
J.~Fleischer, F.~Jegerlehner and O.V.~Tarasov,
Nucl.\ Phys.\ B {\bf 566} (2000) 423
[arXiv:hep-ph/9907327];\\
O.~V.~Tarasov,
Acta Phys.\ Polon.\ B {\bf 29} (1998) 2655
[arXiv:hep-ph/9812250];\\
Z.~Bern, L.J.~Dixon and D.A.~Kosower,
Nucl.\ Phys.\ B {\bf 412} (1994) 751
[arXiv:hep-ph/9306240].

\bibitem{Binoth:1999sp}
T.~Binoth, J.P.~Guillet and G.~Heinrich,
Nucl.\ Phys.\ B {\bf 572} (2000) 361
[arXiv:hep-ph/9911342];\\
Z.~Bern, L.~J.~Dixon and D.~A.~Kosower,
Phys.\ Rev.\ Lett.\  {\bf 70} (1993) 2677
[arXiv:hep-ph/9302280].

\bibitem{Denner:1997ia}
A.~Denner, S.~Dittmaier and M.~Roth,
Nucl.\ Phys.\ B {\bf 519} (1998) 39
[arXiv:hep-ph/9710521].

\bibitem{Binoth:2001vm}
T.~Binoth, J.P.~Guillet, G.~Heinrich and C.~Schubert,
Nucl.\ Phys.\ B {\bf 615} (2001) 385
[arXiv:hep-ph/0106243].

\bibitem{Bauer:2001ig}
C.~Bauer and H.S.~Do,
Comput.\ Phys.\ Commun.\  {\bf 144} (2002) 154
[arXiv:hep-ph/0102231];\\
S.~Weinzierl,
Phys.\ Lett.\ B {\bf 450} (1999) 234
[arXiv:hep-ph/9811365];\\
A.I.~Davydychev,
Phys.\ Lett.\ B {\bf 263} (1991) 107.

\bibitem{aexp} F.V.~Tkachov, in {\it Quarks-82 Proc.~Int.~Seminar},
Eds: A.N.~Tavkhelidze et al. INR, the USSR Acad.~Sci.~Moscow, 1982;
{\it Euclidean asymptotic of Feynman integrals}, preprint INR P-332
(Moscow 1984); Int.~J.~Mod.~Phys. A8 (1993) 2047; 
Sov.~J.~Part.~Nuclei 25 (1994) 649.

\bibitem{Wu:fr}
A.~C.~Wu,
Phys.\ Rev.\ D {\bf 9} (1974) 370.

\bibitem{Kershaw:1971rc}
D.~Kershaw,
Phys.\ Rev.\ D {\bf 5} (1972) 1976.

\bibitem{Vermaseren:2000nd}
J.A.~Vermaseren,
math-ph/0010025.

\end{thebibliography}
\end{document}